\documentclass[twocolumn, trackchanges, times]{aastex62}

\usepackage{amsmath,amstext}
\usepackage[T1]{fontenc}
\usepackage{xcolor}
\usepackage[figure,figure*]{hypcap}
\usepackage{multirow}
\usepackage{outlines}
\newcommand{\trone}{\object{TRAPPIST-1}}
\newcommand{\first}{1}
\newcommand{\second}{2}

\received{Jan. 15, 2018}
\revised{Aug. 28, 2018}
\accepted{Aug. 29, 2018}

\shorttitle{Transmission Spectra of the TRAPPIST-1 Planets}
\shortauthors{Zhang Z. et al.}

\begin{document}

\title{The Near-Infrared Transmission Spectra of TRAPPIST-1 Planets b, c, d, e, f, and g and Stellar Contamination in Multi-Epoch Transit Spectra}

\author{Zhanbo Zhang}
\affiliation{Department of Astronomy/Steward Observatory, The University of Arizona, 933 N. Cherry Avenue, Tucson, AZ 85721, USA}
\affiliation{School of Physics, Peking University, Yiheyuan Rd. 5, Haidian District, Beijing 100871, China}
\author{Yifan Zhou}
\affiliation{Department of Astronomy/Steward Observatory, The University of Arizona, 933 N. Cherry Avenue, Tucson, AZ 85721, USA}
\author{Benjamin V. Rackham}
\affiliation{Department of Astronomy/Steward Observatory, The University of Arizona, 933 N. Cherry Avenue, Tucson, AZ 85721, USA}
\affiliation{National Science Foundation Graduate Research Fellow.}
\affiliation{Earths in Other Solar Systems Team, NASA Nexus for Exoplanet System Science.}
\author{D\'aniel Apai}
\affiliation{Department of Astronomy/Steward Observatory, The University of Arizona, 933 N. Cherry Avenue, Tucson, AZ 85721, USA}
\affiliation{Earths in Other Solar Systems Team, NASA Nexus for Exoplanet System Science.}
\affiliation{Department of Planetary Sciences, The University of Arizona, 1629 E. University Blvd, Tucson, AZ 85721, USA}
\affiliation{Max Planck Institute for Astronomy, K\"onigstuhl 17, Heidelberg, D-69117, Germany}

\begin{abstract}
The seven approximately Earth-sized transiting planets in the \object{TRAPPIST-1} system provide a unique opportunity to explore habitable zone and non-habitable zone small planets within the same system. 
Its habitable zone exoplanets -- due to their favorable transit depths -- are also worlds for which atmospheric transmission spectroscopy is within reach with the Hubble Space Telescope (HST) and with the James Webb Space Telescope (JWST). 
We present here an independent reduction and analysis of two \textit{HST} Wide Field Camera 3 (WFC3) near-infrared transit spectroscopy datasets for six planets (b through g).  
Utilizing our physically-motivated detector charge trap correction and a custom cosmic ray correction routine, we confirm the general shape of the transmission spectra presented by \citet{deWit2016, deWit2018}. 
Our data reduction approach leads to a 25\% increase in the usable data and reduces the risk of confusing astrophysical brightness variations (e.g., flares) with instrumental systematics. 
No prominent absorption features are detected in any individual planet's transmission spectra; by contrast, the combined spectrum of the planets shows a suggestive decrease around 1.4\,$\micron$ similar to an inverted water absorption feature.
Including transit depths from \textit{K2}, the SPECULOOS-South Observatory, and \textit{Spitzer}, we find that the complete transmission spectrum is fully consistent with stellar contamination owing to the transit light source effect.
These spectra demonstrate how stellar contamination can overwhelm planetary absorption features in low-resolution exoplanet transit spectra obtained by \textit{HST} and \textit{JWST} and also highlight the challenges in combining multi-epoch observations for planets around rapidly rotating spotted stars.
\end{abstract}
\keywords{techniques: spectroscopic -- planets and satellites: atmospheres -- stars: late-type -- planets and satellites: individual (TRAPPIST-1) -- planets and satellites: terrestrial planets }

\section{Introduction}

The \trone{} system (\object{2MASSI J23062928-0502285}, \object{2MUCD 12171}) hosts seven known, nearly earth-sized transiting exoplanets \citep[][]{Gillon2016, Gillon2017}. Four of these planets (b, c, d, e) are in or near the liquid water habitable zone \citep[e.g.,][]{Wolf2017,Alberti2017}, although the stellar activity and ultraviolet-radiation of the star \citep[e.g.][]{O'Malley-JamesKaltenegger2017, Bourrier2017_Lya} as well as the formation and initial volatile budget \citep[e.g.,][]{Ciesla2015,Ormel2017} and subsequent volatile loss of the planets \citep[e.g.,][]{Bourrier2017AJ} remain concerns for their habitability. The \trone{} host star -- an M8-type ultracool dwarf at the stellar/sub-stellar boundary -- has a very small radius ($R_*\sim1.14\pm0.04~R_{\mathrm{Jup}}=0.117 \pm 0.004~R_\odot$, \citealt[][]{Filipazzo2015}), leading to exceptionally deep transit depths (0.3-0.8\%) for its small planets. These favorable transit depths, in combination with the relatively bright host star (V=18.8, but J=11.35) and the frequent transits (planet orbital periods between 1.6 and 15 days, \citealt[][]{Gillon2017}), make the \trone{} planetary system exceptionally well suited for follow-up infrared transit spectroscopy. Of particular importance for such observations are photometrically very stable and sensitive infrared space telescopes: the Hubble Space Telescope ($HST$) and the James Webb Space Telescope ($JWST$). High-precision spectroscopy with these facilities may be able to probe atmospheric composition (gas-phase absorbers: O$_3$, scattering, and particulates) in the inner \trone{} planets, including those in the habitable zone \citep[][]{BarstowIrwin2016,Morley2017}.

These studies find that the most prominent absorption features that may be present and detectable in these atmospheres are water, ozone, and carbon-dioxide absorption bands. While the detection of one or more of the features could distinguish between Earth, Venus, or Titan-like atmospheres \citep[][]{Morley2017}, even the lack of features may be interesting: stringent non-detections of absorption features could be interpreted as lack of stratospheric water \citep[e.g.,][]{Madhusudhan2014}, the veiling effect of high-altitude hazes \citep[][]{Kreidberg2014,Yang2015}, or the lack of a significant atmosphere. Recent ambitious \textit{HST} transmission spectroscopy programs showed encouraging results demonstrating that instrumental systematics can be successfully corrected even for very long combined integrations (i.e., very low photon noise) \citep[e.g.][]{Kreidberg2014,Stevenson2014,Morley2017}. 

With the \textit{JWST} guaranteed time observations and early release science observations determined and the community working on the \textit{JWST} Cycle-1 open time proposals, the assessment of the feasibility of the photon-noise limited transit spectroscopy with \textit{HST} and \textit{JWST} is of paramount importance for the field.

Over the past months two new results impact extrapolations from past \textit{HST} programs toward future, even more ambitious \textit{HST} and \textit{JWST} programs. \citet[][]{Zhou2017} demonstrated solid state-physics-based correction algorithm for the \textit{HST} charge trapping processes that introduce the so-called "ramp effect", the dominant \textit{HST} systematics in time-resolved observations. This model offers a more efficient use of the telescope and enables observers to correct for different systematics occurring in different orbits, in contrast to the previously widely utilized empirical correction that assumed identical systematics in all orbits beyond the first. This model has recently begun to be applied in \textit{HST}/WFC3 transmission spectroscopic studies \citep[e.g.,][]{Spake2018}.

However, new study by \citet[][]{Rackham2018} highlighted a major {\em astrophysical} noise source: these authors provided a comprehensive exploration of the impact of stellar heterogeneity on high-precision near-infrared spectroscopy of M-dwarf transiting planets and show that this method may ultimately be limited by the fact that heterogeneous stellar photospheres introduce a spectral contamination into the transmission spectra (i.e., the ``transit light source effect''). In fact, the study by \citet[][]{Rackham2017} showed that repeatable, high-quality visual spectra of the sub-Neptune GJ~1214b (also orbiting an M dwarf host star) are only consistent with {\em stellar contamination} and not with planetary features, providing the first clear example for the effect that may also impact other high quality exoplanet transmission spectra \citep{Apai2017, Pinhas2018}.

Therefore, the central questions that emerge on the atmospheric characterization of the \trone{} (and similar, to-be-discovered M-dwarf habitable planet systems) and could be addressed, at least partly, before $JWST$ are: What are the compositions of the individual atmospheres and is there evidence for differences in the seven atmospheres? and, What effects will limit the precision with which \textit{HST} and \textit{JWST} will be able to probe these atmospheres?

In this study we present an independent reduction and analysis of two recently obtained \textit{HST} infrared spectroscopic datasets published in \citet[][]{deWit2016,deWit2018}.  Our reduction builds on the new and physically motivated detector charge trap correction \citep[][]{Zhou2017}, which provides an improved correction for the primary systematics affecting \textit{HST} high-precision spectroscopy. In addition, we provide a comprehensive assessment of the potential impact of stellar activity 
{on observations and stellar spectral contamination of the transmission spectra due to the heterogeneous photosphere of \trone{}.}

\section{Observations}
\label{S:Observations}
The data presented in this study were obtained in two Hubble Space Telescope (HST) Wide Field Camera 3 (WFC3) programs (GO-14500 and GO-14873, PI: de Wit) targeting the \trone{} system.  In the following we refer to the two programs by Program~\first{} and Program~\second{}, respectively. Program \first{} consists of one visit, executed on May 4, 2016, and covers the overlapping transits of planets \trone{} b and c. The results were initially published in \citet{deWit2016}. Program~\second{} consists of four visits, executed between December 2016 and January 2017, and covers the transits of planets \trone{} d, e, f, and g \citep{deWit2018}. In the two programs, the six inner planets (\trone{} b to g) have been observed at least once during transit. In addition, the observations include two overlapping transits of the planet pairs b \& c and e \& g. For convenience we label the seven transits in chronological order as Transit 1, 2, 3, 4, 5, 6, and 7, as listed in Table~\ref{tab:ObsLog}. As normal for \textit{HST} observations, the phase coverages of transit light curves were limited by Earth occultations.

All transmission spectra were obtained using the WFC3 infrared G141 grism, which covers wavelengths from $1.1\micron$ to $1.7\micron$. The observations utilized state-of-the-art strategies for WFC3 IR transit spectroscopy, including spatial scanning (to avoid saturation and increasing observing efficiency), detector sub-arraying (to avoid memory saturation), and the acquisition of a direct image at the beginning of each orbit to provide an accurate wavelength calibration for the slitless spectra. For each spectroscopic image, the exposure time was $112\,\mathrm{s}$, and the scanning rate was $0.027''/\mathrm{s}$, yielding a scanning length on the detector of $3.02''$ or $\sim25$ pixels. Spatial scans were conducted in bi-directional scanning mode in Program~\first, while Program~\second{} adopted the single directional scanning mode, resulting in slightly different cadences (151 s for Program~\first, 176 s for Program~\second) for observations in the two programs.  Table~\ref{tab:ObsLog} lists the key details of the observations.

The first, third, and fourth visits of Program~\second{} were severely affected by cosmic rays (CR) due to HST's passage through the South Atlantic Anomaly (SAA\footnote{The SAA is the lowest region to which Earth's inner Van Allen Belt extends. SAA passages by \textit{HST} result in enhanced of CR hits on the detector \citep{Deustua2016}.}), which in several of these orbits also negatively affected the \textit{HST} guiding performance, resulting in unrecoverable data. Due to particularly severe CR damage, we had to discard the following data subsets: Orbits 1-4 in Visit 1, Orbits 1 and 2 in Visit 2, and Orbit 5 in Visit 4.

\begin{deluxetable*}{lccccccl}
  \tablecaption{Observation log \label{tab:ObsLog}}
  \tablehead{\colhead{P ID}&
    \colhead{Visit No.}&
    \colhead{Obs. date}&
    \colhead{Planet}&
    \colhead{Transit}&
    \colhead{No. of orbits} &
    \colhead{No. of exposures\tablenotemark{a}} &
    \colhead{Note}}

  \startdata
  14500                  & 0 & 2016-05-04 & b,c & 1,2 & 4 & 74  & 
                                                                                                               \\ \hline
  \multirow{4}{*}{14873} & 1 & 2016-12-04 & d   & 3   & 7 & 114 & Only last three orbits not affected by CRs         \\
                         & 2 & 2016-12-29 & g,e & 4,5 & 5 & 84  & No apparent SAA influence                          \\
                         & 3 & 2017-01-09 & f   & 6   & 6 & 93  & First two orbits discarded with a possible transit \\
                         & 4 & 2017-01-10 & e   & 7   & 5 & 69  & Last orbits discarded including a possible transit \\
  \enddata \tablenotetext{a}{The numbers of exposures exclude those
    that were discarded due to guiding failure or compromised data quality.}
\end{deluxetable*}

\section{Data Reduction}
We downloaded the data from both programs from the Mikulski Archive for Space Telescopes. Our data reduction procedure started from the \texttt{ima} frames produced by the CalWFC3 pipeline Version 3.4. The \texttt{ima} frames are bias, dark, and non-linearity corrected and include all non-destructive reads. Each spectroscopic \texttt{ima} file contains seven read-outs.  We discarded the ``zeroth'' read because the detector was reset during this read \citep{Deustua2016}. Following \citet{Deming2013}, we formed \emph{sub-exposures} by differencing adjacent reads. There were seven major steps in our data reduction procedure: (i) wavelength calibration; (ii) flat field correction; (iii) cosmic ray removal; (iv) image registration; (v) light curve extraction; (vi) ramp effect correction; and (vii) light curve fitting and transmission spectra extraction. Steps (i) to (iv) were applied to individual sub-exposures, while the subsequent steps were applied to the combined data. In the following, we review these key steps.

\subsection{Wavelength Calibration and Flat Field Correction}
We derived wavelength solutions based on the position of the target in the direct images. We adopted up-to-date wavelength calibration coefficients from \citet{Wilkins2014}. The centroids of the target point source in the direct images were determined by fitting two-dimensional Gaussian profiles. In Program~\first, the direct image frame had a different aperture from the spectroscopic frame. For those observations, we adjusted the direct image coordinates accordingly.

We adopted a third-order polynomial function in wavelength for the flat field correction.  For each visit, we calculated separately a wavelength calibration-dependent flat field correction, i.e. differences introduced primarily by the variations in the position of the target on the direct images.  We next applied the correction to individual non-destructive reads.  While applying the flat field correction, we also identified and corrected for low data quality pixels: Pixels in the flat field frame that deviated more than 20\% from unity were flagged (most of these also had non-zero data quality flags, i.e., were also flagged by the CalWFC3 pipeline). We also flagged any additional pixels identified by the CalWFC3 pipeline as bad pixels, hot pixels, pixels with unstable response, or bad or uncertain flat values\footnote{These correspond to data quality flags 4, 16, 32, and 512, respectively \citep{Deustua2016}.}. Finally, we replaced the flagged pixels by interpolating over the neighboring unflagged pixels.

 \subsection{Cosmic Ray Correction}
 Correcting for cosmic rays (CR) was a crucial step in our reduction, particularly for the orbits heavily affected by SAA passage. We identified seven orbits that suffered from SAA passage using the fits header flag and the number of CR hits. On average, there were at least 20 visually apparent 
 CR hits per frame in these orbits. We developed a suite of custom algorithms to identify and remove the CRs and to evaluate the efficiency of the CR-corrections.

 First, we applied iterative bi-directional median filtering to each non-destructive read and identified pixels as CR-affected if they exceeded the median-filtered image level by a threshold of {11 $\sigma$} (determined through the CR removal assessment described below). For each iteration, pixels that were previously marked as CRs were excluded from median filter calculations. We repeated this iterative filtering procedure typically for at least three times (see discussion below for the connection between the algorithm's performance and the number of iterations used).

\begin{figure}
  \centering
   \includegraphics[width=\linewidth]{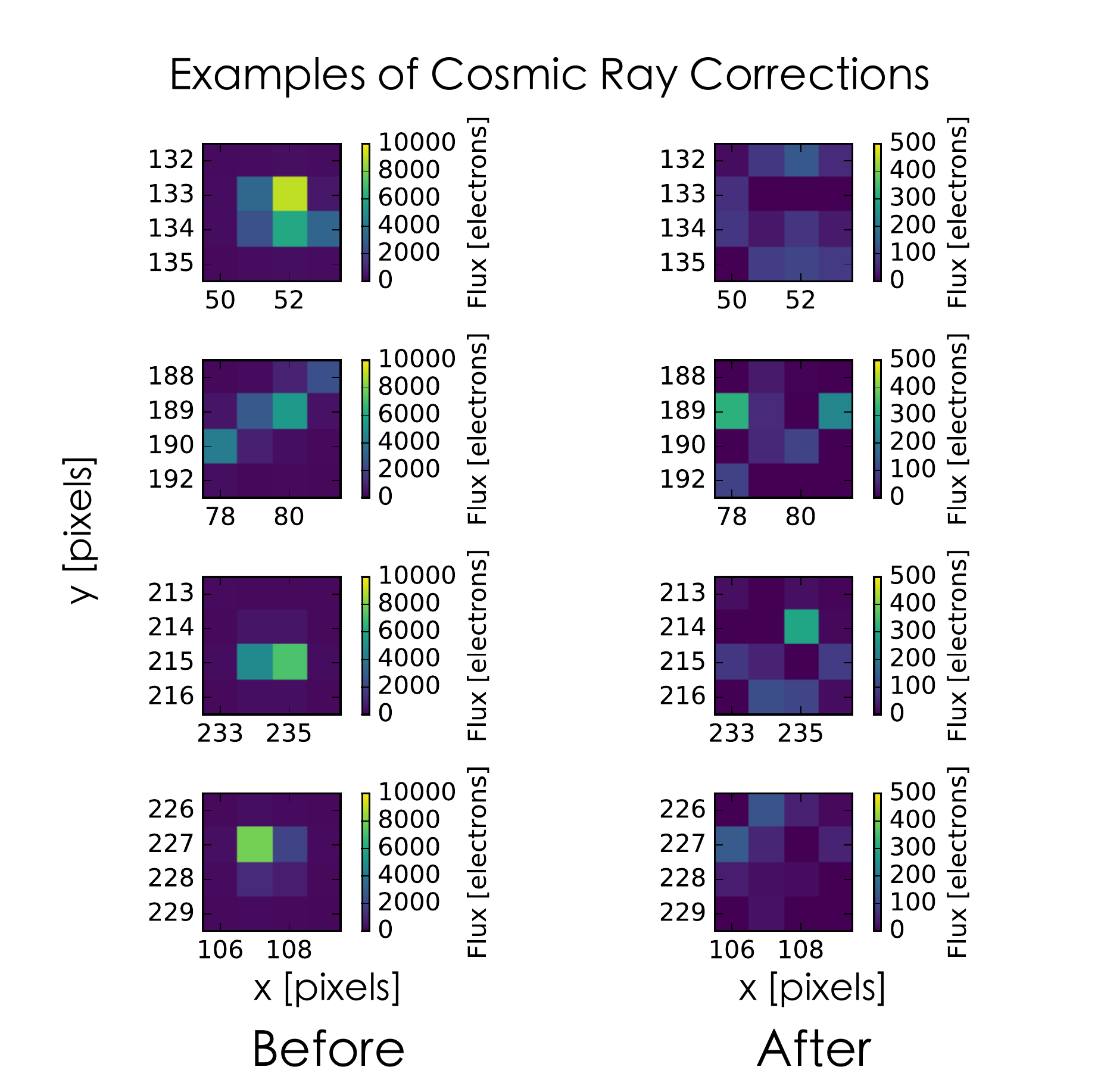}
  \caption{Comparison of image subsets before (\emph{left}) and after
    (\emph{right}) CR corrections. All image subsets are from frame \texttt{idde01koq}.}
  \label{fig:CR}
\end{figure}

 For each identified CR, we replaced the CR-affected pixel value with the weighted average ($f_{\mathrm{replace}}$) of the same pixel in the exposures preceding ($f_{-1}$) and following ($f_1$)) the image, as described by the relation:
 \begin{equation}
   \label{eq:1}
   f_{\mathrm{replace}}=\frac{t_{0}-t_{-1}}{t_{1}-t_{-1}}f_{1}+\frac{t_{1}-t_{0}}{t_{1}-t_{-1}}f_{-1},
 \end{equation}
 in which {$t_{0}$ refers to the time of} the CR-affected exposure and the subscripts $\pm 1$ denote the two adjacent (in time) exposures. The weights for the averaging are effectively the inverse of the time difference between the exposures. Program~\first{} adopted bi-directional scanning mode, i.e., forward and reverse scanning directions were applied alternatively. In this case, the preceding and following exposures had slightly different scanned image regions due to the upstream/downstream effect \citep[][]{Mccullough2012}. To account for this effect, we corrected the CR hits separately for images taken with different scanning directions.  Figure~\ref{fig:CR} shows example images before and after CR removal.

We assessed our CR correction algorithm quantitatively by injecting and removing CR hits with a CR template. We constructed the CR template using one uncorrected frame (observation identifier: iddea1meq) that was taken during an SAA crossing and contaminated by over 1,000 cosmic ray hits of different sizes and amplitudes. We removed the portion of the image with the stellar spectrum (pixel coordinates [135:170, 50:200]) and replaced it with randomly selected copies of regions out of the spectrum.  We set all pixels with a flux below 2,000 e$^{-}$ ($\sim3\times$ sky background) as zero and used the resulting image as the CR template. The CR template had 1,150 CR hits, representing the most severe CR-affected case. We added the template to 20 cleaned frames and applied our CR removal algorithm. We evaluated the performance of our CR identification and correction algorithm based on three criteria: cosmic ray identification rate, false positive rate, and correction efficiency ($\eta_{c}$). The latter we defined on a pixel-to-pixel basis as
\begin{equation}
  \eta_{c} = 1-\Biggl|\frac{f_{\mathrm{corrected}}-f_{\mathrm{clean}}}{f_{\mathrm{dirty}}-f_{\mathrm{clean}}}\Biggr|,
 \end{equation}
 in which $f_\mathrm{clean}$, $f_\mathrm{dirty}$, and $f_\mathrm{corrected}$ refer to the pixel fluxes in the input, CR-injected, and output images, respectively.

 Three parameters influence the effectiveness of the algorithm: the size of the median filter, the CR-identification threshold, and the number of iterations.  We optimized the algorithm through a three-dimensional grid search, with the size of the median filter ranging from 3 to 25 pixels, the thresholds ranging from 3 to 20~$\sigma$, and the number of iterations ranging from 1 to 10. The most effective combination included an 11 pixel median filter, 11~$\sigma$ threshold, and a minimum of 3 iterations. Our algorithm yields identification rates of 98.5\% on average, false positive rates below 1.5\%, and correction efficiencies above 90\% for over 99.9\% of the pixels across the entire image for non-SAA exposures. Within the image region containing the stellar spectrum, the identification rate dropped slightly to approximately 90\%. Typically, there were less than 40 pixels identified as CR hits in the spectrum region in an exposure obtained out of SAA passage, indicating that the total numbers of missed CR pixels and false positive pixels are 5 and 1 pixels, respectively. Since the region we used for the spectral extraction has a total size of $60\times140=$8,400 pixels, we found that unidentified and false positive CR hits had a negligible influence on our results.
  
 In addition, SAA passage severely affected the pointing accuracy of HST for a few orbits in Program~\second. For example, during Orbit 5 of Visit 4, the pointing shifted by $\sim14$ pixels (2\arcsec). Shifts with similar amplitudes were observed in the first four orbits of Visit 1 and the first two orbits of Visit 3 in Program~\second, which all indicate failures in the \textit{HST} fine guidance. In these orbits the light curve also drops by $\sim1\%$. We identify two potential causes for these systematics. First, after a pointing shift, the spectrum moved to the part of detector that was not previously illuminated. This increased the ramp effect induced by charge-trapping, because charge traps in these newly illuminated pixels have not been filled yet \citep{Zhou2017}. Second, WFC3's IR flat field has intrinsic uncertainties of $\sim$1.0\% \citep[][]{Deustua2016}, which introduce light curve systematics for such large pointing shifts. The ramp effect correction would, in principle, be correctable using the RECTE model \citep{Zhou2017} with additional free parameters describing the image drifts. However, the existing science data and calibrations did not provide a viable option for alleviating the increased flat field uncertainty. Therefore, we excluded the first four orbits in Visit~1, the first two orbits in Visit~3, and the last orbit in Visit~5 from the remainder of our analysis.

  We consulted transit times listed in the NASA Exoplanet Archive \citep{Akeson2013} and identified three transits that may lie within the discarded orbits. For completeness, we list these transits in Table~\ref{tab:missedtransits}.

\begin{deluxetable*}{ccccccc}
  \tablecaption{Possible Transits within Discarded Datasets \label{tab:missedtransits}}
  \tablehead{\colhead{Planet}&
    \colhead{Mid-time [UT]}&
    \colhead{Mid-transit time uncertainty}&
    \colhead{Mid-transit time}\tablenotemark{a}&
    \colhead{$T$}&
    \colhead{$a$}\\
    &UT&[Day]&JD&day&[$R_*$]&Possible Visit \& Orbit}
  
  \startdata
  f & 12/04/2017 03:18 & 0.00032  & 2457726.64 & 9.20669  & 68.4  & Orbit 3, Visit 1 \\
  d & 01/09/2017 18:47 & 0.001799 & 2457763.28 & 4.04961  & 39.55 & Orbit 2, Visit 3 \\
  e & 01/10/2017 13:43 & 0.000567 & 2457764.07 & 6.099615 & 51.97 & Orbit 5, Visit 4 \\
  \enddata
  \tablenotetext{a}{From \citet[][]{Wang2017}, the three planets display large transit timing variations (up to the half an hour). The mid-transit times do not take TTVs into account and actual timings could be different by up to 40 minutes from the times given here.}
\end{deluxetable*}

\subsection{Sky Background Removal}

We identified and removed the sky background using a sigma-clipping algorithm.  Pixels within 5$\sigma$ of the median image level after 10 sigma-clip iterations were considered as background. The median value of the background pixels was then subtracted from the image.

\begin{deluxetable}{cccccc}
  \tablecaption{Center Wavelengths of Different Bands \label{tab:wavelength}}
  \tablehead{\colhead{Band}   & 
    \colhead{}                & 
    \colhead{}                & 
    \colhead{Wavelength[\AA]} & 
    \colhead{}                & 
    \colhead{}                                                \\\hline
    Program                   & GO-14500 &   & GO-14873 &   & \\
    Band / Visit                     & 0        & 1 & 2        & 3 & 4}

  \startdata
  1  & 11505 & 11465 & 11377 & 11408 & 11442 \\
  2  & 11970 & 11930 & 11841 & 11873 & 11907 \\
  3  & 12434 & 12394 & 12305 & 12337 & 12371 \\
  4  & 12898 & 12858 & 12770 & 12801 & 12835 \\
  5  & 13363 & 13322 & 13234 & 13265 & 13300 \\
  6  & 13827 & 13787 & 13698 & 13730 & 13764 \\
  7  & 14292 & 14251 & 14163 & 14194 & 14228 \\
  8  & 14756 & 14715 & 14627 & 14658 & 14692 \\
  9  & 15220 & 15179 & 15092 & 15123 & 15157 \\
  10 & 15685 & 15644 & 15556 & 15587 & 15621 \\
  11 & 16149 & 16108 & 16020 & 16051 & 16085 \\
  12 & 16613 & 16572 & 16485 & 16516 & 16550 \\
  \enddata
\end{deluxetable}

\subsection{Image Drift Calibration}

The remaining observations suffered from \textit{HST} pointing drifts in both $x$ and $y$ directions at levels of $0.05$--$0.1$ pixels per orbit. Such drifts, especially in the wavelength dispersion direction, introduced systematic slopes in the spectrally binned light curves when left uncorrected. To correct for the drifts we measured the shifts between each image and the reference image (first image in each dataset) by cross correlation. We then used bi-cubic interpolation to shift and align images to the reference image.

\subsection{Light Curve Extraction}

We generated white-light and spectrally binned light curves.  We summed the CR-cleaned, background subtracted, and aligned sub-exposures back to a total exposure image. We created 12 ten-pixel-wide bands from the scanned area ranging from 1.1$\micron$ to 1.70$\micron$. We then obtained the light curves by summing pixels inside a 60-pixel wide window for every band. The uncertainty of each point on the light curve includes photon noise, dark current, and read-out noise. As wavelength solutions differ slightly for each visit, the central wavelengths of the ten-pixel wide bins vary at levels of ($\sim 0.01\,\micron$) for the different visits. We list the central wavelengths of the bands in Table~\ref{tab:wavelength}. In this way, five raw light curves were derived, each with twelve bins.

We note that the bins applied here are slightly different from those used by \citet{deWit2016,deWit2018}, who use eleven (0.05~\micron-wide) bins in the 1.15 to 1.7 \micron{} range for Transits~1 and 2 and ten (also 0.05~\micron-wide) bins in the 1.15 to 1.65 \micron{} range for the subsequent transits. In general, determining bin sizes with integer number of pixels is more widely adopted in HST/WFC3 transmission spectroscopic studies \citep{Mandell2013, Deming2013, Kreidberg2014}. Considering that 0.05\,\micron{} bin size is not an integer multiple of the spectral resolution unit of G141 grism, without knowing the exact binning and interpolations applied there, we could not use identical bins in the light curve extraction steps. Therefore, to compare our results with those of \citet{deWit2016,deWit2018}, we instead interpolated our pixel-binned transmission spectra. The interpolation had a negligible effect due to the coarse wavelength resolution of the spectra. We discuss this point further in Section~\ref{sec:transspec}.

\subsection{Ramp Effect Correction}

\begin{figure*}[!thb]
  \centering
  \includegraphics[width=\linewidth]{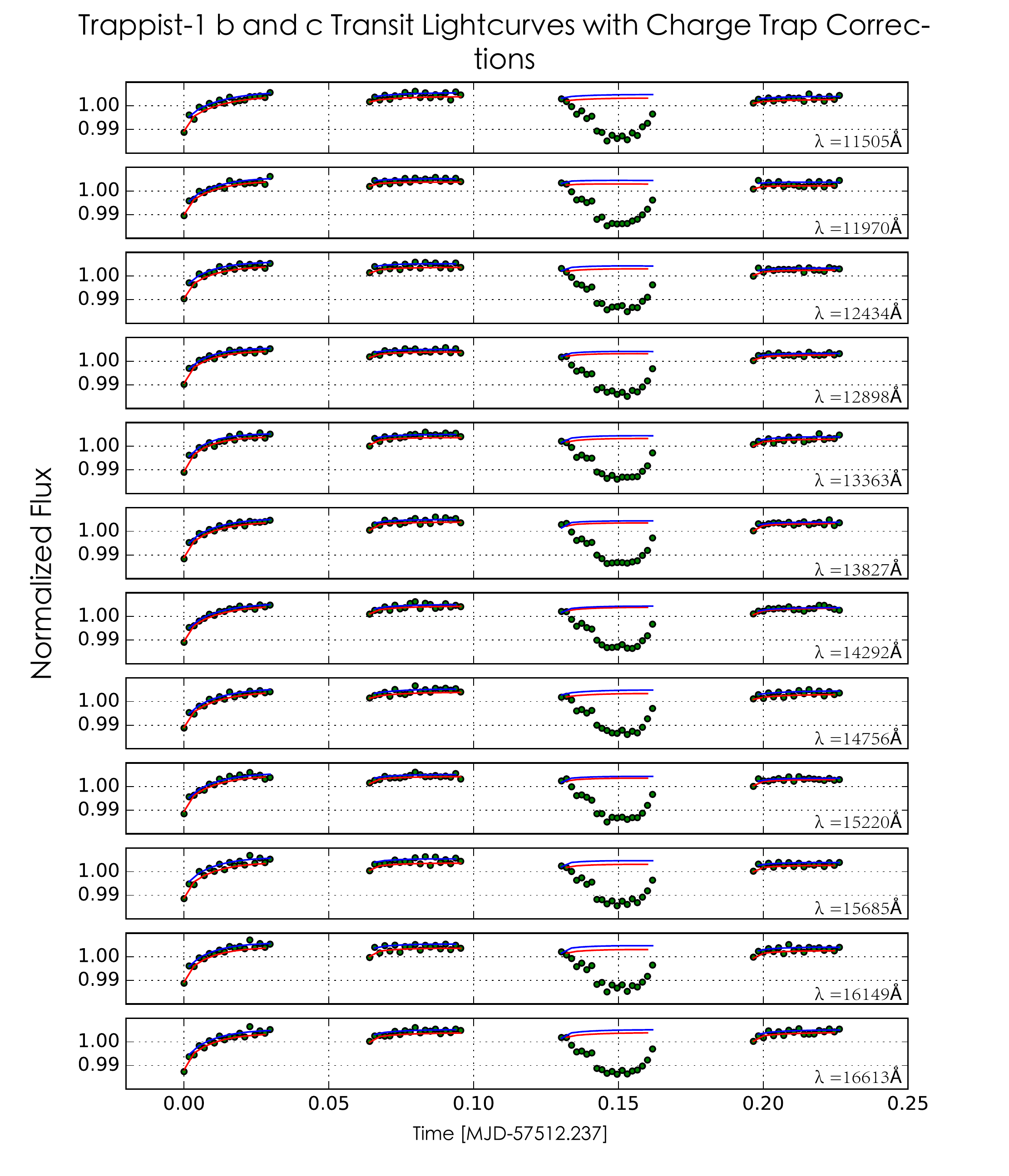}
  \caption{ Light curves (green dots) and the best-fitting RECTE ramp effect correction. The predicted charge trap effects for the two scanning directions are plotted separately (red and blue curves). This figure includes Transits 1 and 2 from planet c and b, respectively.}
  \label{fig:rawlc}
\end{figure*}

The raw light curves show prominent ramp effect systematics (Figure~\ref{fig:rawlc}), typical to HST/WFC3/IR time-resolved observations \citep[e.g.,][]{Berta2012,Apai2013}.  This systematic is caused by two populations of charge carriers that are trapped and then, with some delay, released by impurities in the HgCdTe detectors \citep{Zhou2017}.  We corrected these systematics using the RECTE model \citep{Zhou2017}, which models the history of illumination, trapping, and release for each pixel. This model offers a consistent solution to correct the ramp effect systematics from the perspective of the physical cause, instead of fitting empirically determined exponential/polynomial functions, which are used in most \emph{HST}/WFC3 transiting exoplanet studies to date. The use of this correction is also a major difference between our data reduction and that of \citet{deWit2016}, who used exponential functions to model and correct for the ramp effect. \citet{deWit2018}, which was published after the submission of this paper used another model bearing more resemblance to RECTE model to remove the ramp effects, and we compare them in \ref{sec:deWit2018}.

\citet{Zhou2017} described the charge trapping processes with six parameters ($E_{\mathrm{s,tot}}$, $E_{\mathrm{f,tot}}$, $\eta_{\mathrm{s}}$, $\eta_{\mathrm{f}}$, $\tau_{\mathrm{s}}$, and $\tau_{\mathrm{f}}$ \footnote{The subscripts ``s'' and ``f'' denote two charge trap types, slow and fast, which describe the release speed. Detailed descriptions of the two trap types are provided in \citet{Zhou2017}.}) representing the trap numbers, trapping efficiency, and charge release time for slow and fast charge trap populations. \citet{Zhou2017} found these parameters to vary little in different observations and considered to be intrinsic to the WFC3 detector. Therefore, we fixed these six parameters and provided the adopted values in Table~\ref{tab:recte}. The free parameters that determine the systematic profiles are
\begin{itemize}
  \item $f$: The incoming flux on each pixel as a function of time. We consider it to be a constant here.
  \item $E_{\mathrm{s, 0}}$, $E_{\mathrm{f, 0}}$: The initial numbers of trapped charges.
  \item{ $\Delta E_{\mathrm{s}}$ and $\Delta E_{\mathrm{f}}$: The number of additional charges trapped during inter-orbit gaps due to unintended detector illumination.}
  \item{ $v$: The slope of the visit-long trend.}
\end{itemize}

\begin{deluxetable}{llll}
  \tablecaption{RECTE Model Parameters \label{tab:recte}}
  \tablehead{\colhead{Parameter} & \colhead{Value} & Parameter            & Value}
  
  \startdata
  $E_{\mathrm{s,tot}}$ & 1525.38  & $E_{\mathrm{f,tot}}$ & 162.38   \\
  $\eta_{\mathrm{s}}$  & 0.013318 & $\eta_{\mathrm{f}}$  & 0.008407 \\
  $\tau_{\mathrm{s}}$  & 16300    & $\tau_{\mathrm{f}}$  & 281.463  \\
  \enddata
\end{deluxetable}

We found the best-fit RECTE profiles for the light curves of each band.  While the RECTE algorithm essentially models charge trapping process in individual pixels, it was not feasible (or necessary) to fit the light curves at the single-pixel level. First, the ramp effect at the pixel level is overwhelmed by other systematics, particularly telescope jitter. Second, the accuracy of the single-pixel level ramp effect correction is negatively influenced by photon noise in these data. As \citet[][]{Zhou2017} found no evidence for the charge trap parameters varying between pixels, we therefore adopted an average band-level charge trapping correction instead of a single-pixel level correction.

For observations using bi-directional scanning, exposures conducted in the opposite scanning directions bear an intrinsic flux level difference of $\sim0.5\%$.  For these cases, we assumed different $f$ and $v$ values for the light curves observed in different scanning directions but calculated the charge trapping/release processes for the two scanning directions together.

We found the best-fit parameters using a Markov Chain Monte Carlo (MCMC) with 500 walkers for 600 steps, with the first 300 steps discarded as burn-in. The MCMC runs were performed using the \texttt{emcee} package \citep{Foreman2013}. Examples of the best-fit RECTE profiles are shown in Figure~\ref{fig:rawlc}. In most bands the ratio of the average value of the standard deviation to the average of photon noise is within the range of 0.8--1.2, i.e., our complete procedure (including cosmic ray and charge trapping corrections) robustly reach the photon noise level or very near to it.

For each transit we also derived a broad-band light curve by computing the weighted average of all bands, adopting the inverse variance as the weight.

\subsection{Transit Profile Fitting and Spectrum Extraction}

Our final reduction steps were fitting the transit profiles and extracting the transmission spectra. We first fitted the broadband light curves by generating model transit light curves using the Python package \texttt{batman} \citep[][]{batman}, in which the light curve shape model is based on \citet[][]{Mandell2002}. The fitting procedure was performed with an MCMC algorithm using the \texttt{emcee} package \citep[][]{Foreman2013}. The transit profile model contained 7 parameters, namely transit mid-time $t_{0}$, orbital period $T$, relative planet size $r_{p}/r_{*}$, semi-major axis $a$, eccentricity $e$, inclination $i$, argument of periapsis $\omega$, and quadratic limb darkening coefficients $u_{1},u_{2}$.

\begin{figure}[!tb]
  \centering
  \includegraphics[width=\linewidth]{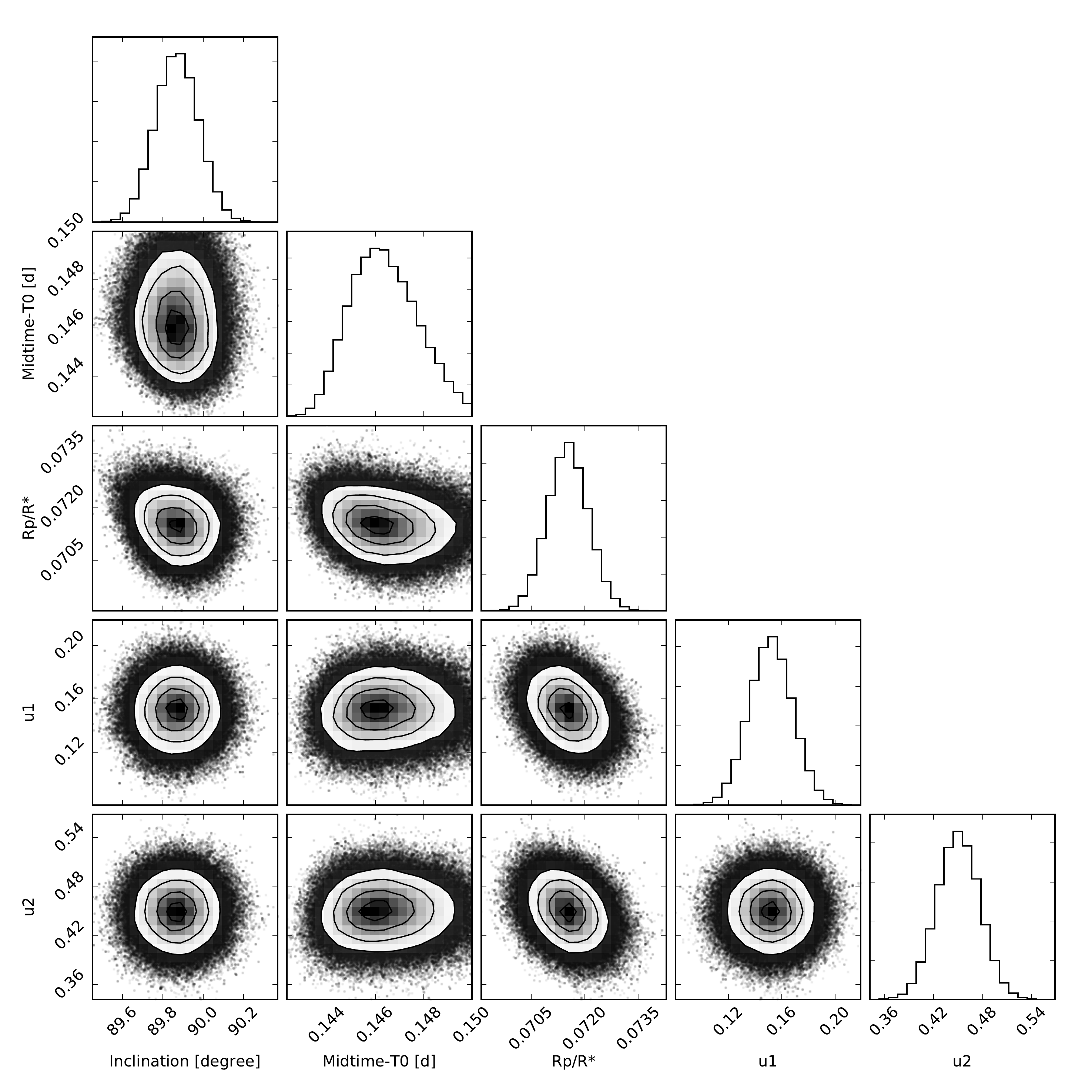}
  \caption{Posterior distributions of transit profile of TRAPPIST-1 e in Transit 7}
  \label{fig:corner}
\end{figure}

We found that due to the lack of ingress or egress data in some transits, the limb darkening coefficients (LDCs) could not be constrained well by the light curves. Consequently, the LDCs obtained in different transits are not always consistent. It is important to note that models predict that LDCs of late-M stars will vary significantly with wavelength in the 1.1 to 1.7\micron{} range and that the transit depths derived are anti-correlated with LDCs (Figure~\ref{fig:corner}). Therefore, errors in LDCs may introduce apparent spectral features in the transmission spectra. To carefully examine the effect of LDCs, we experimented with three different LDC treatments: (i) interpolating LDC values and uncertainties provided in \citet[][]{deWit2016} -- derived using PHOENIX stellar models \citep{Husser2013} -- to our bandpasses and by using these as Gaussian-distributed priors; (ii) fixing the LDC to the best-fit values reported in \citet{deWit2016}; and (iii) independently deriving the LDCs by fitting PHOENIX specific intensity model stellar spectra\footnote{http://phoenix.astro.physik.uni-goettingen.de/} (disk-integrated, multiplied by the HST/G141 bandpass, and normalized) to the HST/G141 out-of-transit spectrum and fitting a quadratic limb darkening law to the limb darkening profile of the best-fit model.  We then fixed the LDCs in the transit fit using the derived values.

The comparison of the results based on the different LDCs showed that the final spectra are only weakly affected by the adopted LDCs: In every band the transit depths derived from the different methods agreed with each other to levels better than $1\sigma$. We adopt the first limb darkening treatment described above as our nominal procedure and present the results from this approach here.

The limited phase coverage of ingress and egress in some visits that complicated the LDC studies also precluded precise measurements of the transit durations. This uncertainty likewise hampered the precise determination of the orbital inclinations. Therefore, we adopted the distributions obtained by \citet[][]{Gillon2017} as priors in our fitting procedure.

We fixed all eccentricities to be zero, as justified by the small values found by \citet[][]{Gillon2017}, which rendered $\omega$ irrelevant.  We also fixed the orbital periods and semi-major axes to the values given in \citet[][]{Gillon2017}. We did not directly adopt external constraints on the mid-transit times as TTVs are large in the system and not yet well understood. The fit parameters are summarized in Table~\ref{Table:TransitFitParameters}. With $r_{p}/r_{*}$, $t_{0}$, $i$, and the LDCs as the only free parameters, we performed an MCMC search. We adopted 500 walkers, and ran them for 2,000 steps, the first 1,000 of which were treated as burn-in. In the evaluation of the fit quality, we did not use the uncertainty estimates directly derived from the pipeline (which are dominated by photon noise), as these did not include the assessment of the residual systematic noise (even though these are found to be very small). Instead, for each light curve in each band of each visit, we opted to calculate the standard deviation of the data in the baseline (pre- and post-transit) and adopted this value as a uniform relative uncertainty applicable to all data points in the light curves.

\begin{figure*}[!thb]
  \centering
  \includegraphics[width=\linewidth]{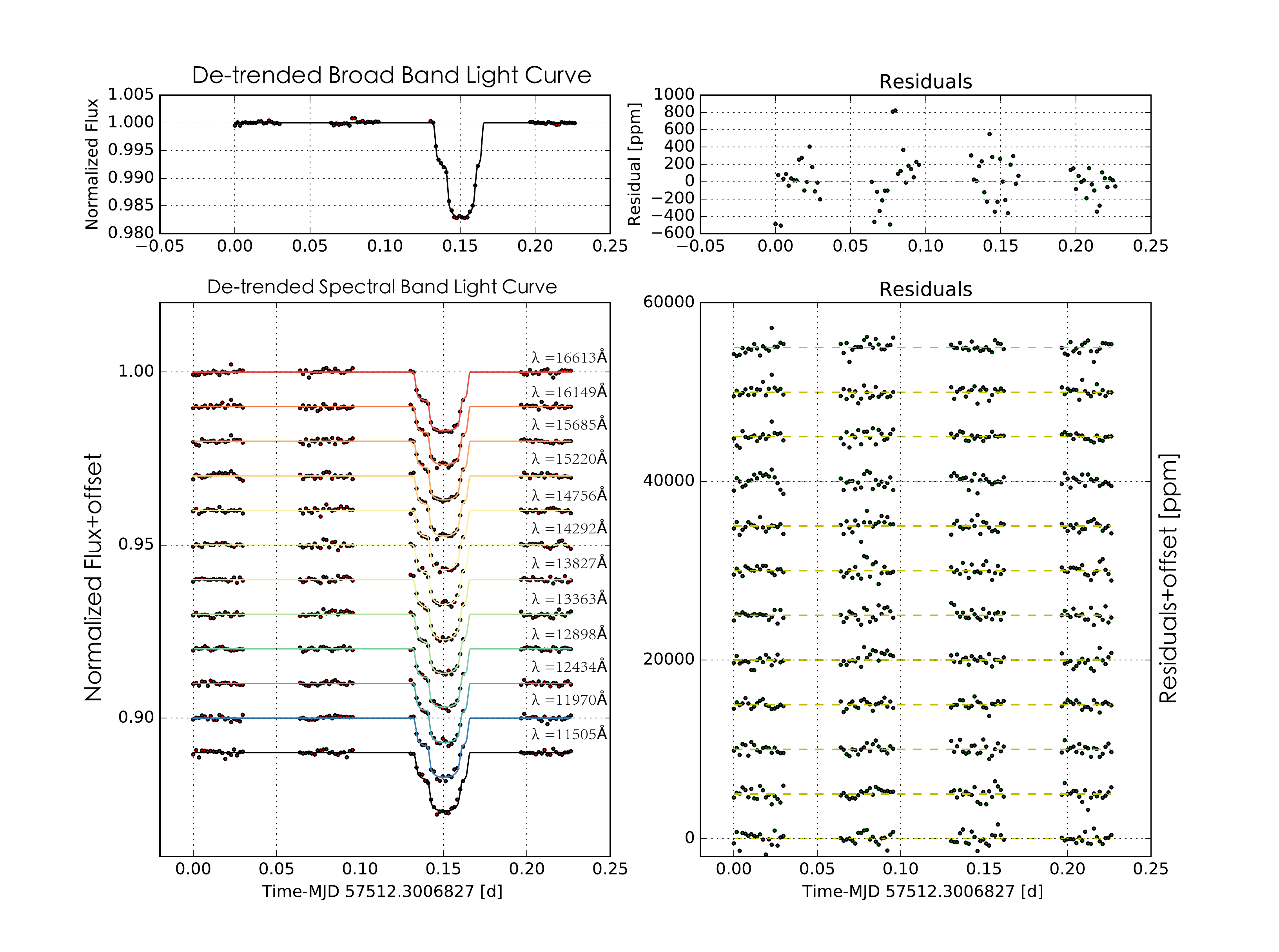}
  \caption{Broadband (\emph{upper}) and spectral band (\emph{lower}) light curve profile fits for planets \trone{} b\&c. The observations and best-fit profiles are shown in the left and the fitting residuals are shown in the right panel.}
  \label{fig:lc}
\end{figure*}

We present a corner plot of the MCMC posterior distributions in Figure~\ref{fig:corner} and an example of transit profile fit in Figure~\ref{fig:lc}.

After fitting the broad-band transit light curves, we fitted the transits in the individual spectral bins in each transit, keeping the mid-transit times fixed to the values found in the broad-band transit fits. To test the reliability of these fits, we carried out a Shapiro-Wilk test on the residuals of the fittings \citep[][]{SP1965}.  Most of our fits passed the test with p-values exceeding 0.1.  We inspected each of the few exceptions (p-values less than 0.1) visually and found that a few outlying data points, probably caused by stellar flares or other activity, were the reason that the transit models did not provide complete fits. We further investigate possible stellar activity in the light curves in Section~\ref{Discussion:StellarActivity}.

Finally, we subtracted the best-fit broadband transit depth value ($r_{p}/r_{*}$) from each spectral bin's transit fit (to determine relative, spectrally dependent differences in the transit depths) to derive the transmission spectra of the seven transits.

\begin{deluxetable*}{ccccc}
  \tablecaption{Parameters used as input in transit profile fitting from \citet[][]{Gillon2016, Gillon2017}\label{tab:transit}}
  \tablehead{\colhead{Planet} & \colhead{Inclination}\tablenotemark{a} & \colhead{$T$}\tablenotemark{b} & \colhead{$a$} \\
                              & [degree]                               & day                            & [$R_*$]}
  
  \startdata
  b & $89.65^{+0.22}_{-0.27}$ & 1.5109   & 20.50    \\
  c & $89.67\pm0.17$          & 2.4218   & 28    \\
  d & $89.75\pm0.16$          & 4.04961  & 39.55 \\
  e & $89.86^{+0.10}_{-0.12}$ & 6.099615 & 51.97 \\
  f & $89.680\pm0.034$        & 9.20669  & 68.4  \\
  g & $89.710\pm0.025$        & 12.35294 & 83.2  \\
  \enddata \tablenotetext{a}{Gaussian distributed priors}
  \tablenotetext{b}{Fixed parameters}
\label{Table:TransitFitParameters}
\end{deluxetable*}

\section{Results}

We measured the transit depths and mid-transit times of all seven transit events in both the broadband and individual spectral bins. In the transit profile fits, we reached an average reduced $\chi^{2}$ of 0.99 and our residuals were typically 1.05 times the photon noise level. In total we derived seven transmission spectra of six planets, including two of \trone{}~e. In addition to the individual planets' spectra, in the following we also present their combined spectrum. We list the results of our broad-band model fits in Table \ref{tab:results} and compare them to the results from the literature \citep[][]{Gillon2017,Gillon2016} and mid transit-times in the Online Exoplanet Archive \citep[][]{Akeson2013}.

\begin{deluxetable*}{cccccccc}
\tablecaption{Comparison of Broad-band Model Fit Results to Literature Values. Literature $r_{p}/r_{*}$ values are from \citet[][]{Gillon2016,Gillon2017,deWit2016,deWit2018}, while mid-transit times are from the Online Exoplanet Archive \citep[][]{Akeson2013}. \label{tab:results}}
  
  \tablehead{
    \colhead{Transit}&
    \colhead{Planet}&
    \colhead{Mid-time}&
    \colhead{Mid-time uncertainty}&
    \colhead{Mid-time}&
    \multicolumn{2}{c}{Literature $r_{p}/r_{*}$} &
    \colhead{Best-fit $r_{p}/r_{*}$} \\
    &&UT& days &$\mathrm{BJD_{TDB}}$& \citet[][]{Gillon2016,Gillon2017} & \citet{deWit2016,deWit2018} & this study}

  \startdata
  1 & c & 2457512.88051 & 0.000352 & 2457512.8807 & $0.0828^{+0.0006}_{-0.0006}$ & $0.0854^{+0.0014}_{-0.0014}$ & $0.0849^{+0.0012}_{-0.0012}$ \\
  2 & b & 2457512.88712 & 0.000176 & 2457512.8876 & $0.0852^{+0.0005}_{-0.0005}$ & $0.0895^{+0.0012}_{-0.0012}$ & $0.0879^{+0.0012}_{-0.0011}$ \\
  3 & d & 2457726.83624 & 0.001232 & 2457726.8400 & $0.0605^{+0.0015}_{-0.0015}$ & $0.0631^{+0.0007}_{-0.0006}$ & $0.0622^{+0.0006}_{-0.0005}$ \\
  4 & g & 2457751.81998 & 0.00105  & 2457751.8397 & $0.0884^{+0.0015}_{-0.0016}$ & $0.0885^{+0.0008}_{-0.0007}$ & $0.0888^{+0.0007}_{-0.0007}$ \\
  5 & e & 2457751.87282 & 0.000545 & 2457751.8701 & $0.0720^{+0.002}_{-0.002}$   & $0.0689^{+0.0007}_{-0.0006}$ & $0.0694^{+0.0005}_{-0.0005}$ \\
  6 & f & 2457763.46460 & 0.00038  & 2457763.4462 & $0.0820^{+0.0014}_{-0.0014}$ & $0.0803^{+0.0011}_{-0.0011}$ & $0.0802^{+0.0004}_{-0.0004}$ \\
  7 & e & 2457764.07205 & 0.00117  & 2457764.0671 & $0.0720^{+0.002}_{-0.002}$   & $0.0707^{+0.0008}_{-0.0007}$ & $0.0715^{+0.0006}_{-0.0006}$ \\
  \enddata
\end{deluxetable*}

\subsection{Broad-Band Transit Depths}
We measured the WFC3 broad-band transit depths of \trone{} b, c, d, e, f, and g with an average precision of 123 ppm.  We note that overlapping Transits 1 and 2 have larger uncertainties in transit depths than the other transits, which are all due to single planets. Most \trone{} planets' transit depths we measured in WFC3 G141 broad-band are consistent with those measured in Spitzer Channel 2 (central wavelength 4.5 $\mu m$) light curves \citep[][]{Gillon2017}. However, our transit model's $r_{p}/r_{*}$ for Transit 1,2,3 ($0.0849\pm0.0012, 0.0879\pm0.0012, 0.0622\pm 0.0005$) are deeper than the corresponding Spitzer transit depth measurement for the same planets ($0.0828\pm0.0006, 0.0852\pm0.0005, 0.0605\pm 0.0015$) by over $1\sigma$.  Differences between the \textit{HST} and Spitzer bands should come as no surprise, as these may be introduced either by planetary absorption features or -- more likely -- by stellar activity and heterogeneity, which we will explore in greater details in Section~\ref{Discussion:StellarActivity}.

\subsection{Mid-Transit Times}
\label{Section:Mid-transitTimes}

Through our transit light curve modeling, we obtained high-precision mid-transit time measurements, which were converted from Modified Julian Dates (MJD) to Barycentric Julian Dates ($\mathrm{BJD_{TDB}}$) using the algorithms described in \citet[][]{Eastman2010}. The results are summarized in Table~\ref{tab:results}.

The mid-transit times have a typical uncertainty of 0.0002~days (or 17~s).  The uncertainties are typically dominated by the limited phase coverage (due to HST's visibility windows), i.e., light curves missing either the ingress or egress. Different phase coverage of the individual transits causes the quality of the constraints on mid-transit times to vary.  Better constraints were achieved for the transits of planets \trone{} b and c because they have better than typical transit phase coverages, owing to their shorter transit durations; for these planets the transit mid-times are constrained with uncertainties of only 0.0001~days (or 8~s).

In contrast, Transit 7---corresponding to a transit of \trone{}~e---lacks both the ingress and ingress, resulting in a greater uncertainty (0.002 day) on the mid-transit time.

We found substantial TTV signals (observed-predicted time differences) in the observed transits when comparing our results with those from the Online Exoplanet Archive, which are calculated assuming strict periodicity \citep[][]{Akeson2013}. In particular, the best-fit mid-transit time for Transit~4 (\trone{}g), occurring on Dec 9, 2016, deviates from that predicted from the best-fit Spitzer transit mid-time \citep[][]{Gillon2017} and the planet's orbital period by $\sim30$ minutes, as was noticed and discussed in \citet[][]{Wang2017}.

\subsection{Transmission Spectra}
\label{sec:transspec}
\begin{figure*}[!thb]
  \centering
  \includegraphics[width=\linewidth]{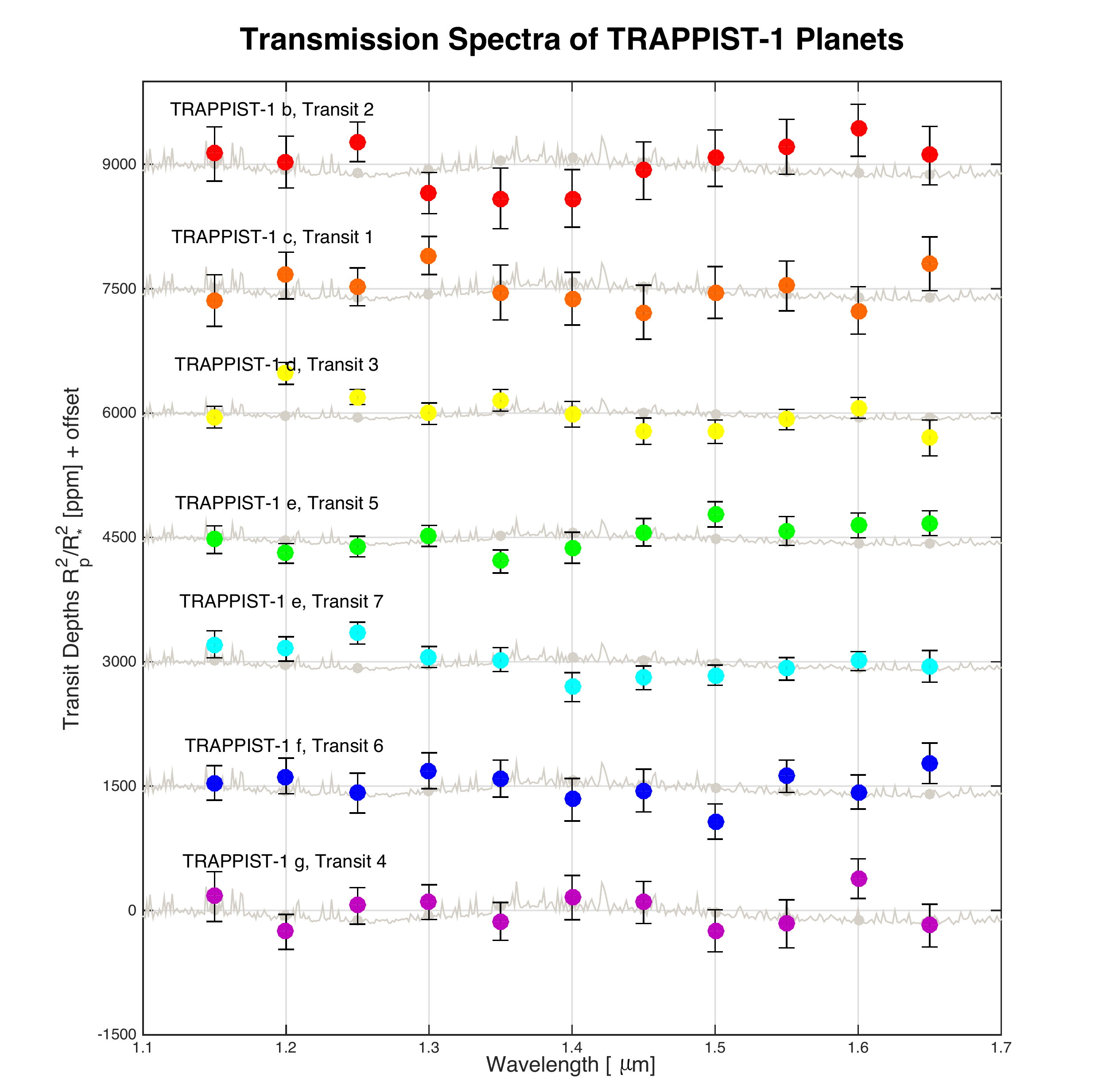}
  \caption{HST/WFC3 G141 transmission spectra for \trone{} b to g (from top to bottom).  Wavelength bands are aligned with those used in \citet[][]{deWit2016}. The observed transmission spectra are plotted in circles. The gray curves are water transmission models \citep[][]{Kempton2017}, scaled to provide the best fit to the data. All spectra have been mean-subtracted and vertical offsets have been  applied for clarity.}
    \label{fig:5}
\end{figure*}

We obtained WFC3/IR G141 transmission spectra for \trone{} b, c, d, e, f, and g. The spectra are shown in Figure~\ref{fig:5}, with the individual spectra offset by arbitrary levels for clarity. Each spectra has 12 bins covering wavelengths from 1.1\micron{} to 1.17\micron, corresponding to a spectral resolution of $\Delta \lambda$=50~\AA~ (or spectral resolving power of $R=\Delta \lambda/\lambda=22-34$, Figure \ref{fig:lc}).

\begin{figure*}[!thb]
  \centering
  \plotone{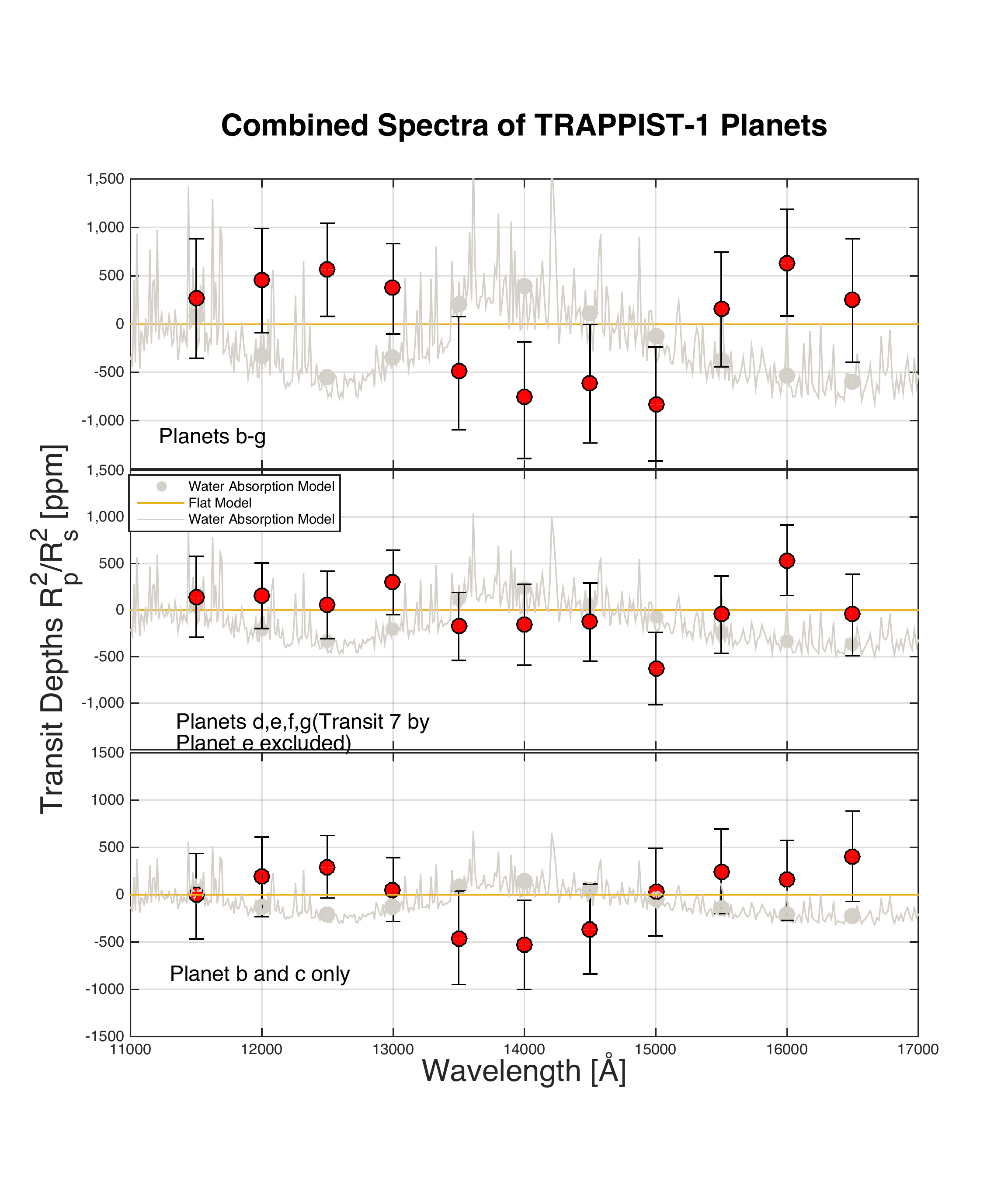}
  \caption{Combined HST/WFC3 G141 transmission spectra for \trone{} planets.  Wavelength bands are aligned with those used in \citet[][]{deWit2016}. For the upper panel, all planets' spectra are included. For the mid panel, Transits~3, 4, 5, and 6 from planets d, g, f, and e are included. For the lower panel, spectra of planet b\&c, whose transits overlapped during observations, are excluded. The gray curves water transmission models \citep[][]{Kempton2017}, scaled to provide the best fit to the data. All spectra have been mean-subtracted.}
  \label{fig:6}
\end{figure*}

Slight differences in the wavelength calibrations of the individual visits resulted in small differences of wavelength solutions among them (Table~\ref{tab:wavelength}). To directly compare our results with those by \citet{deWit2016} and to combine spectra from multiple visits, we interpolated our single-transit transmission spectra to align our wavelength bins with those of \citet{deWit2016}. We adopted the 3rd-order spline interpolation for both transit depths and uncertainties. We justify the interpolation of transit depths in two respects. First, the adopted bin size is coarse and significantly larger than the average difference of the interpolated wavelengths and the original wavelengths. Second, the transmission spectra and their uncertainties have small spectral variations. The interpolation introduced negligible modifications to the transit depths and uncertainties. In order to allow direct comparison with the spectra published in \citet{deWit2016}, we interpolated our transmission spectra to match the wavelength bins used in that study (Figure \ref{fig:5},\ref{fig:6},\ref{fig:7}).

In the following we also explore the combined spectra of the \trone{} planets as well as the combined spectra of different subsets. We combined the spectra by summing the transit depths from the six planets in each of the bands (already aligned to those in \citet{deWit2016}), giving the combined transit depth
\begin{equation}
\label{eq:3}
\left(\frac{R_{p}^{2}}{R_{s}^{2}}\right)_{\mathrm{combined}} = \sum_i\frac{R_{\mathrm{p},i}^2}{R_\mathrm{s}^2},\\
\end{equation}
and uncertainty
\begin{equation}
\label{eq:4}
\delta\left(\frac{R_{p}^{2}}{R_{s}^{2}}\right)_{\mathrm{combined}}=\sqrt{\sum_{i}\bigl[\delta (R_{\mathrm{p},i}^2/R_\mathrm{s}^2)\bigr]^2.}
\end{equation}
For planet e, for which we have two transits, we used the inverse-variance weighted average of the two as the transmission spectrum.

Combined spectra may provide information about shared spectral features: combining seven spectra could increase the signal-to-noise of the spectra by up to a factor of 2.6.
 
The combined spectra of all seven transits is shown in the top panel of Figure~\ref{fig:6}. 

Given the differing quality of the spectra, we also explored combinations of different subsets of the transits. As noted above, the transits of planets b and c overlapped in time and their spectra show an apparent anti-correlation. Additionally, the parameters of Transit 7 (planet e) are determined less precisely than those of the other transits due to its lack ingress or egress coverage. To allow the assessment of the impacts of these data quality differences and possible systematics on the combined spectra, we show two additional spectral combinations in Figure~\ref{fig:6}. The middle panel shows the combined spectra of planets d, e, f, and g (with spectra for planets b, c, and e from Transits 1, 2 and 7 excluded). The lower panel in Figure~\ref{fig:6} shows the combined spectrum of Transits 1 and 2, i.e., planets c and b, respectively.

Water absorption is the most prominent expected spectral feature in the planetary atmospheres in this wavelength range. As such, we show in Figure~\ref{fig:6} a model for water absorption for comparison (light gray lines). For this comparison we adopted a water transmission model calculated using \texttt{ExoTransmit} \citep[][]{Kempton2017}, with an amplitude scaled to provide the best fit to the observed spectra. None of the three combined spectra in Figure~\ref{fig:6} resemble the water absorption spectrum: while the water absorption would result in larger $r_p/r_*$ values around 1.4~\micron{}, all three spectra suggests a {\em decrease} in $r_p/r_*$ values. We conclude that no water {\em absorption} or other {\em planetary} molecular absorption features are visible in any of the individual or combined spectra. In Sections~\ref{Section:WaterLimit} and \ref{Section:StellarContamination} we explore the upper limits we can place on planetary water absorption as well as the impact of stellar contamination on these spectra.

\section{Discussion}

In this and the following sections, we discuss the following key points of our study: comparison of our data reduction to that of \citet[][]{deWit2016} and the resulting data quality; comparisons of multiple transits of the same planet; placing upper limits on water absorption bands; and discussing in detail stellar activity and stellar contamination in the HST/WFC3/IR spectra of the \trone{} planets. Finally, we place these results in the context of future \textit{HST} and \textit{JWST} transit spectroscopy of the \trone{} and similar planetary systems.

\subsection{HST/WFC3 IR transiting exoplanet data reduction comparison}

One of the key differences between our study and that of \citet{deWit2016} is that we used the RECTE model to correct for the HST/WFC3 IR ramp effect, while the \citet[][]{deWit2016} study discarded the first orbits of each visit and relied on an empirical fit to correct the systematics in the subsequent orbits, implicitly assuming those to be identical to the ramp seen in Orbit~2. Here we compare the results of the two data reduction approaches.

First, by using RECTE, we successfully corrected the ramp effect in each visit's first orbit, which was discarded in \citet{deWit2016} as well as in almost all published HST/WFC3 transit spectroscopic observations. As a result, we increased the available useful data and effectively improved the efficiency of the HST observations by about 25\%. This increase translated to a better orbital phase coverage and an improved accuracy of the transit baseline levels. Furthermore, the additional baseline observations also enabled a more thorough exploration of the stellar activity and spectral changes (see Section~\ref{Discussion:StellarActivity}) than would have been possible had those orbits been discarded.

Second, in addition to the increase in the data quantity and efficiency, we seek to compare the resulting data quality. At this point any such comparison must be limited to Visit 1 (from Program~\first), the only visit for which reduced data 

{had been published at the time of submission for this current work\footnote{A comparison with the work of \citet{deWit2018} is provided in Section~\ref{sec:deWit2018}}}. 
With an uncertain astrophysical signal underlying possible residual systematics, such comparisons are not trivial. We proceed here by assuming that, to first order, the astrophysical signal is well understood and purely consists of the planetary transit that follows the analytic models of \citet[][]{Mandell2002}; we will then discuss the limitations of this approach.  

Under the assumptions laid out, the residuals of the observed and modeled light curves contain no systematics and should be photon noise-limited.  Therefore, we use the standard deviation of the light curve fit residuals as a metric to compare the data quality between our reduction and that of \citet{deWit2016}. De Wit et al. kindly provided their detailed results, enabling accurate comparison studies. We note that \citet{deWit2016} report a larger broadband residual standard deviation (240 ppm) than that calculated from data in Figure~1 of \citet{deWit2016} (215 ppm). We conservatively adopt the latter for the comparison. The first three rows of Table~\ref{tab:residuals} compare the standard deviations of residuals between the two studies for Orbits 1, 2, and 4 as well as several orbit combinations.

Several points are notable about this comparison. If our underlying assumptions were correct, the residuals in all orbits should be the same. However, differences in Orbits 1, 2, and 4 are visible even within the same reductions. We attribute these differences in part to the scatter of the standard deviations themselves and, in part, to the fact that physical processes other than the transit are present in the data, as discussed below.

The immediate comparison of the standard deviations of the residuals between the two studies (rows 1--3 of Table~\ref{tab:residuals}) have overall very similar levels, considering the limited number of data points from which the standard deviations are calculated. Data from our reduction have lower standard deviations in Orbit 4 than those from \citet{deWit2016} (136 ppm vs. 159 ppm), but larger residuals in Orbit 2 (342 ppm vs. 257 ppm). For combined residuals of Orbits 2 and 4 (row 5), our results still have slightly larger standard deviation than that of \citet{deWit2016} (266 $\pm$25 ppm vs 215 ppm), corresponding to about a 1$\sigma$ level difference (when assuming similar uncertainties for the standard deviations themselves). We note that Orbit 2's correction quality is worse in both reductions than that of Orbit 4, suggesting that the inherent data quality or other processes also play an important role.

Thus, based on a superficial comparison one may conclude that: (1) both reductions reach similar precision; (2) underlying differences in the data quality are more important than the type of ramp effect correction applied; and, (3) the empirical ramp effect correction performs sometimes slightly better than the physical model for the charge trapping (not considering the benefits of the non-discarded first orbit).

Closer inspection of the residuals in Figure~\ref{Fig:FlaresSpots} reveals isolated groups of outlying data points. These may mark potential small stellar flare events (discussed further in Section~\ref{Discussion:StellarActivity}), such as an event with 4$\sigma$ outliers above the baseline in Orbit 2 of Visit 0. It is instructional to inspect this event: in fact, the outliers contribute most to the increased standard deviation in the residuals in this orbit. By excluding the four data points around suspected flare events, the standard deviations of the combined Orbit 2 and 4 data decreased by 40\% to 195 ppm (last row of Table~\ref{tab:residuals}). If the same data points are excluded from the \citet{deWit2016} data, the resulting standard deviation is, within uncertainties, the same as that resulting from the RECTE fit.  We conclude that the empirical ramp correction may provide a lower standard deviation occasionally because it can potentially fit the ramp effect correction {\em and} stellar flares together.

Thus, in our RECTE-based reduction, the higher standard deviations are an indication that the model applied (charge trapping model + transit) is incomplete; in contrast, the empirical systematics fit and transit has the capability to absorb different sources of astrophysical signal and instrument systematics without distinguishing these, resulting in a slightly lower standard deviation of the residuals. Compared to empirical corrections, the physically-motivated RECTE-based correction is less likely to be skewed by astrophysical events.  This comparison highlights another advantage of the RECTE model over traditional empirical fits: Given the well-determined detector response, it will be less likely to over-correct and remove astrophysical processes (or other types of instrumental systematics).  In fact, in Orbit 4, which has no signs of stellar flares, our RECTE model indeed resulted in a residual standard deviation that is the lowest of all measured here (136 ppm).

\begin{deluxetable}{cccccc}
  \tablecaption{The comparison of standard deviations of residuals between our study and that of \citet[][]{deWit2016}, utilizing different ramp effect corrections. Note that this comparison is only easily interpretable if no astrophysical systematics (e.g., flares, spots) are present. The standard deviations from \citet{deWit2016} have been provided by the authors (priv. comm.). \label{tab:residuals}}
  \tablehead{Orbit & 
    This Study     & 
    \citet[][]{deWit2016} \\
                   & 
    [ppm]          & 
    [ppm]}

  \startdata
  1                    & 225                         & /     \\
  2                    & 342                         & 257   \\
  4                    & 136                         & 159   \\
  1\&2\&4              & 254                         & /     \\
  2\&4                 & 266$\pm25$\tablenotemark{a} & 215 & \\
  2\&4, flare excluded & 195$\pm16$\tablenotemark{a} & 180 & \\
  \enddata

  \tablenotetext{a}{The uncertainties are derived by calculating the standard deviations of the standard deviations of randomly selected 80\% sub-sets of the data.}
\end{deluxetable}

\begin{figure*}[!thb]
  \centering
  \includegraphics[width=\linewidth]{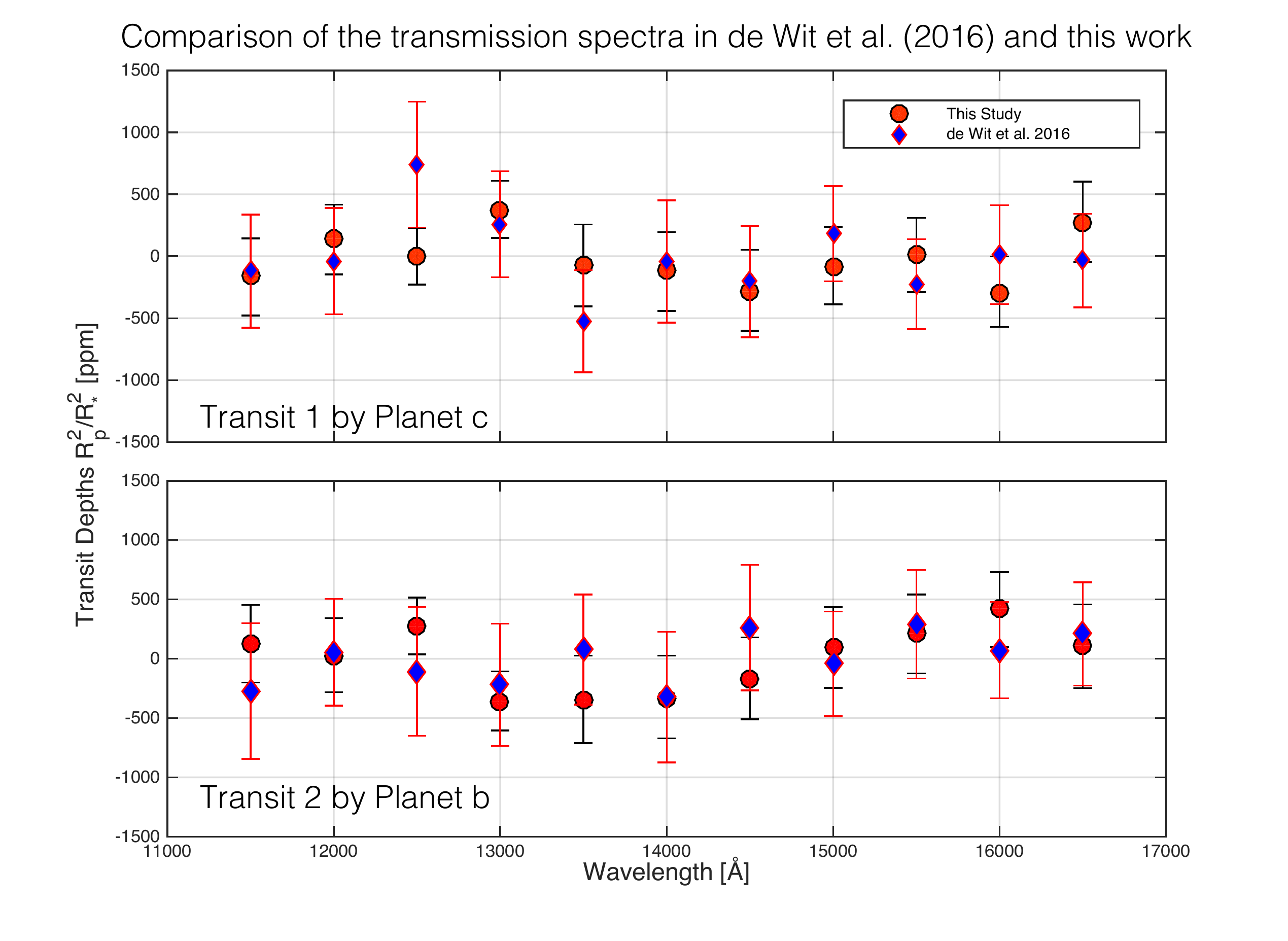}
  \caption{Comparison of the spectra of \trone{} b and c in our work and in \citet[][]{deWit2016}. The two reductions result in spectra that are statistically consistent with each other. All spectra have been mean-subtracted. } 
  \label{fig:7}
\end{figure*}

The transmission spectra from our reduction and that of \citet{deWit2016} agree within uncertainties (Figure~\ref{fig:7}).  For planets~b and c, we performed Kolmogorov-Smirnov tests on the spectra from this study and \citet[][]{deWit2016}.  The tests resulted in p-values of 0.9094 and 0.8286, i.e., they did not show any evidence for the data points being drawn from statistically different parent samples.
 
\subsection{Comparison of Spectra from Two Transits of Planet e}

\begin{figure}[!thb]
  \centering
  \includegraphics[width=\linewidth]{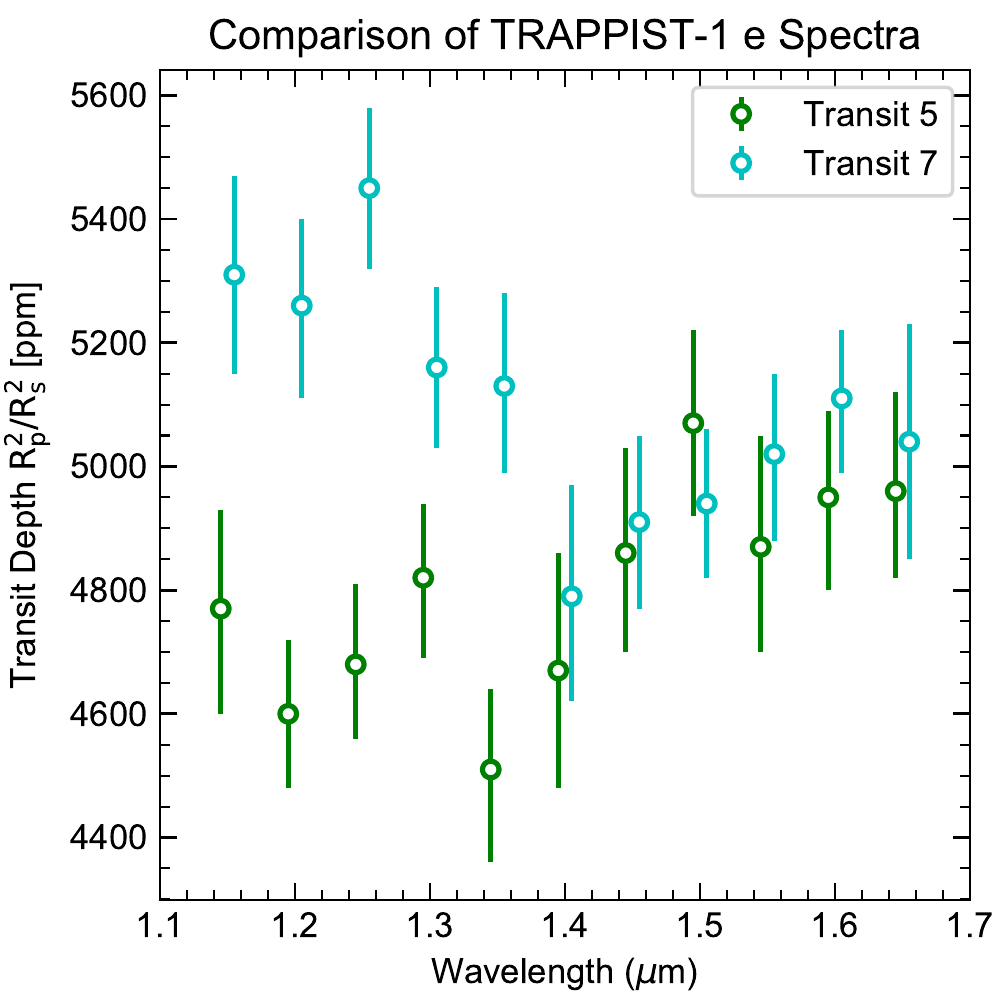}
  \caption{The transmission spectra of planet \trone{} e from two visits.  Green and cyan points are the spectra from Transits~5 and 7, respectively. The spectra are offset in wavelength for clarity.}
 \label{fig:8}
\end{figure}

The fact that our datasets contain two transit events (Transits~5 and 7) for planet \trone{}~e offers an opportunity to examine the similarities between the transmission spectra from the two epochs.  Figure~\ref{fig:8} shows the two spectra of \trone{}~e interpolated to the same wavelength bins. While the two spectra share some features such as average transit depth and non-detection of planetary water absorption, they differ in the blue parts (<1.35\,\micron) of the spectra.
There appears to be a difference in the overall slope between the two transit spectra. 

To quantify the agreement between the two spectra we calculated the bin-by-bin difference between them and report an average difference of 2.4$\sigma$. A KS-test yields a p-value of 0.01, i.e., supporting the conclusion that the two datasets are drawn from different parent populations. Manual inspection confirms that blue parts of the spectra contribute most of the difference (Figure~\ref{fig:8}).
This difference may be an indication of a time-evolving stellar contamination signal, which we discuss further in \S~\ref{sec:singleplanet} and \ref{Section:MultiEpochContamination}.

\subsection{No Evidence for Water Absorption Features in Individual or Combined Spectra}
\label{Section:WaterLimit}
In our data all six planets (\trone{}~b to g) have spectra that show no obvious absorption features in the 1.1 to 1.65 \micron{} wavelength range. We examine the possibility of planets sharing the similar spectral features by combining the spectra from the six planets.

We also compared our observed transmission combined spectra with a scaled water transmission model \citep[][]{Kempton2017}, as shown in Figure \ref{fig:6}. The deviations, as expressed in $\chi^{2}$, are 2.2, 1.3 and 1.0, which are all worse than those of the flat model, (0.85, 0.56 and 0.45, respectively). We attribute this to the contamination from stellar heterogeneity, which we will discuss in more detail in \S~\ref{Section:StellarContamination}.

Water is of the most interest among molecules with absorption features in the WFC3 G141 band-pass. Since no spectral features were observed, we estimate the upper limits of water absorption for each of the planets instead.  Following \citet[][]{Fu2017}, we quantified the water absorption amplitude $A_H$, i.e. the transit depth difference in and out of the water absorption band in terms of scale heights $H$, as
\begin{equation}
  A_{H}=\Delta d\,R_{*}^{2}/2R_{p}H.
\end{equation}
We estimated the scale height of each planet based on TTV masses from \citet[][]{Wang2017}. We set the host star radius to $R_{*}=0.114\pm0.006 R_{\sun}$ \citep[][]{Gillon2016} to calculate the absolute radii of the planets, and the host star effective temperature to $2,559\,\mathrm{K}$ \citep{Gillon2017}. For the mean molecular weight, we adopted 18~amu, corresponding to that of water. We calculated $\mathcal{U}(A_{H})$, the $3 \sigma$ upper bounds of $A_{H}$ for planets \trone{}b-g. 

The results along with essential information are presented in Table~\ref{tab:water}.
We do not find evidence for a planetary water absorption feature in any of the transmission spectra, i.e. $A_{H}$ is consistent with or less than 0 at the $1\sigma$ level in each case. 
However, the wide uncertainties on $A_{H}$ allow for the possibility of significant water absorption, as illustrated by the upper limits. 
Improved precisions on the transmission spectra may place tighter constraints on planetary absorption features, though additional observations will have to contend with the time-resolved activity and photospheric heterogeneity of the host star.

\begin{deluxetable*}{ccccccc}
  \tablecaption{Planetary Parameters for Water Absorption Calculations \label{tab:water}}
  \tablehead{Planet&Mass&Teff&$\Delta d$&Radius&$A_{H}$&$\mathcal{U}(A_{H})$\\
    &$M_{\earth}$&K&ppm&$R_{\Earth}$&&}
  \startdata
  b & $0.79\pm0.27$          & $400.1\pm7.7$ & $-413\pm266$ & $1.086\pm0.035$ & $-9.2\pm6.8$   & 11.2 \\
  c & $1.63\pm0.63$          & $341.9\pm6.6$ & $-212\pm245$ & $1.056\pm0.035$ & $-11.1\pm13.6$ & 29.7 \\ 
  d & $0.33\pm0.15$          & $288.0\pm5.6$ & $17\pm113$   & $0.772\pm0.030$ & $0.2\pm1.0$    & 3.2  \\ 
  e & $0.24^{+0.56}_{-0.24}$ & $251.3\pm4.9$ & $-209\pm129$ & $0.918\pm0.039$ & $-1.9\pm4.6$   & 11.9 \\ 
  e & $0.24^{+0.56}_{-0.24}$ & $251.3\pm4.9$ & $-303\pm126$ & $0.918\pm0.039$ & $-2.8\pm6.6$   & 17   \\  
  f & $0.36\pm0.12$          & $219.0\pm4.2$ & $-130\pm190$ & $1.045\pm0.038$ & $-2.3\pm3.5$   & 8.2  \\
  g & $0.566\pm0.0038$       & $198.6\pm3.8$ & $77\pm197$   & $1.127\pm0.041$ & $2.5\pm6.6$    & 22.3 \\
  \enddata
\end{deluxetable*}

\subsection{TRAPPIST-1 stellar activity}
\label{Discussion:StellarActivity}
\begin{figure*}[!thb]
  \centering
  \plottwo{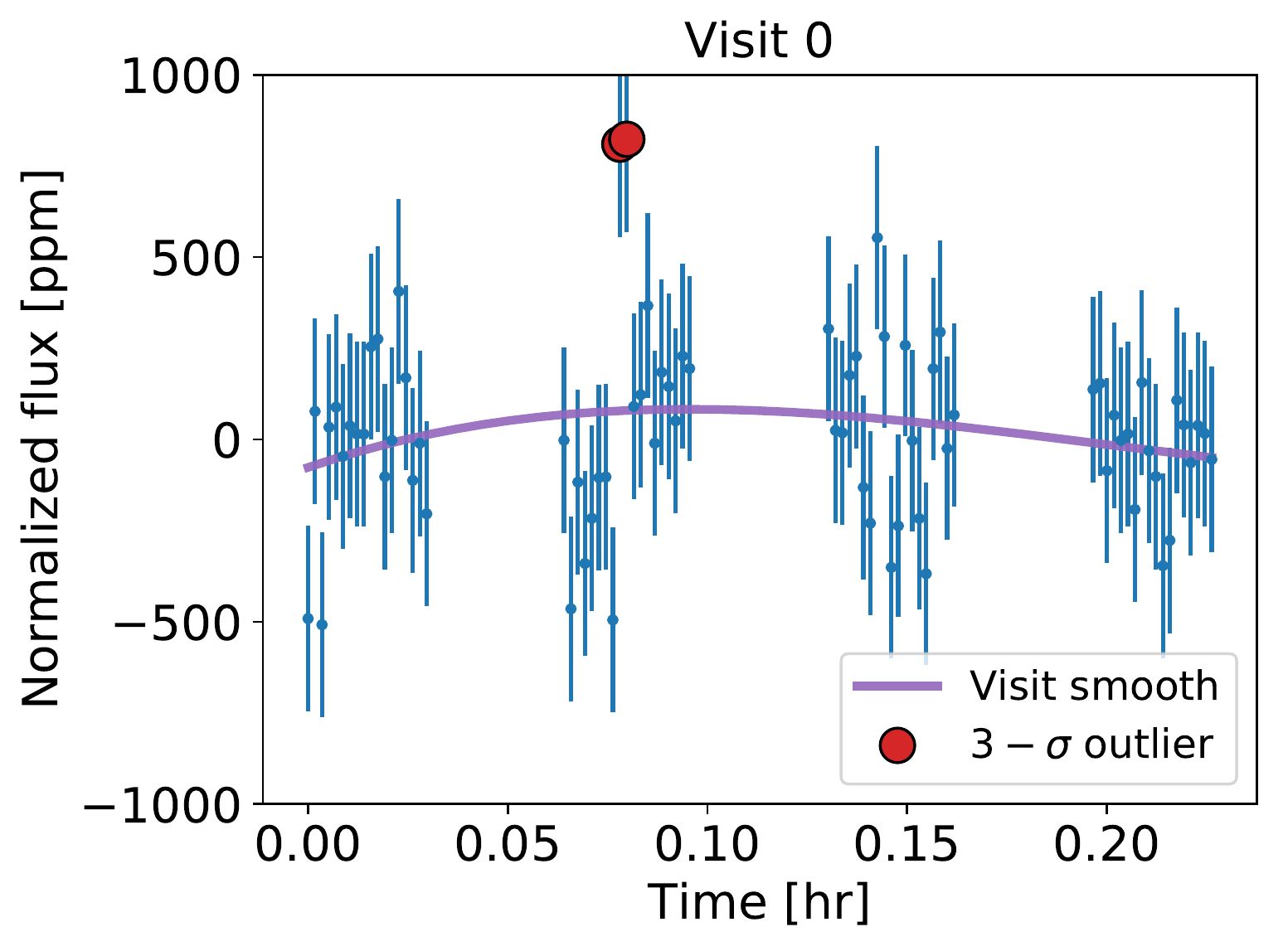}{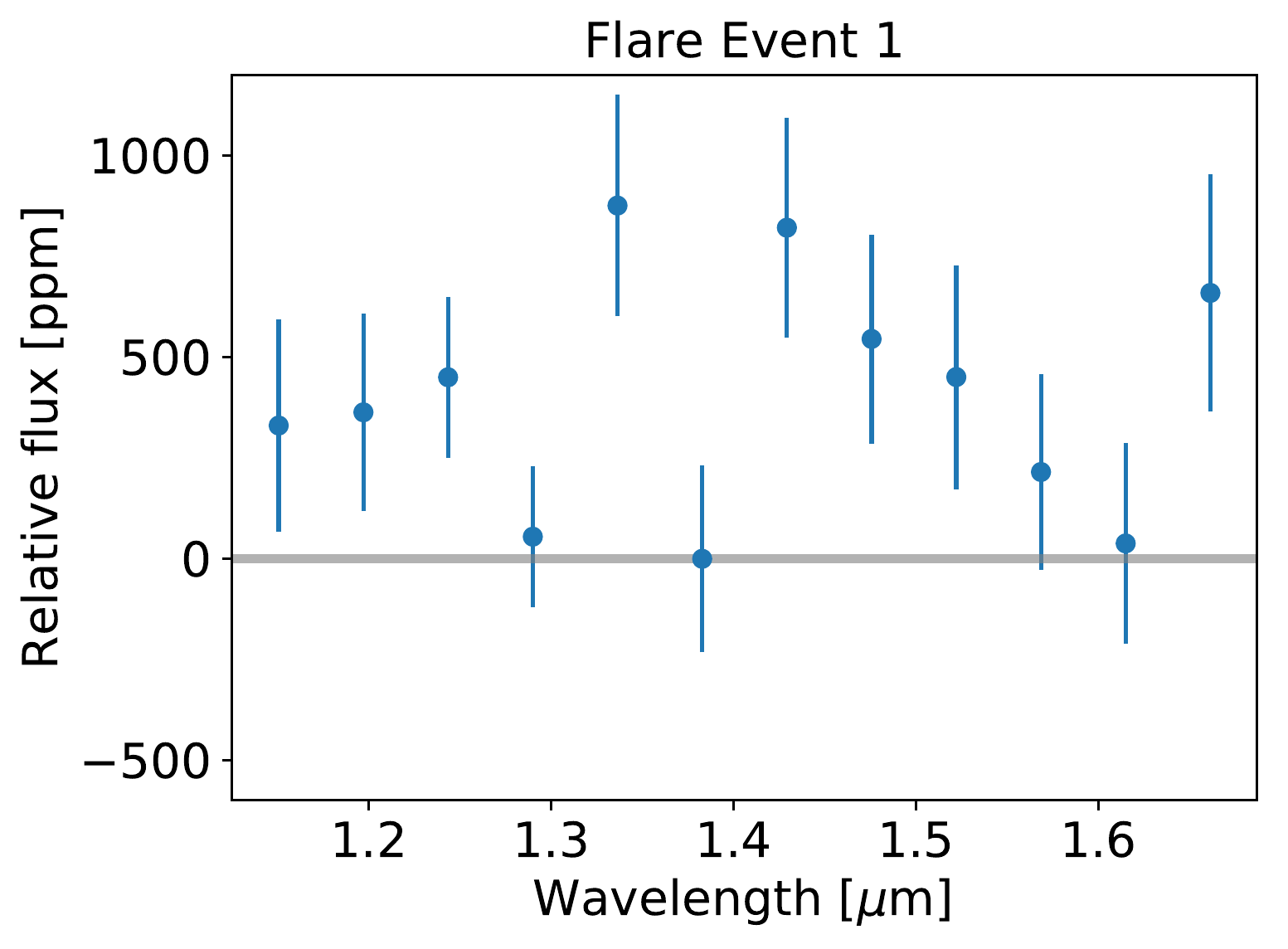}
  \plottwo{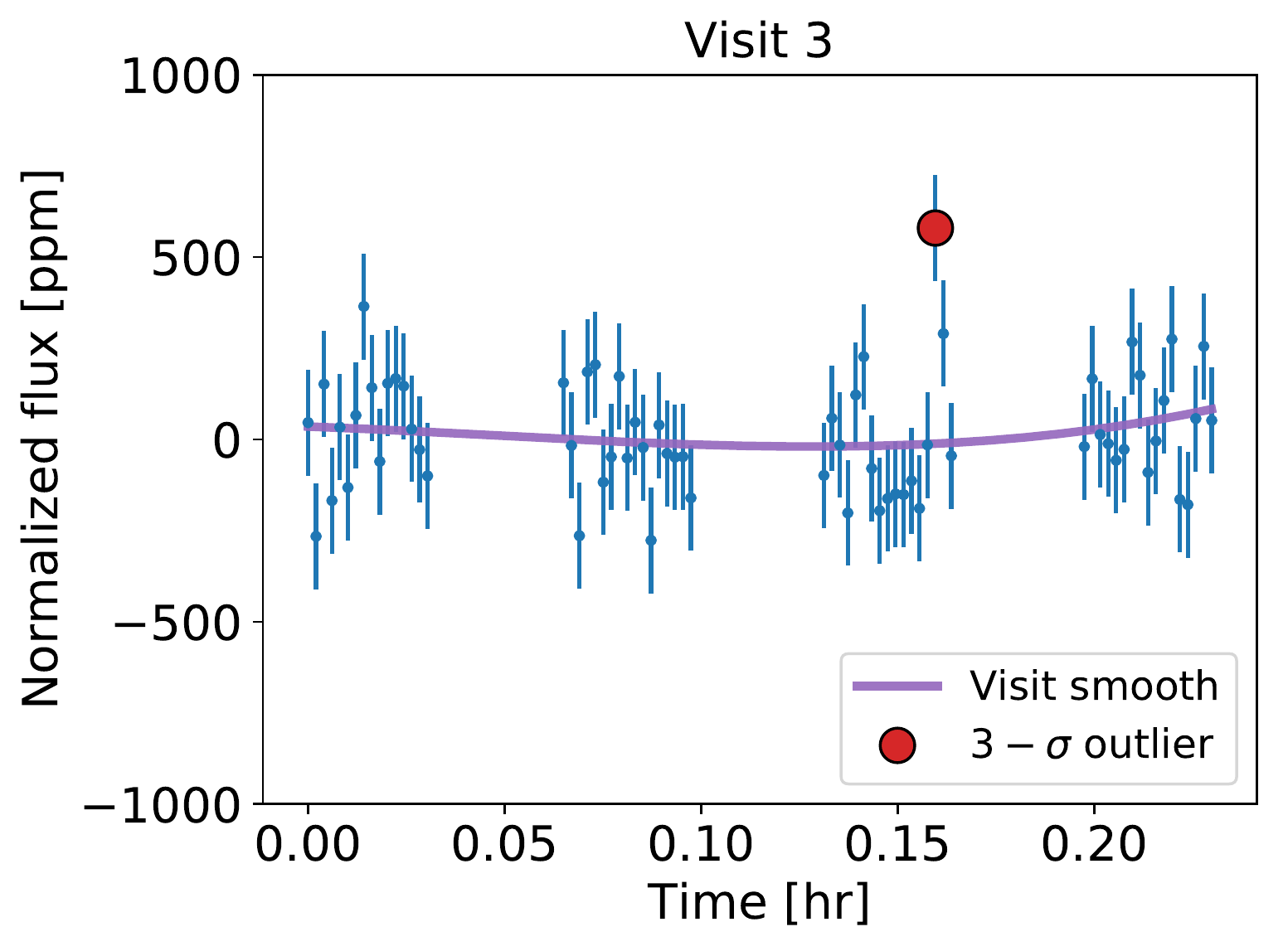}{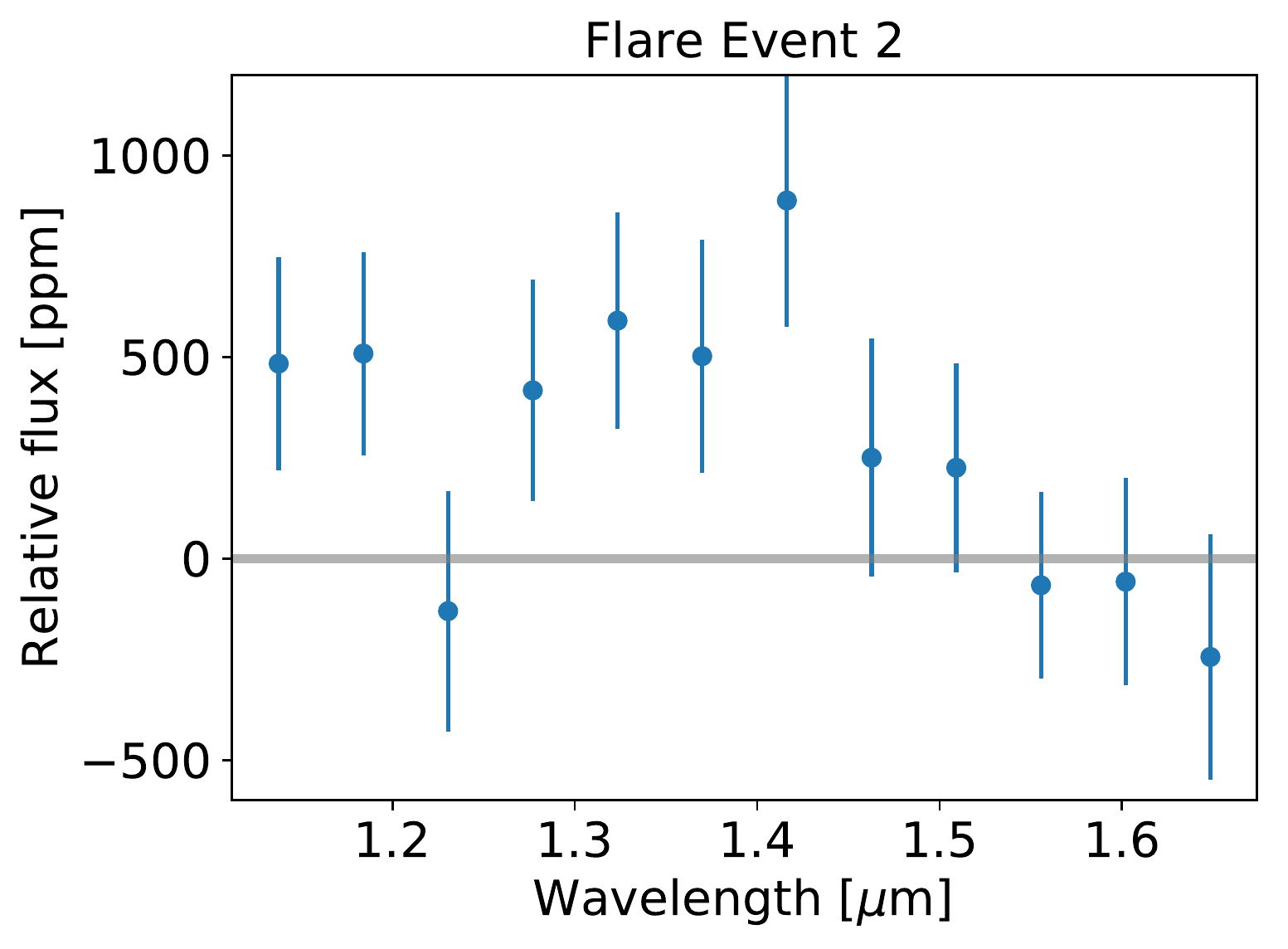}
  \caption{Identified stellar activity events. \emph{Left panel:} Two $3\sigma$ flare events were found in the light curves, one in Visit~0 of Program \first{} and another in Visit~3 of Program \second. The purple lines show the smoothed baseline constructed using third-order spline smoothing. Red circles mark the identified events. \emph{Right panel:} The spectral information of the identified event.}
  \label{Fig:FlaresSpots}
\end{figure*}

We examined \trone's out-of-transit broadband light curves and searched for possible flares and dimming events. We identified the baseline using the \texttt{scipy UnivariateSpline} routine, which smooths the light curves with third order splines. For the spline fit, the numbers of knots were optimized with inverse-standard-deviation weighting. We then computed the weighted average of deviations from unity within each window. Windows that had deviations above $3\sigma$ were selected as candidates for flare events or dimming events. In total, two potential flares were found. We summarize the key properties of these potential events in Table~\ref{tab:activity} and show their broadband light curves in Figure~\ref{Fig:FlaresSpots}. This figure also shows the spectrum of each event.

\begin{deluxetable}{cccccc}
  \tablecaption{ Identified Stellar Activity Information \label{tab:activity}}
  \tablehead{Type&Center Time&Deviation&Visit \& Orbit\\
    &MJD&$\sigma$}
  
  \startdata
  Flare & 57512.4 & 4.06  & Visit 0, Orbit 2 \\
  Flare & 57763.0 & 3.02  & Visit 3, Orbit 3 \\
  \enddata
\end{deluxetable}

The possible events were sampled in only a few read outs and their signal to noise ratio remains low, inhibiting their robust classification. We only note here a na\"ive expectation for these spectra, which is that (micro-)flares, representing hotter-than-photospheric plasma, would show an overall `bluer' continuum. This expectation may be met by these events, but the generally low quality of the spectra does not allow for meaningful characterizations. 

In total, 20 orbits are used in our analysis, amounting to approximately 850 minutes of observations. We estimate the stellar occurrence rate of marginally detectable (micro)flares to be on the order of 
$1/425~\mathrm{min}^{-1}$, i.e., an event every seven hours on average. We note here that the identification of the events leading to this statistic benefited from the use of the RECTE correction, which allowed us to include data from four additional orbits in our analysis and provided a ramp effect correction that was not affected by the flare events.

\section{Stellar contamination of the transmission spectra}
\label{Section:StellarContamination}

\trone{} demonstrates a 1.40-day periodic photometric variability in the $I$+$z$ bandpass with a full amplitude of roughly 1\% \citep[][Extended Data Figure 5]{Gillon2016}. \citet{Rackham2018} found that this observed variability is consistent with rotational modulations due a heterogeneous stellar photosphere with whole-disk spot and faculae covering fractions of $F_{spot} = 8^{+18}_{-7}$\% and $F_{fac} = 54^{+16}_{-46}$\%, respectively. These authors also found that spots and faculae, if present in regions of the stellar disk that are not occulted by the transiting planets, can alter transit depths by roughly 1--15$\times$ the strength of planetary atmospheric features, thus dominating the observed wavelength-dependent variations in transit depth (i.e., the ``{\em transit light source effect}'').

The observed transmission spectrum $D_{\lambda, \mathrm{obs}}$ is the multiplicative combination of the nominal transit depth $D_{\lambda}$ (i.e., the square of the true wavelength-dependent planet-to-star radius ratio) and the stellar contamination spectrum $\epsilon_{\lambda}$ \citep{Rackham2018}.
As the primary purpose of this exercise was to investigate the possible stellar contribution to the transmission spectrum, we assumed an achromatic transit depth $D$ for each planetary spectrum.
Thus, we modeled the observed transmission spectra as
\begin{equation}
D_\mathrm{\lambda, obs} = \epsilon_{\lambda}D
\label{eq:D}
\end{equation}
and assumed a stellar origin for all variations from a flat transmission spectrum.

\citet{Rackham2018} present a formalism for calculating the stellar contamination spectrum in the specific case that no heterogeneities---spots or faculae---are present within the transit chord or, if they are, they can be identified in the light curve and properly taken into account (their Equation~3). Of course, the precision of observations may not allow stellar surface heterogeneities within the transit chord to be reliably detected. In general, the stellar contamination spectrum $\epsilon_{\lambda}$ is given by the spectral ratio of the region occulted by the exoplanet relative to the integrated stellar disk. If we assume the transit chord is composed of the same spectral components as the integrated disk, namely spots, faculae, and immaculate photosphere, but we allow their covering fractions to differ from the whole-disk values, then the generalized stellar contamination spectrum is given by
\begin{equation}
\epsilon_{\lambda} = 
\frac{(1 - f_\mathrm{spot} - f_\mathrm{fac})S_\mathrm{\lambda, phot} 
      + f_\mathrm{spot}S_\mathrm{\lambda, spot} + f_\mathrm{fac}S_\mathrm{\lambda, fac}}
     {(1 - F_\mathrm{spot} - F_\mathrm{fac})S_\mathrm{\lambda, phot} 
      + F_\mathrm{spot}S_\mathrm{\lambda, spot} + F_\mathrm{fac}S_\mathrm{\lambda, fac}},
\label{eq:epsilon}
\end{equation}
in which $S_{\lambda, \mathrm{phot}}$, $S_{\lambda, \mathrm{spot}}$, $S_{\lambda, \mathrm{fac}}$
refer to the spectra of the photosphere, spots, and faculae, respectively;
$F_\mathrm{spot}$ and $F_\mathrm{fac}$ refer to the whole-disk covering fractions of spots and faculae, respectively;
and $f_\mathrm{spot}$ and $f_\mathrm{fac}$ refer to the spot and faculae covering fractions within the transit chord. We adopt this generalized stellar contamination framework in this analysis.

\subsection{Composite Photosphere and Atmospheric Transmission Model}
\label{sec:model-setup}

With this framework, we investigated the possible contribution of photospheric heterogeneities to the observed transmission spectra of the \trone{} planets using the composite photosphere and atmospheric transmission (CPAT) model \citep{Rackham2017}. We used an MCMC approach developed with the \texttt{PyMC} \citep{Patil2010} Python package to fit the CPAT model to our observations. The free parameters of the model and their priors are given in Table~\ref{tab:CPAT_priors}. For the prior on the photosphere temperature $T_\mathrm{phot}$, we adopted the stellar effective temperature $T_\mathrm{eff}$ from \citet{Gillon2017}, with a width $5 \times$ the reported uncertainty to allow the algorithm to thoroughly explore the parameter space. 
We adopted the same uncertainty for priors on the spot and facula temperatures, $T_\mathrm{spot}$ and $T_\mathrm{fac}$, with means given by $T_\mathrm{spot} = 0.86 \times T_\mathrm{phot}$ and $T_\mathrm{fac} = T_\mathrm{phot} + 100$~K, following \citet{Rackham2018}. 
For priors on the spot and faculae covering fractions, both within the transit chord and for the whole disk, we adopted normalized estimates of the covering fractions found for \trone{} by \citet{Rackham2018}.

\begin{deluxetable}{lccc}[!tbp]
  \tabletypesize{\footnotesize}
  \tablecaption{Priors for Stellar Contamination Model Fits \label{tab:CPAT_priors}}
  \tablehead{\colhead{Parameter} & 
    \colhead{Description}        & 
    \colhead{Prior}              & 
    \colhead{Unit} \\}
  \startdata
  $D$               & Nominal transit depth                   & Uniform(0, 100)      & \%  \\
  $T_\mathrm{phot}$ & Photosphere temperature                 & TruncNorm(2559, 250) & K   \\
  $T_\mathrm{spot}$ & Spot temperature                        & TruncNorm(2201, 250) & K   \\
  $T_\mathrm{fac}$  & Facula temperature                      & TruncNorm(2659, 250) & K   \\
  $F_\mathrm{spot}$ & Whole-disk spot covering fraction       & TruncNorm(8, 13)     & \%  \\
  $F_\mathrm{fac}$  & Whole-disk faculae covering fraction    & TruncNorm(54, 31)    & \%  \\
  $f_\mathrm{spot}$ & Transit chord spot covering fraction    & TruncNorm(8, 13)     & \%  \\
  $f_\mathrm{fac}$  & Transit chord faculae covering fraction & TruncNorm(54, 31)    & \%  \\
  $\eta$            & Spectra error inflation factor          & Uniform(1, 100)      & ... \\
  \enddata
  \tablecomments{Uniform(\textit{a}, \textit{b}) distributions are uniform between \textit{a} and \textit{b}. 
  TruncNorm($\mu$, $\sigma$) distributions are normal distributions with means $\mu$ and standard deviations $\sigma$. 
  The priors for $T_\mathrm{phot}$, $T_\mathrm{spot}$, and $T_\mathrm{fac}$ were truncated on the range [1000 K, 3000 K], given by the temperature limits of the DRIFT-PHOENIX model grid. 
  We enforced $T_\mathrm{fac}$ > $T_\mathrm{phot}$ and $T_\mathrm{phot}$ > $T_\mathrm{spot}$ with likelihood penalties. 
  Similarly, the covering fractions were allowed to vary over the range [0 \%, 100 \%] and we enforced $F_\mathrm{spot} + F_\mathrm{fac} <= 100 \%$ and $f_\mathrm{spot} + f_\mathrm{fac} <= 100 \%$ with likelihood penalties.}
\end{deluxetable}

The CPAT model describes the emergent disk-integrated spectrum of the photosphere as the sum of three distinct components, the immaculate photosphere, spots, and faculae, each covering some fraction of the projected stellar disk. We utilized the grid of DRIFT-PHOENIX model stellar spectra \citep{Hauschildt1999, Woitke2003, Woitke2004, Helling2006, Helling2008, Helling2008a, Witte2009, Witte2011} with solar metalicity ([Fe/H] = 0.0) to generate spectra for each component. We parameterized the three components by their temperatures ($T_{\mathrm{phot}}$, $T_{\mathrm{spot}}$, and $T_{\mathrm{fac}}$ for the photosphere, spots, and faculae, respectively), which we allowed to vary, and linearly interpolated between models with different temperatures to produce the component spectra. For all components, we linearly interpolated between models with $\log{g} = 5.0$ and $\log{g} = 5.5$ to produce spectra matching the surface gravity of \trone{} ($\log{g} = 5.21$), which we calculated from the star's mass and radius \citep{Gillon2017}.

While fitting the transmission spectra,
we required that the stellar parameters also produced a disk-integrated stellar spectrum matching the median observed out-of-transit stellar spectrum ($\lambda$ = 1.15--1.70~$\micron$) of \trone{}. We computed the disk-integrated spectrum as

\begin{equation}
S_\mathrm{\lambda, disk} =  (1 - F_\mathrm{spot} - F_\mathrm{fac})S_\mathrm{\lambda, phot} 
+ F_\mathrm{spot}S_\mathrm{\lambda, spot} + F_\mathrm{fac}S_\mathrm{\lambda, fac}
\label{eq:disk}
\end{equation}
ignoring projection effects owing to the positions of photospheric heterogeneities, which are not constrained by our model. Both the observed and model stellar spectra were normalized to the median flux between 1.27 and 1.31~$\micron$ for comparison. To account for discrepancies between the high-precision \textit{HST} observations and stellar models, we multiplied the observational uncertainties by an error inflation factor $\eta$, which was allowed to vary between 1 and 100 (see Table~\ref{tab:CPAT_priors}).

We utilized two variations of this modeling framework. In the first, which we call the \texttt{flat} model here, we set the active region covering fractions within the transit chord to the whole-disk values (i.e., $f_\mathrm{spot}=F_\mathrm{spot}$ and $f_\mathrm{fac}=F_\mathrm{fac}$). Thus, $\epsilon_{\lambda}=1$ (see Equation~\ref{eq:epsilon}) and there was no stellar contribution to the observed transmission spectra. The achromatic transit depth $D$ solely determined the transmission spectra model, while the stellar parameters in the model only affected the fit to the out-of-transit stellar spectrum. In the second framework, which we call the \texttt{contamination} model, we allowed the active region covering fractions within the transit chord to differ from the whole-disk values. In this case, both planetary and stellar parameters affected the model fit to the observed transmission spectra (Equation~\ref{eq:D}) and the stellar parameters (save the transit chord covering fractions) determined the fit to the stellar spectrum. \texttt{Flat} models have 7 free parameters, \texttt{contamination} models have 9, and each fit includes 144 data points---14 for the transmission spectrum and 130 for the stellar spectrum (see \S~\ref{sec:sources}).

For each transmission spectrum that we considered, we performed an MCMC optimization procedure for both of these modeling frameworks. In each procedure, we marginalized over the log likelihood of a multivariate Gaussian. We ran three chains of $5\times10^5$ steps with an additional $5\times10^4$ steps discarded as the burn-in. We checked for convergence using the Gelman-Rubin statistic $\hat{R}$ \citep[][]{Gelman1992} and considered chains to be well-mixed if $\hat{R} < 1.03$.

\subsection{Multi-instrument Transit Measurements}
\label{sec:sources}

We initially performed this analysis using only the \textit{HST} transit depths presented here and the \textit{Spitzer} 4.5~$\micron$ transit depths provided by \citet{Delrez2018}.  After the submission of this manuscript, \citet{Ducrot2018} presented \textit{K2} (0.42--0.9~$\micron$) transit depths and \textit{I+z} (0.8--1.1~$\micron$) transit depths from the SPECULOOS-South Observatory \citep[SSO;][]{SPECULOOS} for each of the planets discussed here.  In the following analysis, we consider the full K2+SSO+HST+Spitzer transmission spectra.  Thus, each transmission spectrum includes 14 transit depths: 11 HST/WFC3 depths from this analysis, along with a \textit{K2}, SSO, and Spitzer 4.5~$\micron$ depth.

We note that \citet{Ducrot2018} examined the impact of stellar contamination from TRAPPIST-1 by comparing the \textit{K2+SOO+HST+Spitzer} spectra of TRAPPIST-1 b+c to a model from a pre-peer-reviewed version of this work \citep[see Figure 4 in][]{Ducrot2018}. They found a 20-$\sigma$ discrepancy between the observed K2 transit depth and the model prediction and claimed that the stellar contamination model can be ``firmly discarded''. For the model that was used in the study of Ducrot et al., we assumed that spots and faculae were present only in the non-occulted stellar disk but not in the transit chords, i.e. the spectra of the transit chords were the same as the immaculate photosphere spectrum. During the reviewing process and prior to the publication of \citet{Ducrot2018}, we updated the model to a more general form in which spots and faculae affect both the non-occulted disk and the transit chords (see Equation~\ref{eq:epsilon}). The analysis of the \textit{K2+SOO+HST+Spitzer} dataset using the more general stellar contamination model is presented here.
Appendix~\ref{sec:HST+Spitzer} provides the results of this
same analysis considering only the HST+Spitzer transmission spectra, along with the accompanying predictions for the \textit{K2} and \textit{I+z} transit depths.
The results of the analyses for the two datasets
are \emph{fully consistent}. Most notably, the best-fit \texttt{contamination} models from the HST+Spitzer analysis offer accurate predictions of the \textit{K2} and \textit{I+z} transit depths for the combined transmission spectra.

We also note that transit depth determinations depend on $i$ and $a/R_{s}$, which may vary between analyses of transit data.
In our analysis we fixed these values, which are hard to constrain with the \text{HST} observations, to those reported by \citet{Gillon2017}.
As the K2, SSO and Spitzer observations all covered a longer baseline, however, \citet{Ducrot2018} and \citet{Delrez2018} both benefited from larger datasets including many repeated transits, which allowed them to fit for more parameters. 
To be more specific, both these works adopted impact parameter as a free parameter in MCMC analyses. 
They also include the stellar mass and radius as parameters either fixed or with a prior distribution, which combine with the orbital periods reported by \citet{Gillon2017} to yield the system scale $a/R_{s}$ for each planet.
For a comparison of the orbit parameters, we refer the reader to Table~\ref{tab:transit}, which contains the parameters that we adopted from \citet{Gillon2017}, and Table~1 of \citet{Delrez2018}, which details the fitted parameter values from that analysis\footnote{The fitted parameters of $i$ and $a/R_{s}$ are not provided by \citet{Ducrot2018}.}.
In short, we find that the inclinations and system scales that we adopt differ on average from those of \citet{Delrez2018} by $0.9\sigma$ and $0.2\sigma$ respectively. 
We note that these subtle differences might lead to minor differences in the following analysis.

\subsection{Fits to single-planet transmission spectra}
\label{sec:singleplanet}

We first fit CPAT models to each of the seven, single-transit transmission spectra presented in this work as well as the weighted mean spectrum of \trone{}~e from Transits~5 and 7. We included the corresponding \textit{K2}, SSO \citep{Ducrot2018}, and Spitzer 4.5~$\micron$ transit depths \citep{Delrez2018} in the fitting procedure. Table~\ref{tab:results_CPAT_single_all_data} summarizes the posterior distributions of the fitted parameters and provides the Akaike Information Criterion \citep{Akaike1974} corrected for small sample sizes \citep[AIC$_\textrm{c}$;][]{Sugiura1978}, Bayesian Information Criterion \citep[BIC;][]{Schwarz1978}, $\chi^{2}$ and its corresponding $p$-values for each fit.
We use the information criteria to evaluate the efficacy of the increased model complexity of the \texttt{contamination} model compared to the \texttt{flat} model.
Following convention \citep[e.g.,][]{Liddle2007}, for both information criteria (IC)---i.e., AIC$_\textrm{c}$ and BIC---we interpret $\Delta \textrm{IC} > +5$ and $\Delta \textrm{IC} > +10$ relative to the best model as `strong' and `decisive' evidence against the current model, respectively. 

Of the single-planet spectra, both information criteria only indicate decisive evidence against the \texttt{flat} model for one case, the \trone{}~d dataset. Considering its $\chi^{2}$ value, the \texttt{flat} model for this dataset is ruled out at 99\% confidence ($p = 0.01$).
For \trone{}~e, of the two transits that were observed in two visits separated by twelve days, one shows evidence to support the \texttt{contamination} model (Transit~7), but the other (Transit~5) does not. This difference result for two transits of the same planet could suggest temporal variability of stellar contamination.
For the remaining datasets, both information criteria generally support the same model---the exception being the \trone{}~c dataset---though they do not both rise to the level of decisive evidence.
In general, the additional complexity of the \texttt{contamination} models results in lower $\chi^{2}$ values, as expected, though the information criteria show that the additional complexity is not decisively warranted by the data for any of the datasets besides that of \trone{}~d.

\begin{deluxetable*}{lcccccccccccccccc}
\tabletypesize{\scriptsize}
\tablecaption{
This table has been updated to reflect the results of the K2+SSO+HST+Spitzer fits.  The original table with the HST+Spitzer fits is available in the Appendix.
Results of Stellar Contamination Model Fits to Single-Planet Spectra\label{tab:results_CPAT_single_all_data}} 
\tablehead{\colhead{Dataset}                    & 
           \colhead{Model}                      & 
           \multicolumn{9}{c}{Fitted Parameter} & 
           \colhead{AIC$_\textrm{c}$}           & 
           \colhead{$\Delta \textrm{AIC}_\textrm{c}$}    & 
           \colhead{BIC}                        & 
           \colhead{$\Delta \textrm{BIC}$}      & 
           \colhead{$\chi^{2}$}                 & 
           \colhead{$p$} \\
           \cline{3-11} 
           \colhead{}                           & 
           \colhead{}                           & 
           \colhead{$D$ (\%)}                   & 
           \colhead{$T_\mathrm{phot}$ (K)}      & 
           \colhead{$T_\mathrm{spot}$ (K)}      & 
           \colhead{$T_\mathrm{fac}$ (K)}       & 
           \colhead{$F_\mathrm{spot}$}          & 
           \colhead{$F_\mathrm{fac}$}           & 
           \colhead{$f_\mathrm{spot}$}          & 
           \colhead{$f_\mathrm{fac}$}           & 
           \colhead{$\eta$}                     & 
           \colhead{}                           & 
           \colhead{}                           & 
           \colhead{}                           & 
           \colhead{}                           & 
           \colhead{}                           & 
           \colhead{}}
         
\startdata
b & \texttt{flat}  & ${0.744}^{+0.005}_{-0.006}$ & ${2118}^{+87}_{-127}$  & ${1962}^{+111}_{-131}$ & ${2974}^{+26}_{-14}$ & ${16}^{+7}_{-15}$  & ${46}^{+5}_{-6}$  & -                 & -                  & ${23}^{+1}_{-2}$ & -736.0 & 6.2 &-716.0& 0.9 & 150.0 & 0.21 \\
b & \texttt{cont.} & ${0.677}^{+0.022}_{-0.023}$ & ${2433}^{+240}_{-245}$ & ${2009}^{+150}_{-99}$  & ${2950}^{+50}_{-25}$ & ${39}^{+10}_{-10}$ & ${49}^{+7}_{-10}$ & ${9}^{+5}_{-9}$   & ${48}^{+8}_{-7}$   & ${23}^{+1}_{-2}$ & -742.2 & - &-716.9& - & 136.4 & 0.45 \\
c & \texttt{flat}  & ${0.705}^{+0.005}_{-0.005}$ & ${2117}^{+87}_{-125}$  & ${1961}^{+106}_{-136}$ & ${2974}^{+26}_{-14}$ & ${16}^{+7}_{-15}$  & ${46}^{+5}_{-6}$  & -                 & -                  & ${23}^{+1}_{-1}$ & -746.0 & 1.9 &-726.1& -3.5 & 141.8 & 0.37 \\
c & \texttt{cont.} & ${0.663}^{+0.021}_{-0.020}$ & ${2271}^{+147}_{-205}$ & ${1960}^{+91}_{-165}$  & ${2964}^{+36}_{-20}$ & ${32}^{+10}_{-10}$ & ${47}^{+6}_{-7}$  & ${8}^{+4}_{-8}$   & ${47}^{+6}_{-7}$   & ${23}^{+1}_{-2}$ & -747.9 & - &-722.6& - & 133.4 & 0.52 \\
d & \texttt{flat}  & ${0.388}^{+0.003}_{-0.003}$ & ${2118}^{+86}_{-127}$  & ${1962}^{+108}_{-133}$ & ${2974}^{+26}_{-14}$ & ${16}^{+7}_{-15}$  & ${46}^{+5}_{-6}$  & -                 & -                  & ${23}^{+1}_{-2}$ & -723.2 & 24.6 &-703.2& 19.3 & 180.9 & 0.01 \\
d & \texttt{cont.} & ${0.309}^{+0.015}_{-0.016}$ & ${2551}^{+253}_{-157}$ & ${2000}^{+107}_{-78}$  & ${2937}^{+63}_{-27}$ & ${48}^{+11}_{-9}$  & ${49}^{+9}_{-12}$ & ${8}^{+4}_{-8}$   & ${52}^{+12}_{-10}$ & ${23}^{+1}_{-2}$ & -747.8 & - &-722.5& - & 151.4 & 0.16 \\
e (T5)\tablenotemark{a} & \texttt{flat}  & ${0.477}^{+0.004}_{-0.004}$ & ${2117}^{+86}_{-126}$  & ${1961}^{+110}_{-131}$ & ${2974}^{+26}_{-14}$ & ${16}^{+7}_{-15}$  & ${46}^{+5}_{-6}$  & -                 & -                  & ${23}^{+1}_{-2}$ & -759.8& -3.2 & -739.9& -8.7 & 141.4 & 0.38 \\
e (T5)\tablenotemark{a} & \texttt{cont.} & ${0.498}^{+0.021}_{-0.020}$ & ${2125}^{+84}_{-120}$  & ${1982}^{+118}_{-124}$ & ${2972}^{+28}_{-15}$ & ${16}^{+8}_{-12}$  & ${47}^{+5}_{-6}$  & ${12}^{+5}_{-12}$ & ${43}^{+5}_{-6}$   & ${23}^{+1}_{-2}$ & -756.6 & - &-731.2& - & 139.5 & 0.38 \\
e (T7)\tablenotemark{b} & \texttt{flat}  & ${0.504}^{+0.004}_{-0.004}$ & ${2116}^{+87}_{-125}$  & ${1960}^{+107}_{-135}$ & ${2975}^{+25}_{-14}$ & ${16}^{+7}_{-15}$  & ${46}^{+5}_{-6}$  & -                 & -                  & ${23}^{+1}_{-2}$ & -743.3 & 12.6 &-723.4& 7.1 & 158.9 & 0.10 \\
e (T7)\tablenotemark{b} & \texttt{cont.} & ${0.449}^{+0.021}_{-0.021}$ & ${2559}^{+163}_{-118}$ & ${2032}^{+100}_{-75}$  & ${2937}^{+63}_{-27}$ & ${46}^{+9}_{-9}$   & ${51}^{+8}_{-12}$ & ${10}^{+4}_{-10}$ & ${38}^{+11}_{-8}$  & ${23}^{+1}_{-2}$ & -755.9 & - &-730.5& - & 139.7 & 0.37 \\
e\tablenotemark{c}      & \texttt{flat}  & ${0.493}^{+0.003}_{-0.003}$ & ${2116}^{+85}_{-127}$  & ${1960}^{+108}_{-133}$ & ${2974}^{+26}_{-14}$ & ${16}^{+7}_{-15}$  & ${46}^{+5}_{-6}$  & -                 & -                  & ${23}^{+1}_{-2}$ & -767.5 & -5.6 &-747.5& -10.9 & 142.0 & 0.37 \\
e\tablenotemark{c}      & \texttt{cont.} & ${0.480}^{+0.022}_{-0.023}$ & ${2267}^{+149}_{-209}$ & ${1981}^{+102}_{-172}$ & ${2965}^{+35}_{-20}$ & ${30}^{+10}_{-11}$ & ${48}^{+6}_{-8}$  & ${9}^{+4}_{-9}$   & ${43}^{+6}_{-6}$   & ${23}^{+1}_{-2}$ & -761.9 & - &-736.6& - & 140.6 & 0.35 \\
f & \texttt{flat}  & ${0.641}^{+0.005}_{-0.005}$ & ${2117}^{+87}_{-125}$  & ${1961}^{+111}_{-130}$ & ${2974}^{+26}_{-14}$ & ${16}^{+7}_{-15}$  & ${46}^{+5}_{-6}$  & -                 & -                  & ${23}^{+1}_{-2}$ & -746.7 & 0 &-726.7& -5.3 & 144.0 & 0.33 \\
f & \texttt{cont.} & ${0.627}^{+0.026}_{-0.027}$ & ${2395}^{+199}_{-254}$ & ${2022}^{+159}_{-106}$ & ${2959}^{+41}_{-23}$ & ${35}^{+11}_{-11}$ & ${50}^{+6}_{-9}$  & ${10}^{+5}_{-10}$ & ${40}^{+8}_{-7}$   & ${23}^{+1}_{-2}$ & -746.7 & - &-721.4& - & 135.8 & 0.46 \\
g & \texttt{flat}  & ${0.774}^{+0.006}_{-0.006}$ & ${2116}^{+87}_{-126}$  & ${1960}^{+109}_{-131}$ & ${2974}^{+26}_{-14}$ & ${16}^{+7}_{-15}$  & ${46}^{+5}_{-6}$  & -                 & -                  & ${23}^{+1}_{-2}$ & -750.4 & -7.8&-730.4& -13.1 & 137.3 & 0.48 \\
g & \texttt{cont.} & ${0.755}^{+0.027}_{-0.027}$ & ${2176}^{+109}_{-173}$ & ${1987}^{+122}_{-133}$ & ${2969}^{+31}_{-17}$ & ${23}^{+12}_{-13}$ & ${48}^{+5}_{-6}$  & ${10}^{+4}_{-10}$ & ${47}^{+6}_{-7}$   & ${23}^{+1}_{-2}$ & -742.6 & - &-717.3& - & 139.1 & 0.39 \\
\enddata

\tablecomments{
Posterior distributions are provided for 16 MCMC optimization procedures resulting from two model frameworks and eight datasets. 
The model frameworks are the \texttt{flat} model, in which the transmission spectrum is unaffected by photospheric features, and the \texttt{contamination} model (identified as \texttt{cont.} here), in which the covering fractions of spots and faculae are allowed to differ from the whole-disk covering fractions. 
The eight datasets are the seven individual transits and the combined \trone{}~e dataset. 
Medians and 68\% confidence intervals of the fitted parameters are quoted. 
The Akaike Information Criterion corrected for small sample sizes, Bayesian Information Criterion, $\chi^{2}$, and corresponding p-value for each model are provided. 
Each model has 144 data points (14 for the transmission spectrum and 130 for the stellar spectrum). 
\texttt{Flat} models have 137 degrees of freedom and \texttt{contamination} models have 135.}
\tablenotetext{a}{Transit 5}
\tablenotetext{b}{Transit 7}
\tablenotetext{c}{Weighted mean of Transits 5 and 7}
\end{deluxetable*}

\subsection{Fits to combined transmission spectra}

The observed effect of a stellar contamination signal scales with the transit depth (Equation~\ref{eq:D}).  
The impact of stellar contamination is therefore more readily observable in the spectra of exoplanets with deeper transit depths.  In the case of \trone{}, assuming that a steady-state
stellar contamination signal similarly affects all the individual transmission spectra, we can co-add the individual transmission spectra to increase the SNR.  An examination of the combined transit spectrum can reveal if the regions probed by the transit chords have different spectra from the average spectrum of the stellar disk.

Thus, in addition to the single-planet spectra, we fit both model frameworks to seven combinations of \trone{} transmission spectra. The first combination is the sum of transit depths for all \trone{} planets observed with HST, b--g, using the weighted mean spectrum of \trone{}~e from Transits~5 and 7. The resulting spectrum utilizes all of the available data and is, in effect, what one would observe if \trone{}~b--g transited simultaneously. This approach probes for shared spectral features, similar to the analysis of the double transit of \trone{}~b and c by \citet{deWit2016}. In this case, we are primarily interested in a stellar contribution that affects all transmission spectra similarly, such as surface active regions that are outside of all the planetary transit chords. This can be more easily studied in the combined spectra because the stellar contamination signal combines multiplicatively with any planetary transmission spectrum.  The remaining combinations exclude the contribution from one of the six planets in turn, allowing us to examine the effect of the individual planets on the combined result.  For each combination, we also included the sum of the corresponding \textit{K2} and SSO \citep{Ducrot2018} and Spitzer 4.5~$\micron$ transit depths \citep{Delrez2018} in the fitting procedure and added the uncertainties of the individual transit depths in quadrature.

\begin{figure*}[!htbp]
\label{fig:CPAT_oot_spectra_all_data}
\includegraphics[width=\linewidth]{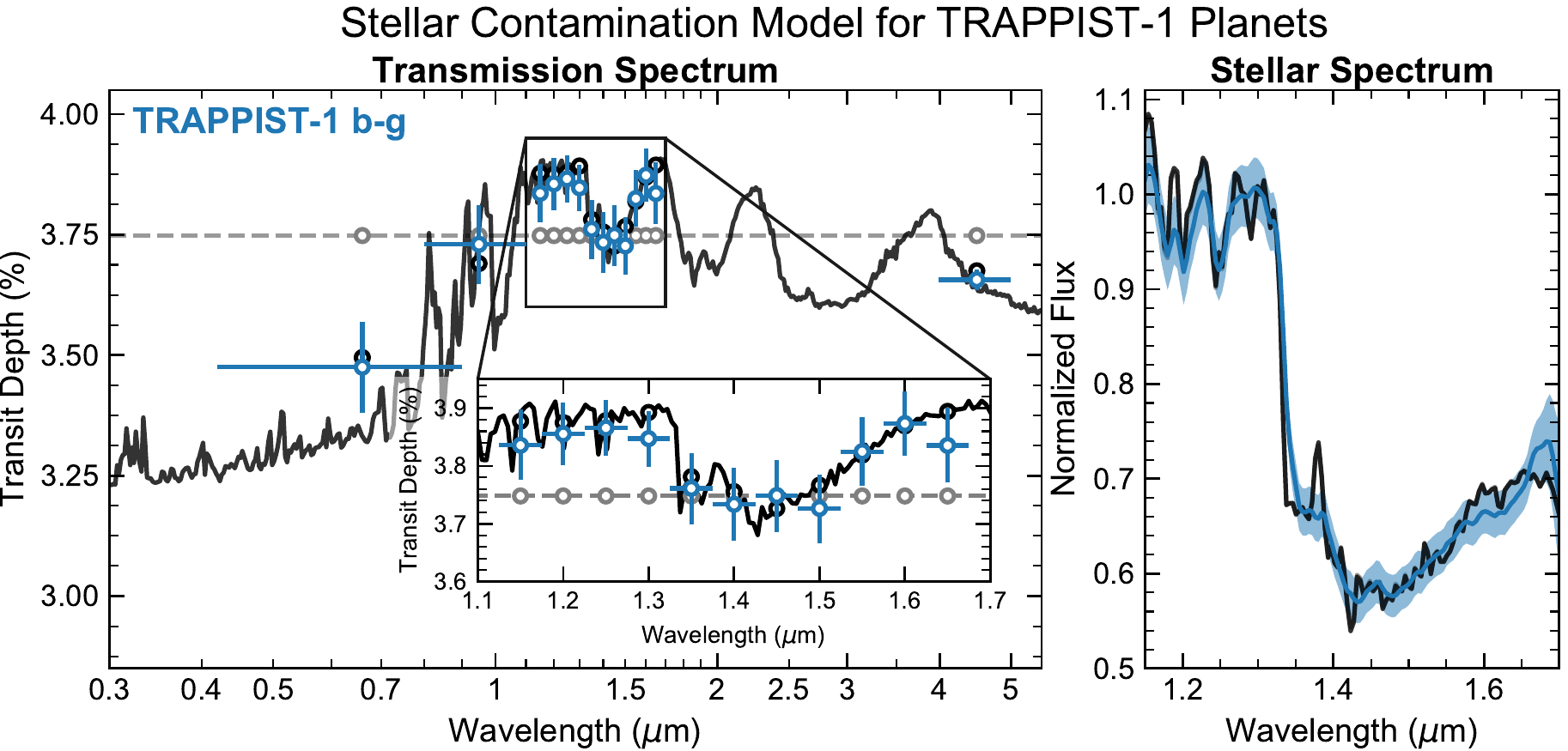}
\caption{
Stellar contamination model jointly fit to K2+SSO+HST+Spitzer \trone{} combined transmission spectra and observed \textit{HST} stellar spectrum.
{The left panel shows the combined transmission spectrum (\textit{blue points}) and best-fitting \texttt{contamination} and \texttt{flat} models (\textit{black solid} and \textit{gray dashed lines}, respectively). The inset panel highlights the HST/WFC3 G141 data.}
The right {panel shows} the observed HST/WFC3 G141 out-of-transit stellar spectrum of \trone{} (\textit{{blue line})} with a scaled uncertainty determined by the MCMC optimization procedure (\textit{shaded region}).
{The best-fit disk-integrated model stellar spectra for the \texttt{contamination} (\textit{black lines}) and \texttt{flat} models (\textit{gray lines}) are indistinguishable.}
}
\end{figure*}

\begin{figure*}[!htbp]
\label{fig:CPAT_fits_five_planets_all_data}
\includegraphics[width=\linewidth]{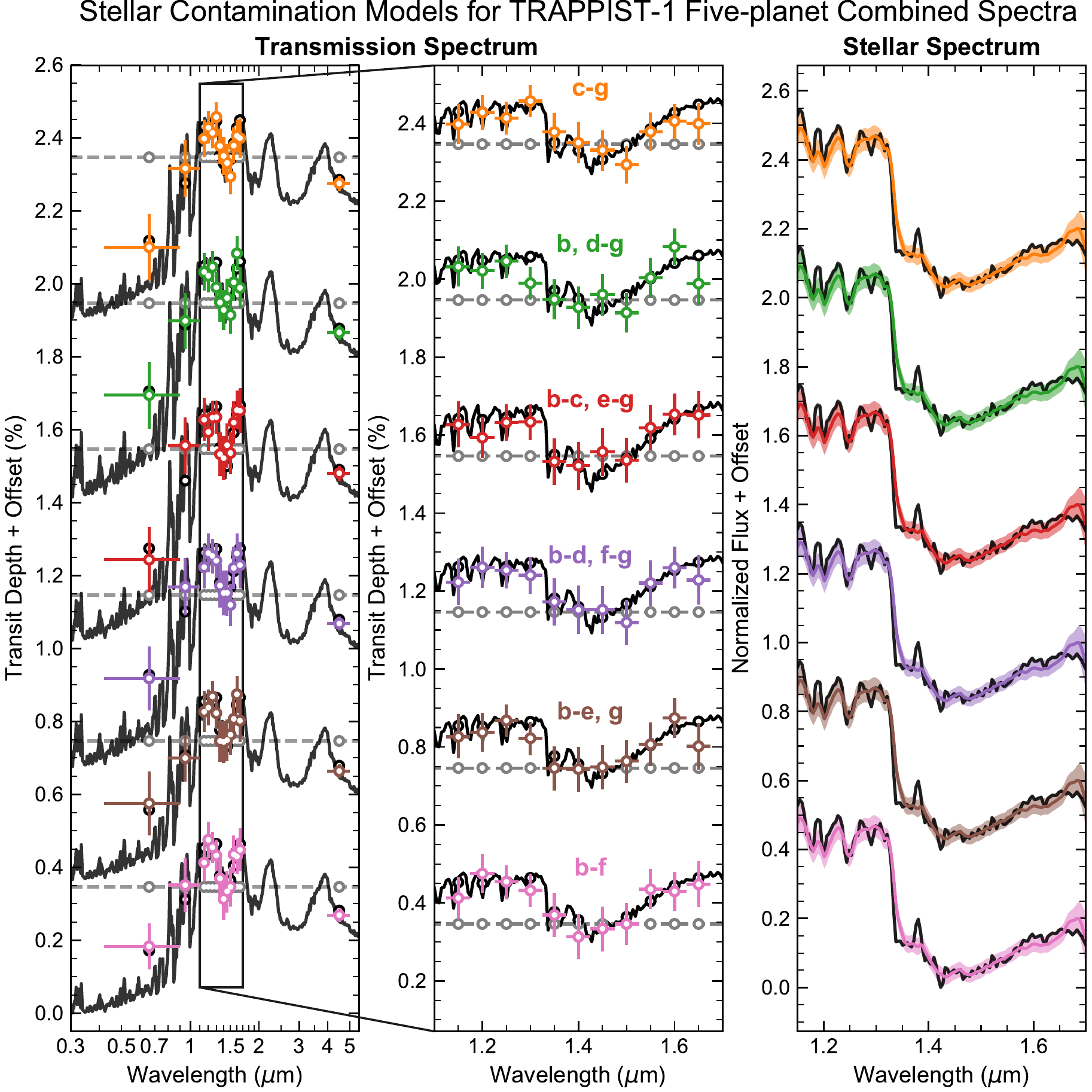}
\caption{
Stellar contamination models jointly fit to K2+SSO+HST+Spitzer \trone{} five-planet combined transmission spectra and observed \textit{HST} stellar spectrum.
Data and models are offset for clarity. 
The left panel shows the combined transmission spectra and models, the middle panel highlights the HST/WFC3 G141 transmission spectrum, and the right panel shows the HST/WFC3 G141 out-of-transit stellar spectrum. 
The best-fit disk-integrated model stellar spectra for the \texttt{contamination} (\textit{black lines}) and \texttt{flat} models (\textit{gray lines}) are indistinguishable.
The figure elements are the same as those in Figure~\ref{fig:CPAT_oot_spectra_all_data}.
}
\end{figure*}

Figure~\ref{fig:CPAT_oot_spectra_all_data} shows the \trone{} b--g combined transmission spectrum, the out-of-transit stellar spectrum, and the best-fit \texttt{flat} and \texttt{contamination} models. The \textit{K2}, SSO, and Spitzer transit depths are  3.6$\sigma$, 1.0$\sigma$, and 7.1$\sigma$ below the mean of the combined HST transit depth, respectively. The combined HST transmission spectrum displays a notable decrease in transit depth around 1.4~$\micron$, which coincides with a strong water absorption band. This ``inverted'' water feature is the opposite of the water absorption signature commonly observed in transiting exoplanet atmospheres \citep[e.g.,][]{Sing2016}. The offsets between instruments and the apparent 1.4~$\micron$ decrease are also evident in all five-planet combined transmission spectra (Figure~\ref{fig:CPAT_fits_five_planets_all_data}), which illustrates that it is not due solely to the spectrum of an individual planet. We find that the offsets between instruments and the apparent 1.4~$\micron$ decrease are \emph{both} reproduced well by the \texttt{contamination} model for each combined spectrum.

Table~\ref{tab:results_CPAT_combined_all_data} provides the complete results of the model fits to the combined spectra. 
For each combination, the AIC$_\textrm{c}$ and BIC both prefer the \texttt{contamination} model. 
According to the information criteria, each dataset provides decisive evidence against the \texttt{flat} model ($\Delta \textrm{AIC}_\textrm{c} > +10$, $\Delta \textrm{BIC} > +10$). 
In each case, the information criteria show that the data warrant the inclusion of the additional parameters in the \texttt{contamination} models.

\subsubsection{Impact of the stellar spectrum}

\begin{figure}
\label{fig:stellar_spectra}
\includegraphics[width=\linewidth]{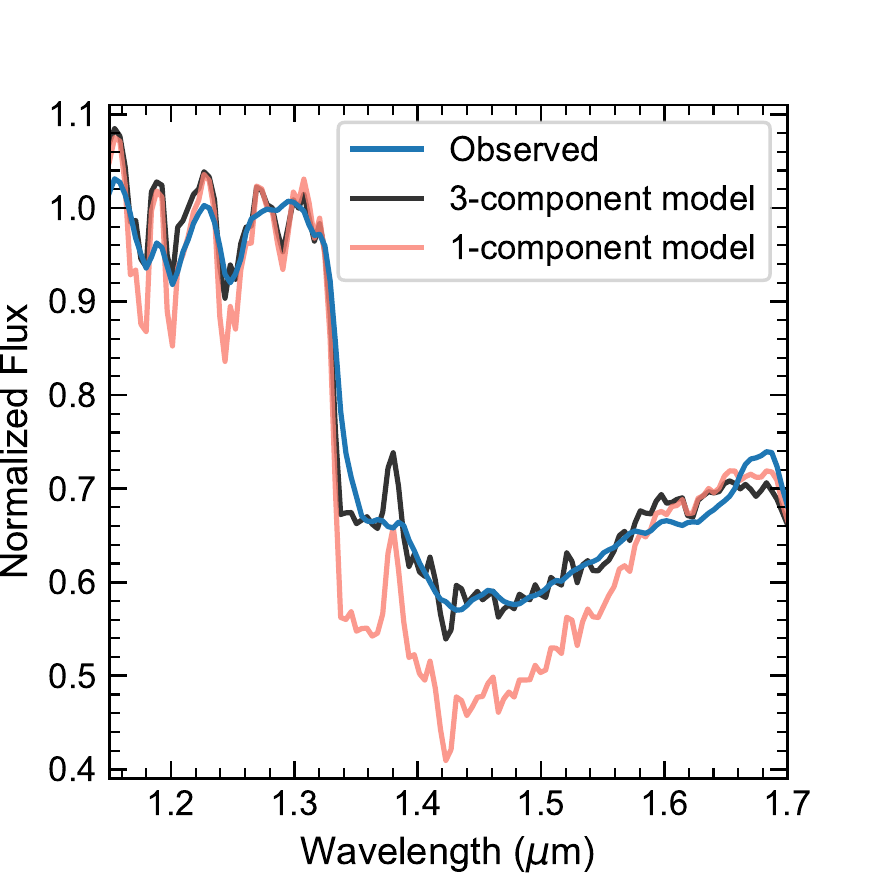}
\caption{
Comparison of the observed \trone{} HST/WFC3 G141 spectrum to models.
The observed out-of-transit stellar spectrum is shown in blue; 
the uncertainties on the observed spectrum are smaller that the line width.
The black line shows the best-fit disk-integrated model spectrum from the \texttt{contamination} model fit to the combined \trone{} b--g dataset (Table~\ref{tab:results_CPAT_combined_all_data}).
All other three-component fitted models from this analysis are indistinguishable.
By contrast, the DRIFT-PHOENIX model stellar spectrum for a star with $T=2559$~K and $\log{g}=5.21$ (salmon line) provides a relatively poor fit to the data.}
\end{figure}

For all model fits, both to the individual and combined transmission spectra, the fitted value of $\eta$ is ${23}^{+1}_{-2}$ (Tables~\ref{tab:results_CPAT_single_all_data} and \ref{tab:results_CPAT_combined_all_data}).
Notably, this is the case for all \texttt{flat} models, in which the stellar parameters are determined by the out-of-transit stellar spectrum alone and have no effect on the fit to the transmission spectra.
This implies that the observational errors of the HST/WFC3 G141 spectrum of \trone{} must be largely inflated to match the DRIFT-PHOENIX model spectra, even when using one composed of multiple component spectra, which should provide more flexibility to the model fits.
Inspection of the right panel of Figure~\ref{fig:CPAT_oot_spectra_all_data} shows that the data and model agree on the general shape of the spectrum but disagree on many details and the continuum level for the bluest and reddest wavelengths.
For further clarity, Figure~\ref{fig:stellar_spectra} illustrates these same spectra but with the true uncertainty on the observations shown.
Also shown is a DRIFT-PHOENIX model spectrum for a star with \trone{}'s effective temperature and surface gravity \citep[$T=2559$~K, $\log{g}=5.21$;][]{Gillon2017}.
This single-component model disagrees with the data significantly between roughly 1.34~$\micron$ and 1.6~$\micron$, which demonstrates why even the \texttt{flat} models prefer multi-component stellar spectra, as shown by the fitted values of the $F_\mathrm{spot}$ and $F_\mathrm{fac}$ parameters in Tables~\ref{tab:results_CPAT_single_all_data} and \ref{tab:results_CPAT_combined_all_data}.

The stellar spectrum contains many more data points than the transmission spectrum.
Therefore, disagreement between the observed and model stellar spectra could strongly influence our results.
In principle, the $\eta$ parameter guards against this by effectively de-weighting the stellar spectrum in  the MCMC optimization procedure.
Nonetheless, we investigated how the inclusion of the stellar spectrum in the model framework affects the results using the \trone{}~b--g dataset.
Considering only the transmission spectrum, the \texttt{flat} model has 14 data points and 1 fitted parameter, giving 13 degrees of freedom.
The $\chi^{2}$ value for this model is 51.8, indicating that it is conclusively ruled out ($p < 10^{-5}$).
The $\textrm{AIC}_\textrm{c}$ and BIC for this model are -129.3 and -129.0, respectively.
By contrast, the \texttt{contamination} model has 8 fitted parameters and thus gives 6 degrees of freedom when considering only the 14 data points of the transmission spectrum.
The $\chi^{2}$ value for the \texttt{contamination} model is 4.2, indicating that it is not ruled out by the data ($p = 0.64$).
Additionally, the $\textrm{AIC}_\textrm{c}$ and BIC for this model are -135.2 and -158.1, respectively.
A comparison of the information criteria for the two models yields $\Delta \textrm{AIC}_\textrm{c} = +5.9$ and $\Delta \textrm{BIC} = +29.1$ in favor of the \texttt{contamination} model.
Thus, relative to the \texttt{contamination} model, the \texttt{flat} model is ruled out decisively by the BIC and strongly by the $\textrm{AIC}_\textrm{c}$, which more stringently penalizes model complexity.
This exercise shows that the additional complexity of the \texttt{contamination} model is warranted by the transmission spectrum alone.
For the remainder of this analysis, however, we opt to utilize the additional information in the stellar spectrum and discuss the results taking into account both the transmission and stellar spectra.

\subsubsection{Impact of instrumental offsets}

Given the significant offsets between the HST transit depths and those from other instruments, one might reasonably wonder how strongly the model results rely on the offsets between the instruments.  To investigate this, we conducted this same analysis on the combined \trone{}~b--g dataset using only the HST data for the transmission spectrum.  In this case, the information criteria prefer the \texttt{flat} model ($\Delta \textrm{AIC}_\textrm{c} > +2$, $\Delta \textrm{BIC} > +8$).  In other words, the additional complexity of the \texttt{contamination} model is not warranted by the HST spectra alone, even though it results in smaller $\chi^2$ values; the offsets between the HST and Spitzer measurements are integral to the interpretation of stellar contamination impacting the transmission spectra.  However, we also note that the \texttt{contamination} models fit to the HST+Spitzer dataset accurately predicted the \textit{K2} and SSO transit depths (see Appendix~\ref{sec:HST+Spitzer}).  The best-fit parameters do not differ significantly between the analyses of the HST+Spitzer datasets and the full K2+SSO+HST+Spitzer datasets. It is unlikely that systematics among four instruments mimic an astrophysical signal. Therefore we argue for an astrophysical origin and against instrumental systematics as the source of the overall shape of the \trone{} combined transmission spectrum.

Additionally, one might also wonder if any offset induced by instrumental systematics could be mistakenly interpreted as evidence for stellar contamination. In other words, can any combination of offsets be fit well by a stellar contamination model?
To examine this point, we conducted the same analysis on two hypothetical variants of the \trone{}~b--g dataset.
In the first we held the HST and Spitzer transit depths to their measured values and perturbed the K2 and SSO depths to 50\% of their measured values, while keeping the original measurement uncertainties for all data points.
The $\chi^{2}$ values for the \texttt{flat} and \texttt{contamination} models are 1115 and 392 for 137 and 135 degrees of freedom, respectively, indicating that both models are conclusively ruled out ($p << 10^{-10}$).
In the second variant we perturbed the K2 and SSO depths to 200\% of their measured values, while keeping the original HST and Spitzer points. The MCMC chains do not converge for the \texttt{contamination} model in this case ($\hat{R}=1.05$ for $D$ and $\hat{R}=1.07$ for $F_\mathrm{spot}$, $f_\mathrm{fac}$, and $f_\mathrm{spot}$), illustrating the difficulty of fitting such a spectrum with a stellar contamination model. Nonetheless, taking the fits at face value, the $\chi^{2}$ values (3219 and 1436 for the \texttt{flat} and \texttt{contamination} models, respectively) indicate that both models are again conclusively ruled out ($p << 10^{-10}$).
This exercise shows that arbitrary offsets cannot be fit well by the additional parameters afforded by the stellar contamination model. 
Instead, the observed offset between HST and Spitzer transit depths provides a specific prediction for the K2 and SSO depths (see Appendix~\ref{sec:HST+Spitzer} for details), which is borne out by the observations.

\begin{deluxetable*}{lcccccccccccccccc}
\tabletypesize{\scriptsize}
\tablecaption{
Results of Stellar Contamination Model Fits to Combined Spectra\label{tab:results_CPAT_combined_all_data}} 
\tablehead{\colhead{Combination}                & 
           \colhead{Model}                      & 
           \multicolumn{9}{c}{Fitted Parameter} & 
           \colhead{AIC$_\textrm{c}$}           & 
           \colhead{$\Delta$AIC$_\textrm{c}$}   & 
           \colhead{BIC}                        & 
           \colhead{$\Delta$BIC}                & 
           \colhead{$\chi^{2}$}                 & 
           \colhead{$p$} \\
           \cline{3-11} 
           \colhead{}                           & 
           \colhead{}                           & 
           \colhead{$D$ (\%)}                   & 
           \colhead{$T_\mathrm{phot}$ (K)}      & 
           \colhead{$T_\mathrm{spot}$ (K)}      & 
           \colhead{$T_\mathrm{fac}$ (K)}       & 
           \colhead{$F_\mathrm{spot}$}          & 
           \colhead{$F_\mathrm{fac}$}           & 
           \colhead{$f_\mathrm{spot}$}          & 
           \colhead{$f_\mathrm{fac}$}           & 
           \colhead{$\eta$}                     & 
           \colhead{}                           & 
           \colhead{}                           &
           \colhead{}                           & 
           \colhead{}                           & 
           \colhead{}                           & 
           \colhead{}}
         
\startdata
b--g       & \texttt{flat}  & ${3.748}^{+0.013}_{-0.013}$ & ${2117}^{+84}_{-128}$  & ${1961}^{+108}_{-132}$ & ${2974}^{+26}_{-14}$ & ${16}^{+7}_{-15}$ & ${46}^{+5}_{-6}$  & -                 & -                & ${23}^{+1}_{-2}$ & -687.7 & 39.1 & -667.8 & 23.6 & 177.8 & 0.01 \\
b--g       & \texttt{cont.} & ${3.467}^{+0.058}_{-0.059}$ & ${2425}^{+168}_{-178}$ & ${2006}^{+127}_{-93}$  & ${2957}^{+43}_{-25}$ & ${38}^{+8}_{-8}$  & ${48}^{+6}_{-8}$  & ${10}^{+4}_{-10}$ & ${45}^{+6}_{-6}$ & ${23}^{+1}_{-2}$ & -726.8 & -    & -701.4 & -    & 130.7 & 0.59 \\
c--g       & \texttt{flat}  & ${3.000}^{+0.012}_{-0.011}$ & ${2116}^{+89}_{-123}$  & ${1960}^{+113}_{-128}$ & ${2974}^{+26}_{-14}$ & ${16}^{+7}_{-15}$ & ${46}^{+5}_{-6}$  & -                 & -                & ${23}^{+1}_{-2}$ & -705.0 & 26.2 & -685.0 & 20.8 & 164.8 & 0.05 \\
c--g       & \texttt{cont.} & ${2.794}^{+0.054}_{-0.051}$ & ${2419}^{+179}_{-192}$ & ${2013}^{+133}_{-98}$  & ${2954}^{+46}_{-25}$ & ${38}^{+8}_{-9}$  & ${49}^{+6}_{-9}$  & ${10}^{+5}_{-10}$ & ${45}^{+6}_{-6}$ & ${23}^{+1}_{-2}$ & -731.2 & -    & -705.9 & -    & 130.4 & 0.59 \\
b, d--g    & \texttt{flat}  & ${3.043}^{+0.012}_{-0.012}$ & ${2117}^{+86}_{-125}$  & ${1961}^{+107}_{-134}$ & ${2974}^{+26}_{-14}$ & ${16}^{+7}_{-15}$ & ${46}^{+5}_{-6}$  & -                 & -                & ${23}^{+1}_{-1}$ & -697.0 & 29.1 & -677.1 & 23.6 & 172.0 & 0.02 \\
b, d--g    & \texttt{cont.} & ${2.808}^{+0.053}_{-0.054}$ & ${2480}^{+206}_{-162}$ & ${2033}^{+119}_{-85}$  & ${2949}^{+51}_{-27}$ & ${39}^{+8}_{-8}$  & ${50}^{+7}_{-10}$ & ${11}^{+5}_{-11}$ & ${45}^{+7}_{-6}$ & ${23}^{+1}_{-2}$ & -726.1 & -    & -700.7 & -    & 134.5 & 0.50 \\
b--c, e--g & \texttt{flat}  & ${3.366}^{+0.012}_{-0.013}$ & ${2118}^{+84}_{-129}$  & ${1961}^{+106}_{-136}$ & ${2974}^{+26}_{-14}$ & ${16}^{+7}_{-15}$ & ${46}^{+5}_{-6}$  & -                 & -                & ${23}^{+1}_{-2}$ & -700.0 & 26.8 & -680.1 & 21.3 & 166.4 & 0.04 \\
b--c, e--g & \texttt{cont.} & ${3.165}^{+0.055}_{-0.055}$ & ${2425}^{+173}_{-211}$ & ${2014}^{+146}_{-101}$ & ${2954}^{+46}_{-25}$ & ${37}^{+8}_{-9}$  & ${49}^{+7}_{-9}$  & ${10}^{+5}_{-10}$ & ${44}^{+6}_{-6}$ & ${23}^{+1}_{-2}$ & -726.8 & -    & -701.4 & -    & 131.2 & 0.58 \\
b--d, f--g & \texttt{flat}  & ${3.254}^{+0.012}_{-0.012}$ & ${2117}^{+87}_{-124}$  & ${1961}^{+107}_{-133}$ & ${2974}^{+26}_{-14}$ & ${16}^{+7}_{-15}$ & ${46}^{+5}_{-6}$  & -                 & -                & ${23}^{+1}_{-2}$ & -693.6 & 34.5 & -673.6 & 29.1 & 172.7 & 0.02 \\
b--d, f--g & \texttt{cont.} & ${2.999}^{+0.052}_{-0.052}$ & ${2444}^{+181}_{-183}$ & ${2015}^{+125}_{-92}$  & ${2952}^{+48}_{-26}$ & ${39}^{+8}_{-9}$  & ${49}^{+7}_{-9}$  & ${10}^{+5}_{-10}$ & ${46}^{+6}_{-6}$ & ${23}^{+1}_{-2}$ & -728.1 & -    & -702.7 & -    & 130.1 & 0.60 \\
b--e, g    & \texttt{flat}  & ${3.105}^{+0.012}_{-0.012}$ & ${2117}^{+87}_{-125}$  & ${1960}^{+106}_{-133}$ & ${2974}^{+26}_{-14}$ & ${16}^{+7}_{-15}$ & ${46}^{+5}_{-6}$  & -                 & -                & ${23}^{+1}_{-2}$ & -693.6 & 36.8 & -673.6 & 31.4 & 174.6 & 0.02 \\
b--e, g    & \texttt{cont.} & ${2.849}^{+0.051}_{-0.051}$ & ${2407}^{+147}_{-191}$ & ${2001}^{+125}_{-95}$  & ${2956}^{+44}_{-24}$ & ${38}^{+8}_{-9}$  & ${48}^{+6}_{-8}$  & ${9}^{+4}_{-9}$   & ${46}^{+6}_{-6}$ & ${23}^{+1}_{-2}$ & -730.4 & -    & -705.0 & -    & 130.3 & 0.60 \\
b--f       & \texttt{flat}  & ${2.970}^{+0.011}_{-0.012}$ & ${2117}^{+90}_{-124}$  & ${1960}^{+108}_{-133}$ & ${2974}^{+26}_{-14}$ & ${16}^{+7}_{-15}$ & ${46}^{+5}_{-6}$  & -                 & -                & ${23}^{+1}_{-2}$ & -690.6 & 40.3 & -670.6 & 35.9 & 178.5 & 0.01 \\
b--f       & \texttt{cont.} & ${2.721}^{+0.051}_{-0.049}$ & ${2390}^{+141}_{-164}$ & ${1987}^{+119}_{-96}$  & ${2959}^{+41}_{-23}$ & ${38}^{+8}_{-8}$  & ${47}^{+6}_{-7}$  & ${9}^{+4}_{-9}$   & ${46}^{+5}_{-6}$ & ${23}^{+1}_{-2}$ & -731.9 & -    & -706.5 & -    & 129.8 & 0.61 \\
\enddata

\tablecomments{
The combined spectra are listed in the first column. 
The combination b--g utilizes all of the current data, including the weighted mean of the two transmission spectra of \trone{}~e. 
The others are five-planet combinations, which in turn exclude the spectrum of a single planet. 
The remaining table elements are the same as those in Table~\ref{tab:results_CPAT_single_all_data}.}
\end{deluxetable*}

\subsubsection{Physical interpretation}

Returning to the interpretation of the fits to the full K2+SSO+HST+Spitzer transmission spectra, we note that the posterior distributions for the stellar parameters ($T_\mathrm{phot}$, $T_\mathrm{spot}$, $T_\mathrm{fac}$, $F_\mathrm{spot}$, $F_\mathrm{fac}$, $f_\mathrm{spot}$, and $f_\mathrm{fac}$) in the \texttt{contamination} models listed in Table~\ref{tab:results_CPAT_combined_all_data} are consistent for all combined transmission spectra. 
They broadly agree on the same general picture for \trone{}, namely a heterogeneous photosphere comprised of three components: a hot component or faculae ($T \sim 3000$~K) covering $\sim 50 \%$ of the projected stellar disk, a cool component or spots ($T \sim 2000$~K) covering $\sim 40 \%$, and an intermediate component or immaculate photosphere ($T \sim 2400$~K) covering the remaining $\sim 10 \%$ of the disk. 
Within the region transited by the \trone{} planets, spots are less prevalent, covering only $\sim 10 \%$ of the transit chord, which results in a spectral mismatch between the disk-integrated stellar spectrum and the light source for the transit spectroscopy observations.

\begin{figure*}[htbp]
\label{fig:CPAT_corner_all_data}
\plottwo{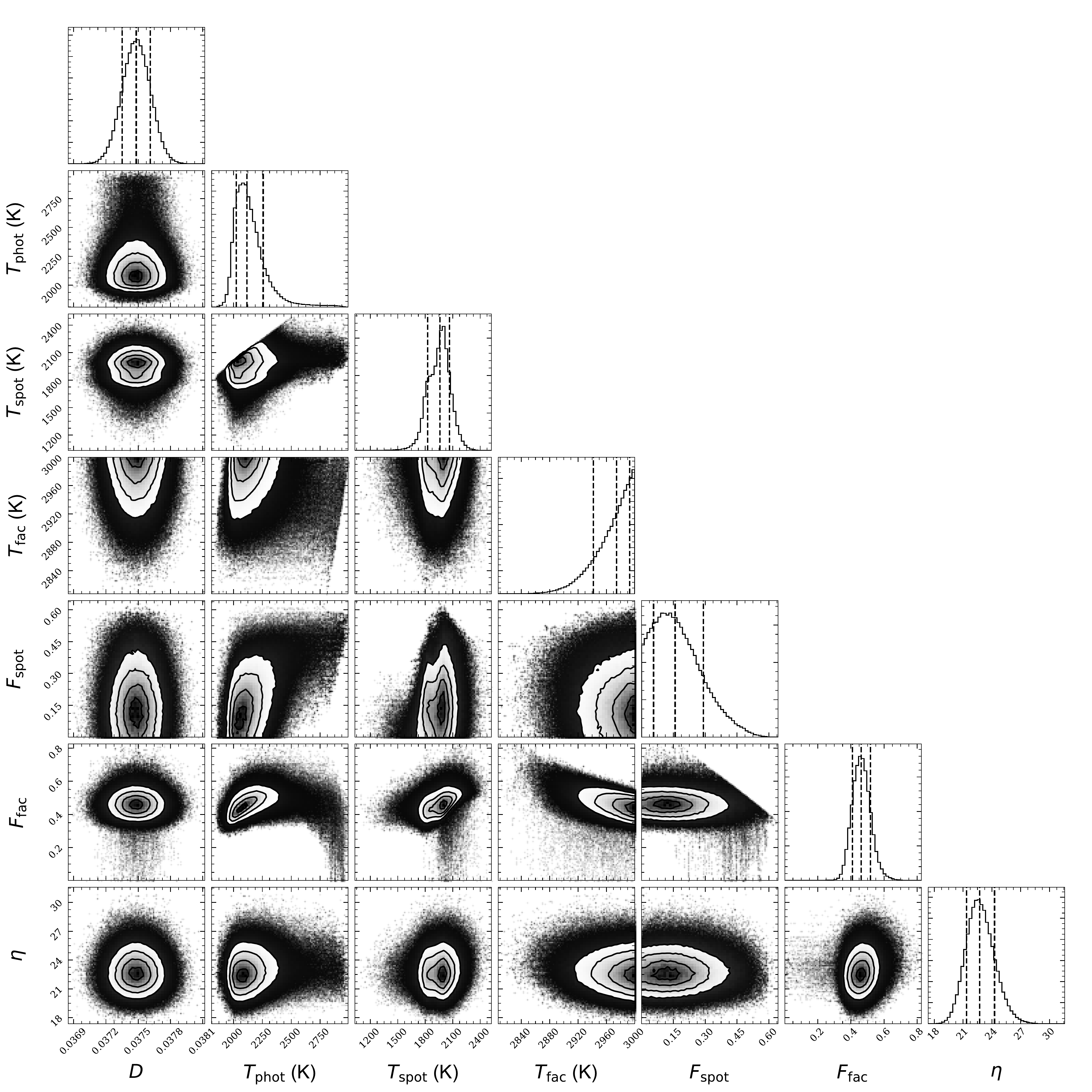}{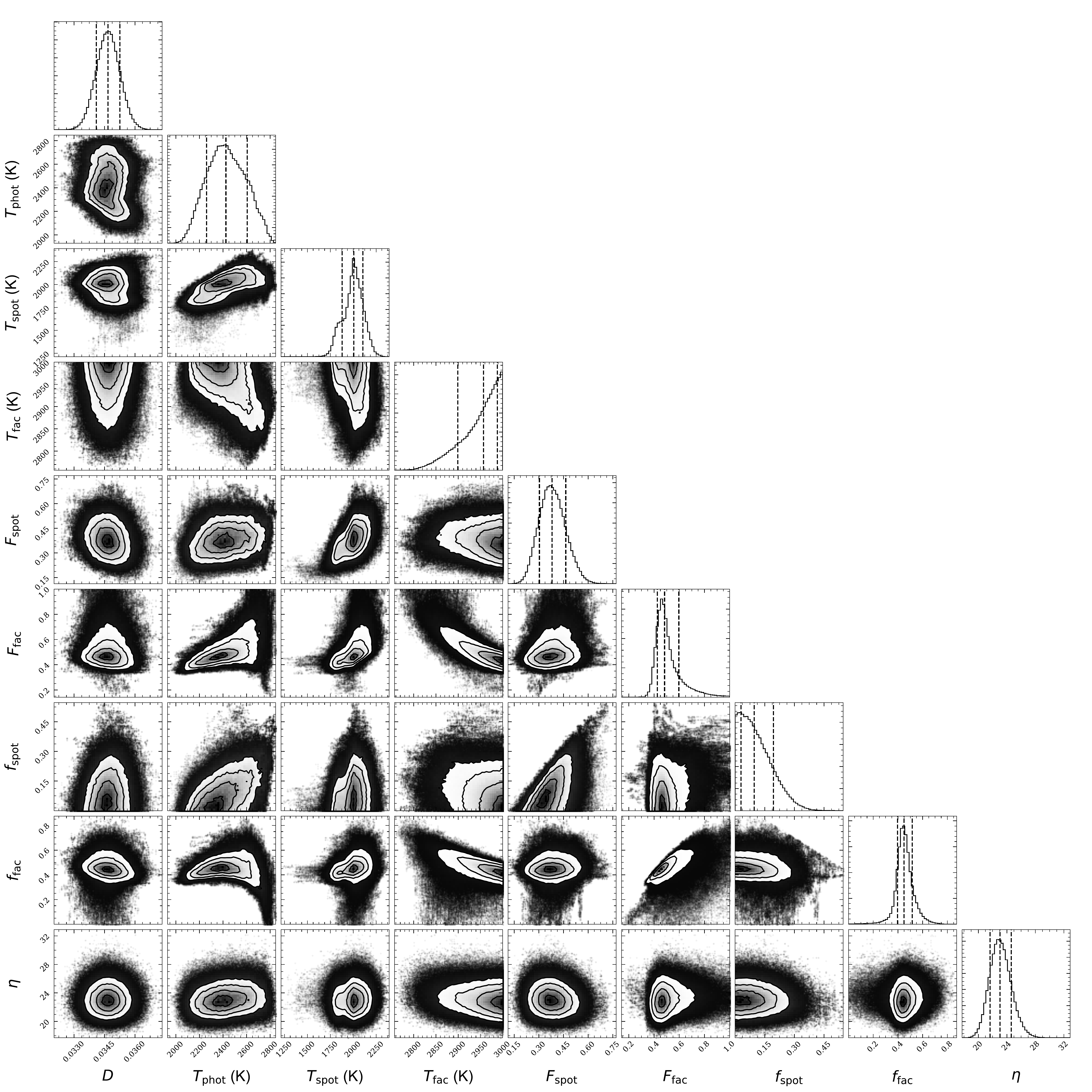}
\caption{
Posterior distributions of free parameters in CPAT model fits to observed transmission spectrum of \trone{} b--g. Results are shown for 
the \texttt{flat} and \texttt{contamination} models in the left and right panels, respectively.}
\end{figure*}

Figure~\ref{fig:CPAT_corner_all_data} illustrates examples of the joint posterior distributions for both modeling frameworks.  
For the \texttt{contamination} model, the distributions of two parameters, $f_\mathrm{spot}$ and $T_\mathrm{fac}$, deserve special notice. 
$f_\mathrm{spot}$ piles on the lower bound of the allowed parameter space in the \texttt{contamination} models and is consistent at $1\sigma$ confidence with no spots being present within the transit chord in each case.
The facula temperature is notable because it piles on the upper bound of the allowed parameter space, even in the \texttt{flat} models. 
This suggests that the HST spectrum of \trone{} shows evidence for a high-temperature component beyond the limit of the DRIFT-PHOENIX model grid, a result in broad agreement with the bright spots proposed by \citet{Morris2018} from an analysis of K2 and Spitzer photometry. 
Given the limitations of the model grid to sample high-enough values for $T_\mathrm{fac}$, the fitted values of $F_\mathrm{fac}$ and $f_\mathrm{fac}$ are likely overestimated.

The discrepancy between $f_\mathrm{spot}$ and $F_\mathrm{spot}$ suggests that the 28\% of the projected stellar disk (or 56\% of a half-hemisphere) covered by the transit chords \citep{Delrez2018} is less spotted than the whole disk and seems to point to the presence of an active region not represented within the transit chord.
As we discuss in Section~\ref{sec:context}, this arrangement bears some resemblance to the active high latitudes and circumpolar spot structures observed on fully convective M dwarfs \citep{Phan-Bao2009, Barnes2015, Barnes2017} and polar spots observed on active earlier-type stars \citep[see][and references therein]{Strassmeier2009}. Still, while the information criteria show that models that allow for stellar contamination in the transmission spectra are strongly preferred to those that assume no stellar contamination, the tendency of $T_\mathrm{fac}$ and $f_\mathrm{spot}$ to pile on the boundaries of their allowed ranges highlights the limitations of this analysis (discussed in more detail in Section~\ref{sec:limitations}) and cautions against over-interpreting the specific component temperatures or covering fractions determined by this analysis.

Nonetheless, these results demonstrate that, in principle, the features in the \trone{} transmission spectra can result from a heterogeneous stellar photosphere and not from transmission through the planetary atmospheres. Of course, both stellar and planetary signals can contribute to the observed spectra, and future efforts to jointly constrain the contributions of each source may bear fruit. However, these results urge caution for planetary interpretations of observed features in near-infrared transmission spectra from low-mass host stars. We consider this point more broadly in Section~\ref{sec:outlook}.
\subsection{TRAPPIST-1 results in context}\label{sec:context}

We present here a relatively simple model to explain a complicated physical phenomenon, namely the heterogeneity of the surface of \trone{} and the its effect on the transmission spectra of the \trone{} planets. Despite the limitations of the approach, we show that models that allow for imprints of stellar features in the combined \trone{} transmission spectra are strongly preferred to those that do not. The model fits point to large spot and faculae covering fractions, suggesting a highly heterogeneous photosphere for \trone{}. In the following paragraphs, we provide further context for interpreting this result.

Photometric variability shows that M dwarfs have highly heterogeneous surfaces. \citet{McQuillan2013} measured rotation periods and variability amplitudes for 1570 M dwarfs in the \textit{Kepler} sample with masses between 0.3 and 0.5~$M_{\sun}$. They found amplitudes ranging from 1.0 to 140.8 mmag in the \textit{Kepler} bandpass, with a median amplitude of 7.6~mmag or 0.70\% \citep[][Table 2]{McQuillan2013}. \citet{Newton2016} measured rotation periods for 387 field mid-to-late M dwarfs ($M < 0.35 M_{\sun}$) in the MEarth bandpass (roughly $i$+$z$) and found that they typically vary in brightness by 1\%--2\% as they rotate. Using an ensemble of model M-dwarf photospheres with randomly distributed active regions, \citet{Rackham2018} found that a 1\% I-band variability amplitude corresponds to spot covering fractions $F_{spot}=14^{+16}_{-7}\%$ and faculae covering fractions $F_{fac}=63^{+5}_{-25}\%$ across all M-dwarf spectral types, assuming a typical spot radius of $2 \degr$. Importantly, they found that the relation between active region coverages and variability amplitudes is not linear, as variability monitoring only traces the non-axisymmetric component of stellar surface features \citep[see also][]{Jackson2012}. Therefore, the typical amplitudes of M dwarfs in the \textit{Kepler} and MEarth samples indicate that variable M dwarfs can be covered in large part by active regions.

As noted above, \trone{} demonstrates photometric variability of roughly 1\% in the $I$+$z$ bandpass. This has generally been interpreted as active regions rotating into and out of view \citep{Gillon2016, Vida2017}, though \citet{Morris2018} have suggested that the variability of \trone{} may be driven by a magnetic activity timescale rather than a rotation period. Assuming the variability owes to rotation, \citet{Rackham2018} found that it could be caused by spot and faculae with covering fractions of $F_{spot}=8^{+18}_{-7}\%$ and $F_{fac}=54^{+16}_{-46}\%$, respectively. We used this information as priors in the current analysis, so the results presented here are not completely independent. However, we point out these previous results to show that the active region coverages found in this analysis are neither unexpected for \trone{} given its variability level nor unique among the population of M dwarfs.

The difference between the best-fit values that we find for $F_\mathrm{spot}$ and $f_\mathrm{spot}$ for the combined transmission spectra argues for variation in the latitudinal distribution of spots for \trone{}. This should not be surprising: Doppler imaging \citep{Vogt1983} of early M dwarfs provides examples of both stars with spots emerging at preferential latitudes and those with spots at all latitudes \citep{Barnes2001, Barnes2004}. 
However, the few Doppler images that exist for fully convective M dwarfs suggest that spots may emerge preferentially at high latitudes.
To date, six mid-to-late M dwarfs have been studied with this technique: 
\object{V374 Peg} \citep[M4V,][]{Morin2008}, 
\object{G 164-31} \citep[M4V,][]{Phan-Bao2009}, 
\object{GJ 791.2A} \citep[M4.5V,][]{Barnes2015, Barnes2017}, 
the binary \object{GJ 65A} and \object{GJ 65B} \citep[M5.5V and M6V,][]{Barnes2017}, and 
\object{LP 944-20} \citep[M9V][]{Barnes2015}.
Three of these six targets (G 164-31, GJ 65A, and LP 944-20) only display spots at high-latitude or circumpolar regions.
Additionally, GJ 791.2A shows a high-latitude circumpolar spot structure as well as low-latitude spots.
For earlier spectral types, which can be studied more readily with Doppler imaging techniques, polar spots are commonly observed on active stars \citep[see][and references therein]{Strassmeier2009} and may be stable and long-lived \citep{Jeffers2007}.

In the case of TRAPPIST-1, the combined transit chords of the seven \trone{} planets probe a substantial portion (28\%) of the projected stellar disk \citep[][]{Delrez2018}. 
However, given the frequent and prominent latitudinal variations in active region occurrence observed in M dwarfs, it is unsurprising that our analysis suggests significant differences between the 28\% equatorial zone and the polar regions of TRAPPIST-1. 
Therefore, while we cannot precisely constrain the latitudes of active regions in the current analysis, we suggest that features like the active high latitudes observed on fully convective M dwarfs or polar spots observed on active earlier-type stars may account for the different spot covering fractions that we fit to the transit chord and whole disk of \trone{}.

Spectral template fitting provides another approach for studying the heterogeneity of stellar surfaces. Observations of TiO molecular bands in dwarf stars with photospheres warmer than $T {\sim} 3500$~K have been used to constrain spot sizes and covering fractions \citep{Vogt1979, Vogt1981, Ramsey1980}. Using spectra of inactive G, K, and M dwarfs as templates, \citet{Neff1995} and later \citet{ONeal1996, ONeal1998} showed that TiO bands in medium-resolution ($R \sim 10,000$) spectra of active G and K stars point to covering fractions of cool spots as large as 64\%. \citet{Gully-Santiago2017} studied the T-Tauri star \object{LkCa 4} and found that spectra features in high-resolution near-infrared spectra were produced by hot (${\sim} 4100$~K) and cool photospheric components (${\sim}$~2700--3000~K), with the cool component covering ${\sim} 80 \%$ of the stellar surface. TiO bands in $R {\sim} 1,000$ spectra of a large sample of stars in the Pleiades show that spot covering fractions of ${\sim} 50\%$ are common among K and M stars at ${\sim} 125$~Myr \citep{Fang2016}.

In this work, we find that the low-resolution \textit{HST} spectrum of \trone{} is fit best by a template combined from three DRIFT-PHOENIX model stellar spectra. This is the case even for our \texttt{flat} models, in which the spot and faculae parameters are determined by fitting the stellar spectrum alone. While \trone{} is both later-type and older than the stars in the examples referenced above, these previous efforts show that the inferred heterogeneity for \trone{} is not extreme but rather typical of active stars.

\subsection{Comparison with \citet{Morris2018b}}

Recently \citet{Morris2018b} investigated the problem of stellar contamination from TRAPPIST-1 using a novel transit light curve ``self-contamination'' technique \citep{Morris2018_technique}. They determined $R_\mathrm{p} / R_\mathrm{s}$ for TRAPPIST-1 b-h (individually) using \textit{Spitzer} transit light curves based on the durations of the ingress/egress and the durations from the mid-ingress to mid-egress, for which stellar heterogeneity has a negligible effect, and compared it to the $R_\mathrm{p} / R_\mathrm{s}$ values determined by the transit depths. They found consistent $R_\mathrm{p} / R_\mathrm{s}$ measurements from the two methods and claimed a non-detection of stellar contamination. However, the $R_\mathrm{p} / R_\mathrm{s}$ values calculated by the light curve self-contamination method have large uncertainties, e.g. $\sim 10\%$ vs. $1.4\%$ in this work for TRAPPIST-1 b, and therefore do not have the sensitivity to probe the level of stellar contamination in transmission spectra that we study here. 
In this light, the non-detection of \citet{Morris2018b} should be more properly viewed as weak upper limit on stellar contamination, rather than evidence for the lack of it. 
Put simply, none of the individual transit depths (\textit{HST, Spitzer, K2, SOO}) that we used in this work would satisfy the criterion used by \citet{Morris2018b} for rejecting the null hypothesis (non-detection of stellar contamination), although we show that the combined transmission spectra strongly favor a stellar contamination model.

\subsection{Limitations of the model} \label{sec:limitations}

While we build upon previous attempts to characterize the contribution of stellar heterogeneity to transmission spectra by imposing the constraint provided by the out-of-transit stellar spectrum, this model has several notable limitations. We utilize disk-integrated stellar spectra for the spectral components, which may neglect unique spectral features that can emerge from magnetically active regions \citep[e.g.][]{Norris2017}. Similarly, we do not consider any chromospheric contribution to the stellar contamination spectrum. We neglect the effect of limb darkening and consider the photosphere to be static during the transit events, though the observed activity and relatively short rotation period of \trone{} may allow for photospheric evolution on a time scale important for transit observations. We note that future efforts could improve on this initial analysis in these respects and more.

Nonetheless, despite the model limitations, we find that models including stellar heterogeneity effects provide significantly better fits to the combined \trone{} transmission spectra than flat planetary transmission models. Interestingly, the spot and faculae covering fractions that we infer from the transmission spectra have considerably lower uncertainties than those inferred through modeling the star's rotational variability \citep{Rackham2018}. While the rotational variability of \trone{} demonstrates the presence of photospheric heterogeneities, its magnitude poorly constrains their covering fractions. We find that the observed transmission spectra place tighter constraints on the stellar heterogeneity.

This interpretation of the spectra also offers testable predictions. We find the best-fit models to display notable decreases in the visual transit depths of the \trone{} planets, which represent an observational challenge given the faintness of this ultra-cool dwarf in the visual, but may be observable with large ground-based telescopes. 
Broadband photometry over repeated transits can also overcome this challenge, and in Appendix~\ref{sec:HST+Spitzer} we show that stellar contamination models fit to the HST+Spitzer datasets accurately predict the recently measured \textit{K2} (0.42--0.9~$\micron$) and \textit{I+z} (0.8--1.1~$\micron$) combined transit depths \citep{Ducrot2018}.
Looking forward, we find ``inverse water features'' akin to the 1.4~$\micron$ feature in the unexplored spectral region between 1.7 and 4~$\micron$, which can be studied with \textit{JWST}.

\section{Outlook: \textit{HST} and \textit{JWST} Transit Spectroscopy}
\label{sec:outlook}

\subsection{Stellar Contamination as Dominant Astrophysical Noise}

The contamination seen in the combined \trone{} data emerges due to the {\em transit light source effect}: the difference between the baseline stellar spectrum (disk-integrated) and the spectrum of the transit chord (actual light source for the transmission spectroscopy), as described by \citet[][]{Rackham2018} in detail. The presence of $\sim$200~ppm-level (inverted) water feature as stellar contamination in the HST/WFC3 spectra should serve as a red flag for investigators planning on major future HST/WFC3 or \textit{JWST} transit spectroscopy campaigns on \trone{} or similar host stars. The possible range of stellar spectral contaminations have been discussed by \citet[][]{Rackham2018}, and here we will just briefly discuss two aspects not addressed there: connection with atmospheric retrievals and emission spectroscopy.

Atmospheric retrievals use a Bayesian exploration of possible atmospheric models to identify best-fitting models and derive confidence intervals (and posterior probability distributions) for the atmospheric pressure-temperature structures, molecular abundances, and cloud properties. These models have been shown to be powerful in characterizing atmospheres (e.g., \citealt[][]{Madhusudhan2009,Benneke2012,Waldmann2015}) and are widely anticipated to be the primary tools for interpreting high-quality transmission spectra from \textit{HST} and \textit{JWST} \citep[e.g.,][]{Greene2016,Morley2017}. The stellar contamination seen in the \trone{} planets, in the sub-Neptune GJ~1214b \citep[][]{Rackham2017}, and those predicted by \citet[][]{Rackham2018} for most M-dwarf stars highlight the importance of including this effect in retrievals. Indeed, without modeling and correcting for the relatively strong ($\sim$200~ppm-level) inverted water absorption feature introduced by the stellar contamination, one cannot hope to accurately measure the water abundance (probably $<$80~ppm levels, \citealt[][]{Morley2017}) in the transmission spectra of small planets. 
Recent efforts to include the effects of stellar contamination in retrievals, however, show promise in separating stellar and planetary contributions to transmission spectra (\citealt[][Bixel et al., submitted]{Pinhas2018, Espinoza2018}).			

We point out, furthermore, that stellar emission (eclipse) spectroscopy will be affected much less by stellar contamination, as the planet is its own light source for emission or eclipse spectroscopy. In fact, for the limiting case of infinitely slowly rotating stars with a constant (non-evolving) starspot/facular pattern, there will be no stellar contamination. Therefore, \textit{JWST} emission spectroscopy of the \trone{} planets is likely to be much easier to interpret than \textit{HST} and \textit{JWST} transmission spectroscopy. We caution, however, that for rapidly rotating stars -- such as \trone{} -- the evolution of the stellar spectrum due to its heterogeneity \citep[e.g.,][]{Apai2013,Yang2015} may also pose a non-negligible astrophysical systematic.

\subsection{Combining Multi-Epoch Transit Data from Planets Orbiting Heterogeneous Host Stars}
\label{Section:MultiEpochContamination}
A central question in \textit{HST} and \textit{JWST} transiting exoplanet observations is whether data from multiple transits can be combined. If the instrumental systematics can be robustly corrected {\em and} the astrophysical systematics are negligible, data can be co-added. This approach has been used in the HST/WFC3/IR transmission spectroscopy study of the sub-Neptune GJ~1214b by \citet[][]{Kreidberg2014}, where observations from 12 transits of the planet were co-added. In fact, the overall success of the empirical ramp effect correction and the physically-motivated RECTE model are very encouraging and suggest that photon-noise-limited performance could be achievable for large, multi-epoch campaigns.

However, the astrophysical noise -- stellar contamination through the transit light source effect \citep[][]{Rackham2018} -- emerges as a major concern. Not only will the stellar contamination features often dominate the planetary absorption features in amplitude (as explored in detail in \citealt[][]{Rackham2018}), but the contamination itself may change as the star rotates and its starspot/faculae coverage evolves over the course of a transit observation.

With a rotational period of 1.4$\pm$0.05 days \citep{Gillon2016}, \trone{} is a relatively rapidly rotating host star at the stellar/substellar boundary. It is worthwhile to briefly outline the implications of this rapid rotation by comparing four different timescales:  starspot evolution, planetary orbits, stellar rotation, and the lengths of planetary transit observations. 

Stellar contamination in transit spectra will depend on the heterogeneity of the projected stellar disk during the observations. A straightforward mechanism that alters the brightness and spectral distribution on the visible hemisphere of a star is the appearance, evolution, and disappearance of starspots and surrounding faculae. Although we have a limited understanding of starspot lifetimes, it seems reasonable to assume that their evolution may occur over timescales of 10--20 days, based on the similar processes observed in the Sun \citep{Bradshaw2014}. This timescale is generally longer than the orbital periods of the \trone{} planets. Therefore, we tentatively conclude that starspot evolution may only have a limited impact on stellar contamination variations in the case that consecutive planetary transits can be observed; nonetheless, it will have a major impact in co-adding data from transits separated by months or years.

The stellar rotation, however, will introduce a more problematic source of time-varying stellar contamination. The rotational period of \trone{} is shorter than the orbital periods of all but the closest planet (b), which means that even during subsequent transits of the same planet, a different (essentially random) stellar longitude will face the observer, i.e., the stellar contamination will be different between even the closest transits of the same planet. 

The rapid rotation of \trone{} poses another interesting problem: the rotational period (1.4$\pm$0.05~d, \citealt[][]{Gillon2016}) is comparable to the duration of a typical transit observation (5 \textit{HST} orbits or about 1/3 of day). This means that even during a single transit the star will rotate by about $60^\circ$. Thus, about 1/3 of the visible stellar hemisphere will be different between the beginning and the end of the transit. This rapid change means that stellar contamination will not only be different between different transits, but will change even during a single transit, making potential starspot modeling and correction even more difficult. We note here that the same effect may also complicate planet eclipse spectroscopy, albeit to a lesser extent. 

With its rapid rotation, it may be tempting see \trone{} as a worst-case scenario in terms of stellar contamination; however, it should be considered as a typical example of very late-M dwarf or brown dwarf planet hosts. Photometric monitoring programs have found that many brown dwarfs are much faster rotators, with rotational periods of a few hours being not uncommon (\citealt[][]{Metchev2015}). Spectroscopic monitoring of brown dwarfs (with the same HST/WFC3/IR grism as used for this study) showed that most, if not all, brown dwarfs have heterogeneous atmospheres  \citep[][]{Buenzli2014}, a result that was reinforced by a larger but less sensitive ground-based broad-band photometry survey \citep[][]{Radigan2014} and by a large and unbiased Spitzer photometric survey \citep[][]{Metchev2015}. At least some, and perhaps most, of these brightness and spectral modulations are attributable to the re-arrangement of condensate clouds driven by global circulation patterns \citep[][]{Apai2017}.

Therefore, rotational modulations and stellar heterogeneity evolution should be expected to be very common for ultracool planet hosts---the coolest host stars and brown dwarfs. The rapidly evolving stellar contamination in the transit spectra of exoplanets orbiting such ultracool hosts will be a major challenge to obtaining the multi-epoch data necessary to build up the data quality required for atmospheric abundance analysis \citep[][]{Morley2017}. In fact, this goal may only be achievable if we can develop a robust understanding of ultracool dwarf heterogeneities and derive a reliable contamination correction method for this astrophysical noise source.

\subsection{Comparison with \citet{deWit2018}}
\label{sec:deWit2018}

\citet{deWit2018} was published after the first submission this work. They reduced and analyzed data from program GO-14873 independently. They adopted a ramp removal method that is similar to ours and different from that in \citet{deWit2016}. We identify a few minor differences between the reduction processes, none of which led to significant differences in the reduced data, but the differences in the analysis led to different interpretations
\begin{enumerate}
\item
The most crucial difference between our study and that of \citet{deWit2018} is the interpretations of the transmission spectra.  \citet{deWit2018} compared observed spectra with planetary models and flat lines, while we compared the observations with flat spectra, planetary atmosphere model spectra with water absorption features, and stellar contamination model spectra. We agree with \citet{deWit2018} that flat spectra are preferred over spectra with planetary water absorption features; but we also show that stellar contamination models are favored over flat spectra for the combined \trone{} transmission spectra. 

\item
  Each visit in \citet{deWit2018} was designed to include two transit events, and the three events identified in Table~\ref{tab:missedtransits} are genuine transits. While we discarded some datasets heavily affected by severe cosmic ray events and resulting guiding failure due to SAA passages, \citet{deWit2018} attempted to recover all the data. Although Visit 1 data -- that covered the transit by TRAPPIST-1 f  -- was successfully recovered, it was not possible to recover the other two events.
  
\item
 \citet{deWit2018} aligned every single read-out, while we applied the alignment on each \texttt{ima} file frame. As for the cosmic-ray correction, in \citet{deWit2018} both spatial and temporal interpolation were used to replace the pixels affected by cosmic rays. 

\item
\citet{deWit2018} adopted a ramp correction formula that resembles the  correction predicted by the RECTE model \citep{Zhou2017}. They report an average standard deviation of residuals in the broadband light curve of 220~ppm, while we reached 198~ppm---very similar values given the uncertainties. For the individual (narrow-band) bins, they report average uncertainties of 545, 526, 493, and 494~ppm for Visits 1 to 4, respectively; by comparison, we find average uncertainties of 500, 550, 487, and 454~ppm for the same datasets. Thus, the residuals from the two reductions are very similar.

\end{enumerate}

\section{Summary}

The \trone{} system offers exceptionally deep transits and a relatively bright host star for multiple approximately Earth-sized planets. We present here HST/WFC3/IR near-infrared grism spectroscopy of the \trone{} transiting planets based on re-reduction of archival data \citep{deWit2016, deWit2018} covering a total of seven transits.

The key findings of our study are as follows:

\begin{enumerate}

\item
We detected transits for the six inner planets: \trone{} b, c, d, e, f, and g. Our data includes two transits for planet e.

\item
We provide improved broadband transit depths and mid-transit times for each of the six planets. We find evidence for transit timing variations for planets f and g.

\item
  Comparisons of the transit light curves of planets b and c between those published by \citet[][]{deWit2016} and our study show the benefits of the RECTE charge trap correction: increased orbital phase coverage, higher observing efficiency, and similar or better systematics correction. Compared to the empirical correction, the RECTE model is less affected by astrophysical signals (e.g., flares). In addition, RECTE can be combined with systematics marginalization methods \citep[e.g.,][]{Gibson2014, Wakeford2016, Sheppard2017} or methods based on Gaussian processes \citep[e.g.,][]{Beatty2017, Evans2017, Nikolov2018} to provide a physically-based model for the ramp effect component.

\item
Our data reduction reaches a typical precision of about 230-340~ppm for individual planets and, after excluding possible flare events, a precision of 180-210 ppm.

\item
We note {two} short-duration brightening events in the broadband light curves, which we identify as candidate flare events. Flare events can complicate the transit light curve fits and can be confused with higher instrumental systematics levels by traditional empirical fits.

\item
No significant planetary absorption features are present in the individual transit spectra of the six planets.

\item
In the combined spectra of the six planets there is no obvious evidence for water or other absorption features.

\item
In the combined transmission spectra, we identify a suggestive decrease in transit depth at 1.4~$\micron$ relative to the adjacent continuum.
We find that this feature in itself does not provide decisive evidence for a heterogeneous stellar photosphere impacting the transmission spectra, though it is consistent with stellar contamination models fit to the K2+SSO+HST+Spitzer dataset covering a wider wavelength range.

\item
  
We present spectral fits to the combined K2, SSO, HST/WFC3/IR grism and Spitzer 4.5~\micron{} transit depths assuming an intrinsically flat planetary spectrum but allowing for the presence of starspots and faculae in the star. 
We find evidence for stellar contamination in the \trone{}~d transmission spectrum and, more notably, in the combined transmission spectrum of \trone{}~b--g. 
The model interpretation for the combined spectrum is robust for all five-planet combinations excluding the impact of a single planet in turn.
The composite photosphere produced by the stellar contamination models also matches the out-of-transit spectrum of TRAPPIST-1 as well as starspot and faculae populations inferred from its observed photometric variability level.

\item
The facular and starspot covering fractions required by the stellar contamination model are consistent with those expected for late-M type stars and their optical/near-infrared photometric variability, demonstrating that stellar contamination from the transit light source effect \citep[][]{Rackham2018} poses a major challenge to HST and JWST high-precision exoplanet transit spectroscopy.

\item
We also point out that co-adding transit spectra from multiple epochs for planets orbiting rapidly rotating late-type host stars will be complicated by rapidly changing stellar contamination: for \trone{} the contamination will even significantly change during the length of a single transit observation.
\end{enumerate}

In summary, our study provides an independent reduction (based on a physically motivated charge trap model) of the first transmission spectra of Earth-sized habitable-zone transiting exoplanets, providing a nearly complete spectral library of the known planets in the \trone{} system. Our findings, however, highlight stellar contamination as a dominant astrophysical noise source and illustrate the challenge in combining multi-epoch spectroscopy for planets orbiting rapidly rotating host stars.

\acknowledgments This research has made use of the NASA Exoplanet Archive, which is operated by the California Institute of Technology, under contract with the National Aeronautics and Space Administration under the Exoplanet Exploration Program. B.R. acknowledges support from the National Science Foundation Graduate Research Fellowship Program under Grant No. DGE-1143953. D.A. acknowledges support from the Max Planck Institute for Astronomy, Heidelberg, for a sabbatical visit. The results reported herein benefited from collaborations and/or information exchange within NASA's Nexus for Exoplanet System Science (NExSS) research coordination network sponsored by NASA's Science Mission Directorate.  Support for Programs 14241 and 15060 was provided by NASA through a grant from the Space Telescope Science Institute, which is operated by the Association of Universities for Research in Astronomy, Incorporated, under NASA contract NAS5-26555. Based on observations made with the NASA/ESA Hubble Space Telescope, obtained in GO programs 14500 and 14873 at the Space Telescope Science Institute.

\facilities{HST/WFC3, Spitzer, SPECULOOS-South Observatory, Kepler} 
\software{Numpy, Scipy, Matplotlib, Astropy, emcee, batman, pymc}

\bibliographystyle{yahapj}
\bibliography{references}

\appendix
\section{Transmission Spectra}
\twocolumngrid
\startlongtable
\begin{deluxetable}{cccc}
  \tablecaption{Transmission Spectra \label{tab:transit_spectra}}
  \tablehead{ \colhead{Wavelength}&
    \colhead{Transit}& \colhead{Upper}&
    \colhead{Lower} \\
    \colhead{\AA}&
    \colhead{depth}&
    \colhead{error}&
    \colhead{error}}
  \startdata
  \multicolumn{4}{c}{Transit 1, TRAPPIST-1 c} \\
  11505 &	0.00709 & 0.00031 & 0.00032         \\
  11970 &	0.0074  & 0.00028 & 0.00029         \\
  12434 &	0.00719 & 0.00024 & 0.00024         \\
  12898 &	0.00762 & 0.00021 & 0.00021         \\
  13363 &	0.00728 & 0.00033 & 0.00032         \\
  13827 &	0.00707 & 0.00031 & 0.00032         \\
  14292 &	0.00707 & 0.00034 & 0.00033         \\
  14756 &	0.00686 & 0.00032 & 0.00029         \\
  15220 &	0.00741 & 0.00031 & 0.00032         \\
  15685 &	0.00704 & 0.00029 & 0.00028         \\
  16149 &	0.00702 & 0.0003  & 0.00028         \\
  16613 & 0.00774 & 0.00034 & 0.00033 \\\hline
  \multicolumn{4}{c}{Transit 2, TRAPPIST-1 b} \\
  11505 &	0.00783 & 0.00032 & 0.00033         \\
  11970 &	0.00771 & 0.00032 & 0.00031         \\
  12434 &	0.00803 & 0.00025 & 0.00025         \\
  12898 &	0.00745 & 0.00023 & 0.00022         \\
  13363 &	0.00731 & 0.00036 & 0.00035         \\
  13827 &	0.00728 & 0.00035 & 0.00034         \\
  14292 &	0.00745 & 0.00037 & 0.00036         \\
  14756 &	0.00781 & 0.00032 & 0.00033         \\
  15220 &	0.00776 & 0.00036 & 0.00035         \\
  15685 &	0.00806 & 0.00031 & 0.00032         \\
  16149 &	0.0081  & 0.00031 & 0.00034         \\
  16613 & 0.00769 & 0.00035 & 0.00036 \\\hline
  \multicolumn{4}{c}{Transit 3, TRAPPIST-1 d} \\
  11465 & 0.00378 & 0.00012 & 0.00012 \\
  11930 & 0.00444 & 0.00013 & 0.00015 \\
  12394 & 0.00419 & 0.00009 & 0.00009 \\
  12858 & 0.00401 & 0.00011 & 0.00013 \\
  13322 & 0.00392 & 0.00014 & 0.00014 \\
  13787 & 0.00424 & 0.00013 & 0.00013 \\
  14251 & 0.00356 & 0.00018 & 0.00018 \\
  14715 & 0.00391 & 0.00014 & 0.00014 \\
  15179 & 0.0036  & 0.00015 & 0.00014 \\
  15644 & 0.00401 & 0.00011 & 0.00011 \\
  16108 & 0.00393 & 0.00014 & 0.00014 \\
  16572 & 0.00361 & 0.00023 & 0.00024 \\ \hline
  \multicolumn{4}{c}{Transit 4, TRAPPIST-1 g}\\
  11377 & 0.00824 & 0.00035 & 0.00038 \\
  11841 & 0.0076  & 0.00021 & 0.00023 \\
  12305 & 0.00769 & 0.00021 & 0.00023 \\
  12770 & 0.00803 & 0.00022 & 0.00021 \\
  13234 & 0.00773 & 0.00021 & 0.00021 \\
  13698 & 0.00776 & 0.00025 & 0.00024 \\
  14163 & 0.00803 & 0.00027 & 0.00027 \\
  14627 & 0.00781 & 0.00025 & 0.00025 \\
  15092 & 0.00753 & 0.00026 & 0.00026 \\
  15556 & 0.00771 & 0.00028 & 0.0003  \\
  16020 & 0.00821 & 0.00024 & 0.00023 \\
  16485 & 0.00769 & 0.00025 & 0.00026 \\\hline
  \multicolumn{4}{c}{Transit 5, TRAPPIST-1 e}\\
  11377 & 0.00476 & 0.0002  & 0.0002  \\
  11841 & 0.00466 & 0.00012 & 0.00012 \\
  12305 & 0.00458 & 0.00012 & 0.00012 \\
  12770 & 0.00483 & 0.00013 & 0.00012 \\
  13234 & 0.00469 & 0.00012 & 0.00014 \\
  13698 & 0.00448 & 0.00015 & 0.00016 \\
  14163 & 0.00479 & 0.0002  & 0.00019 \\
  14627 & 0.00489 & 0.00015 & 0.00015 \\
  15092 & 0.00508 & 0.00016 & 0.00016 \\
  15556 & 0.00484 & 0.00018 & 0.00017 \\
  16020 & 0.00496 & 0.00014 & 0.00015 \\
  16485 & 0.00497 & 0.00016 & 0.00014 \\\hline
  \multicolumn{4}{c}{Transit 6, TRAPPIST-1 f}\\
  11408 & 0.00637 & 0.00023 & 0.00022\\
  11873   & 0.00664 & 0.00021 & 0.00019\\
  12337   & 0.00642 & 0.00024 & 0.00024\\
  12801   & 0.0065  & 0.00023 & 0.00022\\
  13265   & 0.00676 & 0.00022 & 0.00021\\
  13730   & 0.00634 & 0.00024 & 0.00024\\
  14194   & 0.00638 & 0.00026 & 0.00027\\
  14658   & 0.00634 & 0.00026 & 0.00024\\
  15123   & 0.00605 & 0.0002  & 0.0002\\
  15587   & 0.00667 & 0.0002  & 0.00019\\
  16051   & 0.00635 & 0.00021 & 0.00021\\
  16516 & 0.00681 & 0.00025 & 0.00025\\\hline
  \multicolumn{4}{c}{Transit 7, TRAPPIST-1 e}\\
  11442 &	0.00546 & 0.00016 & 0.00016 \\
  11907 &	0.00517 & 0.00014 & 0.00016 \\
  12371 &	0.00552 & 0.00013 & 0.00013\\
  12835 &	0.00514 & 0.00013 & 0.00013\\
  13300 &	0.00526 & 0.00013 & 0.00013\\
  13764 &	0.00487 & 0.00017 & 0.00017\\
  14228 &	0.00483 & 0.00017 & 0.00017\\
  14692 &	0.00494 & 0.00013 & 0.00013\\
  15157 &	0.00494 & 0.00013 & 0.00013\\
  15621 &	0.00506 & 0.00013 & 0.00014\\
  16085 &	0.00511 & 0.00012 & 0.00011\\
  16550 &	0.00503 & 0.00021 & 0.00021\\
  \enddata
\end{deluxetable}

\startlongtable
\begin{deluxetable}{ccc}
  \tablecaption{Transmission Spectra in the Interpolated Bins \label{tab:transit_spectra_interp}}
  \tablehead{ 
    \colhead{Transit depth}&
    \colhead{Upper error}&
    \colhead{Lower error}}
  \startdata
  \multicolumn{3}{c}{Transit 1, TRAPPIST-c}\\  
  0.00708 & 0.00031 & 0.00032 \\
  0.00738 & 0.00028 & 0.00029 \\
  0.00723 & 0.00023 & 0.00023 \\
  0.00761 & 0.00023 & 0.00023 \\
  0.00717 & 0.00033 & 0.00033 \\
  0.0071  & 0.00031 & 0.00032 \\
  0.00692 & 0.00034 & 0.00031 \\
  0.00716 & 0.00031 & 0.00031 \\
  0.00725 & 0.0003  & 0.0003  \\
  0.00694 & 0.00029 & 0.00027 \\
  0.00751 & 0.00033 & 0.00032 \\\hline
  \multicolumn{3}{c}{Transit 2, TRAPPIST-1 b}\\
  0.00784 &	0.00032 &	0.00033\\
0.00774 &	0.00032 &	0.00031\\
0.00799 &	0.00024 &	0.00024\\
0.00737 &	0.00025 &	0.00025\\
0.0073  & 0.00037       & 0.00036\\
0.0073  & 0.00035       & 0.00034\\
0.00764 &	0.00035 &	0.00035\\
0.00779 &	0.00034 &	0.00034\\
0.00793 &	0.00033 &	0.00033\\
0.00814 &	0.0003  & 0.00033\\
0.00783 &	0.00034 &	0.00036\\\hline
\multicolumn{3}{c}{Transit 3, TRAPPIST-1 d}\\
0.00389 &	0.00013 &	0.00013\\
0.00442 &	0.00013 &	0.00014\\
0.00413 &	0.00009 &	0.00009\\
0.00394 &	0.00013 &	0.00014\\
0.00409 &	0.00013 &	0.00013\\
0.00393 &	0.00016 &	0.00015\\
0.00372 &	0.00016 &	0.00016\\
0.00371 &	0.00014 &	0.00014\\
0.00386 &	0.00012 &	0.00012\\
0.004   & 0.00013       & 0.00012\\
0.00364 &	0.00021 &	0.00022\\\hline
\multicolumn{3}{c}{Transit 4, TRAPPIST-1 g}\\
0.00799 &	0.00029 &	0.00031\\
0.00756 &	0.0002  & 0.00022\\
0.00787 &	0.00022 &	0.00022\\
0.00791 &	0.00021 &	0.00021\\
0.00768 &	0.00023 &	0.00023\\
0.00797 &	0.00027 &	0.00026\\
0.00791 &	0.00025 &	0.00025\\
0.00756 &	0.00026 &	0.00025\\
0.00766 &	0.00028 &	0.0003\\
0.0082  & 0.00024       & 0.00024\\
0.00764 &	0.00025 &	0.00027\\\hline
\multicolumn{3}{c}{Transit 5, TRAPPIST-1 e}\\
0.00477 &	0.00016 &	0.00017\\
0.0046  & 0.00012       & 0.00012\\
0.00468 &	0.00013 &	0.00012\\
0.00482 &	0.00012 &	0.00013\\
0.00451 &	0.00013 &	0.00015\\
0.00467 &	0.00019 &	0.00019\\
0.00486 &	0.00017 &	0.00016\\
0.00507 &	0.00015 &	0.00015\\
0.00487 &	0.00018 &	0.00017\\
0.00495 &	0.00014 &	0.00015\\
0.00496 &	0.00016 &	0.00014\\\hline
\multicolumn{3}{c}{Transit 6, TRAPPIST-1 f}\\
0.00651 &	0.00021 &	0.0002\\
0.00659 &	0.00022 &	0.00021\\
0.00639 &	0.00024 &	0.00024\\
0.00666 &	0.00022 &	0.00021\\
0.00656 &	0.00023 &	0.00022\\
0.00631 &	0.00025 &	0.00026\\
0.00641 &	0.00026 &	0.00025\\
0.00604 &	0.00022 &	0.00021\\
0.00659 &	0.0002  & 0.00019\\
0.0064  & 0.00021       & 0.0002\\
0.00675 &	0.00025 &	0.00024\\\hline
\multicolumn{3}{c}{Transit 7, TRAPPIST-1 d}\\
0.00531 &	0.00016 &	0.00016\\
0.00526 &	0.00014 &	0.00015\\
0.00545 &	0.00013 &	0.00013\\
0.00516 &	0.00013 &	0.00013\\
0.00513 &	0.00015 &	0.00014\\
0.00479 &	0.00018 &	0.00017\\
0.00491 &	0.00014 &	0.00014\\
0.00494 &	0.00012 &	0.00012\\
0.00502 &	0.00013 &	0.00014\\
0.00511 &	0.00011 &	0.00012\\
0.00504 &	0.00019 &	0.00019\\
  \enddata
\tablecomments{The central wavelengths of the eleven bins are an arithmetic series from 11500 \AA{} to 16500 \AA.}
\end{deluxetable}

\clearpage
\onecolumngrid
\section{Supplementary Figures}

\begin{figure*}[!htbp]
  \centering
  \plotone{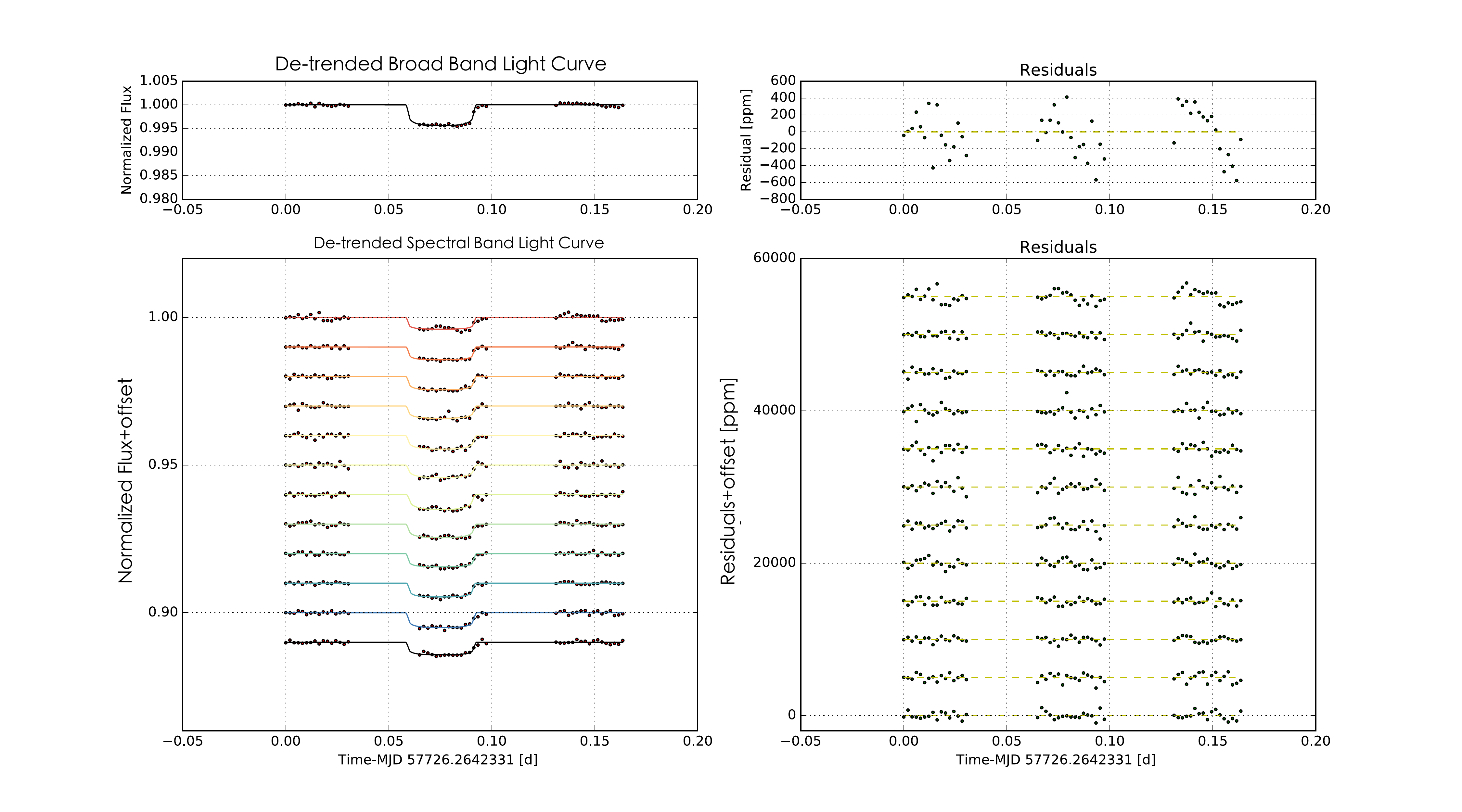}
  \caption{Same as Figure \ref{fig:lc} for Visit 1, broadband and spectral band fit for planet \trone{} d.}
  \label{fig:12}
\end{figure*}

\begin{figure*}[!htbp]
  \centering
  \plotone{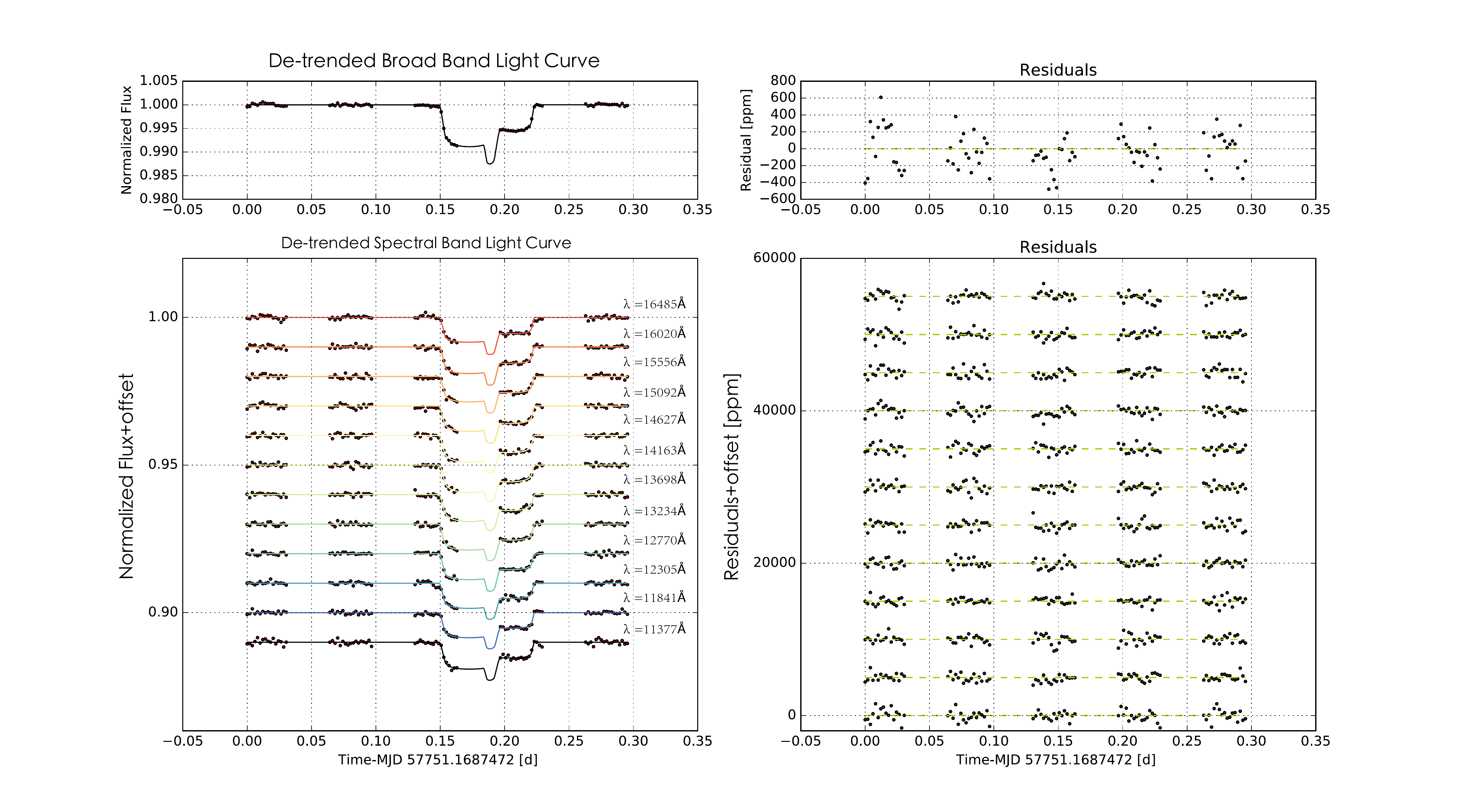}
  \caption{Same as Figure \ref{fig:lc} for Visit 2, broadband and spectral band fit for planet \trone{} g, e.} 
   \label{fig:13}
\end{figure*}

\begin{figure*}[!htpb]
  \centering
  \plotone{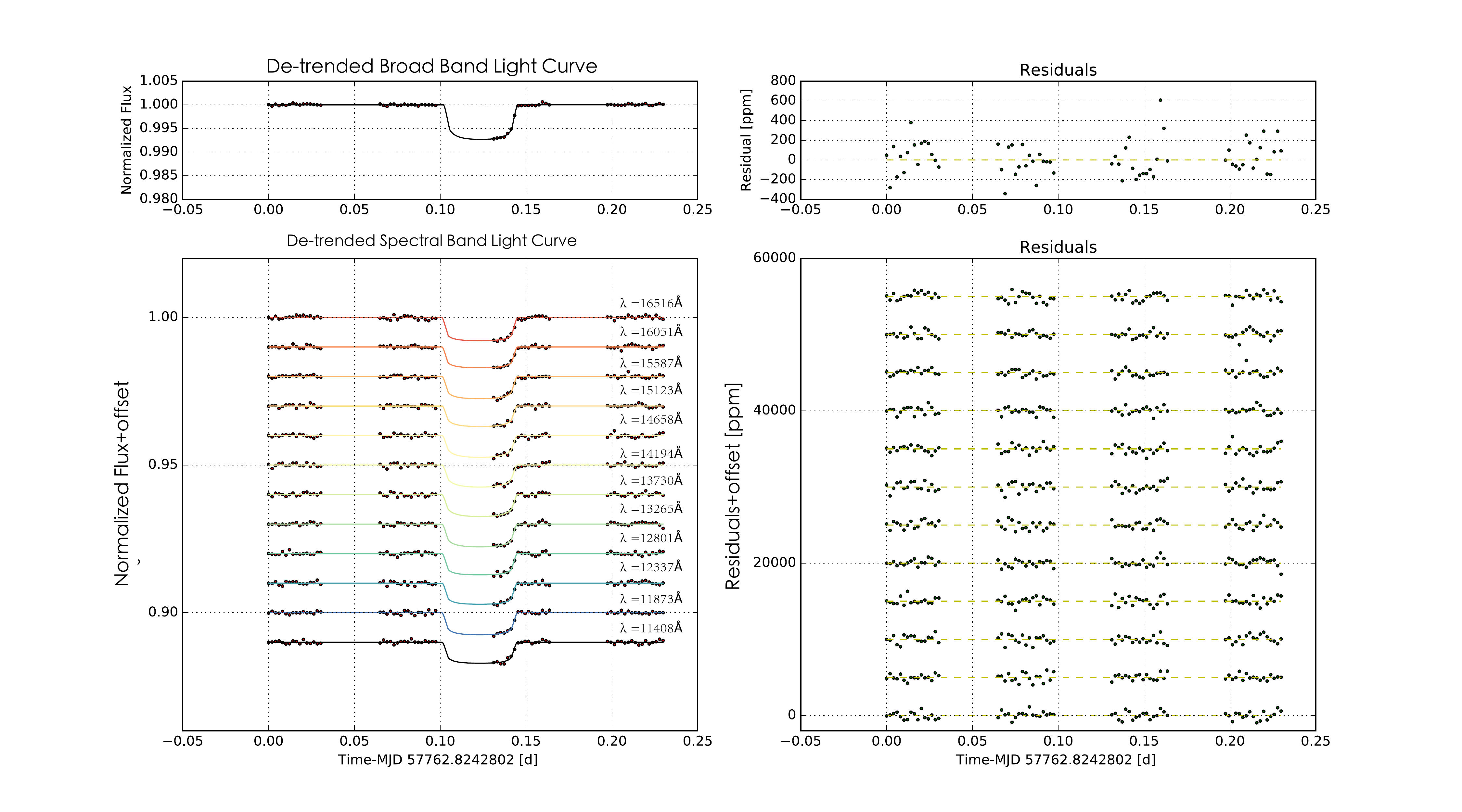}
  \caption{Same as Figure \ref{fig:lc} for Visit 3, broadband and spectral band fit for planet \trone{} f.}
  \label{fig:14}
\end{figure*}

\begin{figure*}[!htbp]
  \centering
  \plotone{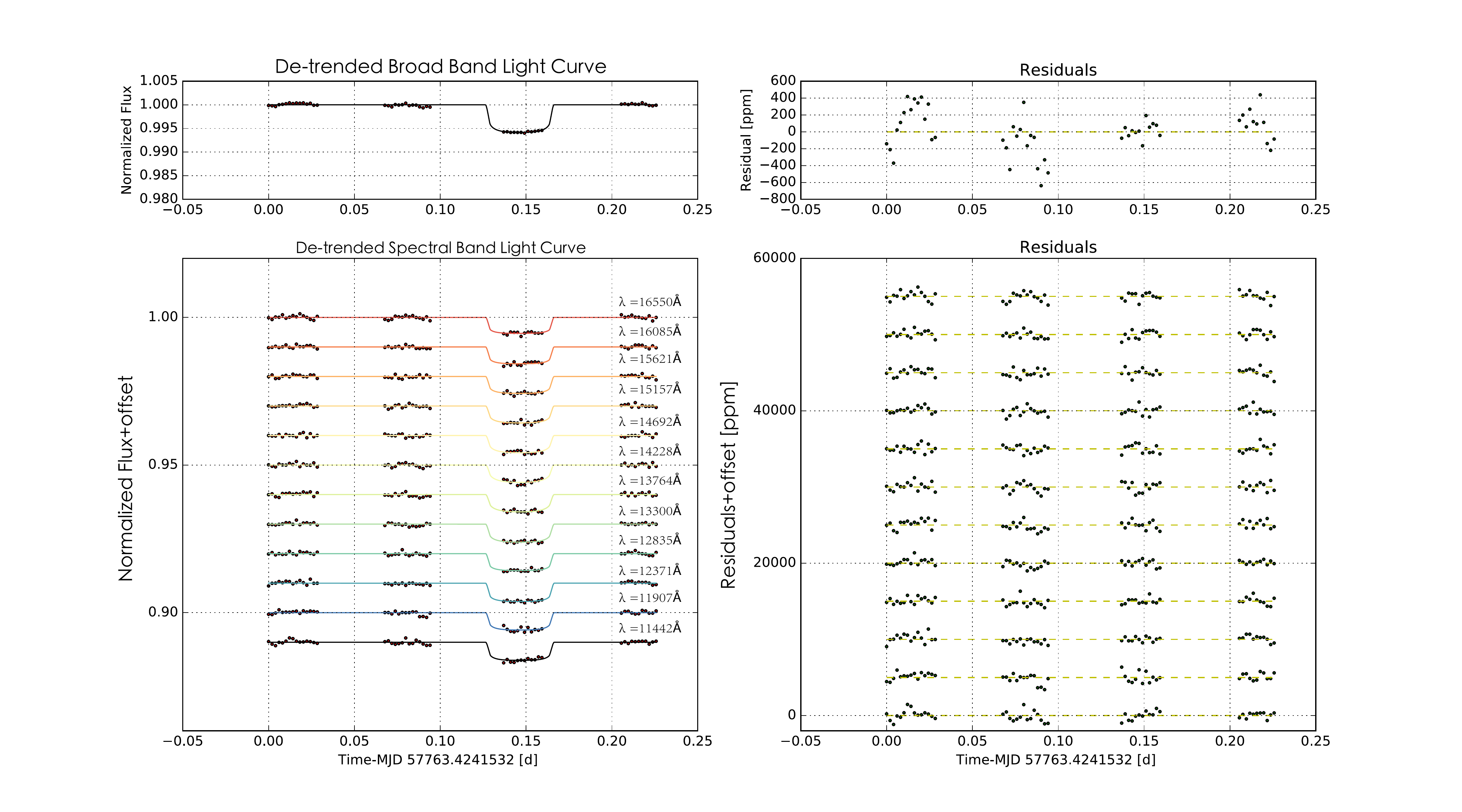}
   \caption{Same as Figure \ref{fig:lc} for Visit 4, broadband and spectral band fit for planet \trone{} e.}
   \label{fig:15}
 \end{figure*}

\clearpage

\section{Analysis of HST+Spitzer Transmission Spectra}
\label{sec:HST+Spitzer}

\begin{figure*}[!htbp]
\label{fig:CPAT_oot_spectra}
\includegraphics[width=\linewidth]{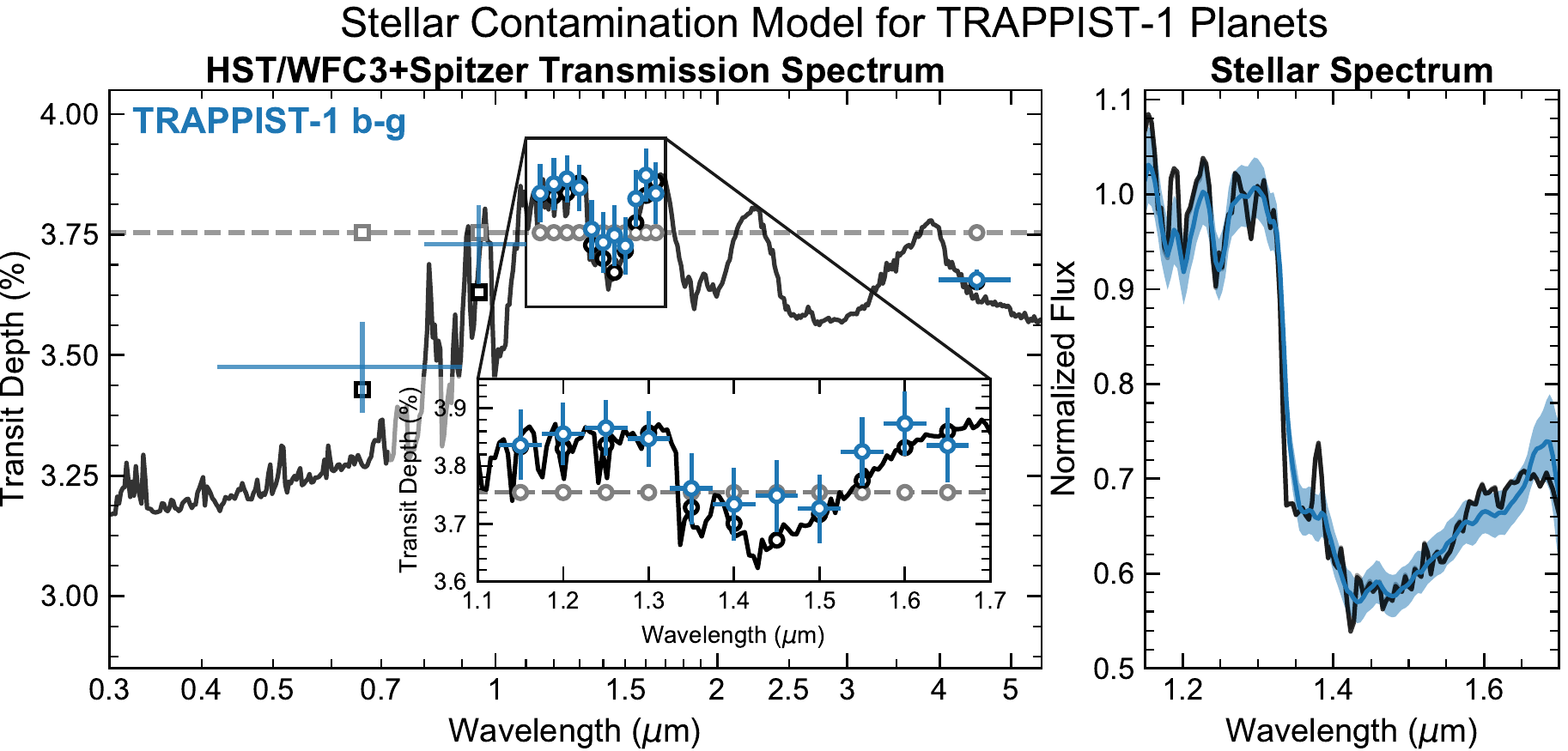}
\caption{
Stellar contamination model jointly fit to HST+Spitzer \trone{} combined transmission spectra and observed \textit{HST} stellar spectrum.
The unfitted \textit{K2} (0.42--0.9~$\micron$) and \textit{I+z} (0.8--1.1~$\micron$) combined transit depths \citep{Ducrot2018} are overplotted as transparent blue crosses, along with the predictions for the \texttt{flat} and \texttt{contamination} models (gray and black squares, respectively) for comparison.
The remaining figure elements are the same as those in Figure~\ref{fig:CPAT_oot_spectra_all_data}.
}
\end{figure*}

\begin{figure*}[!htbp]
\label{fig:CPAT_fits_five_planets}
\includegraphics[width=\linewidth]{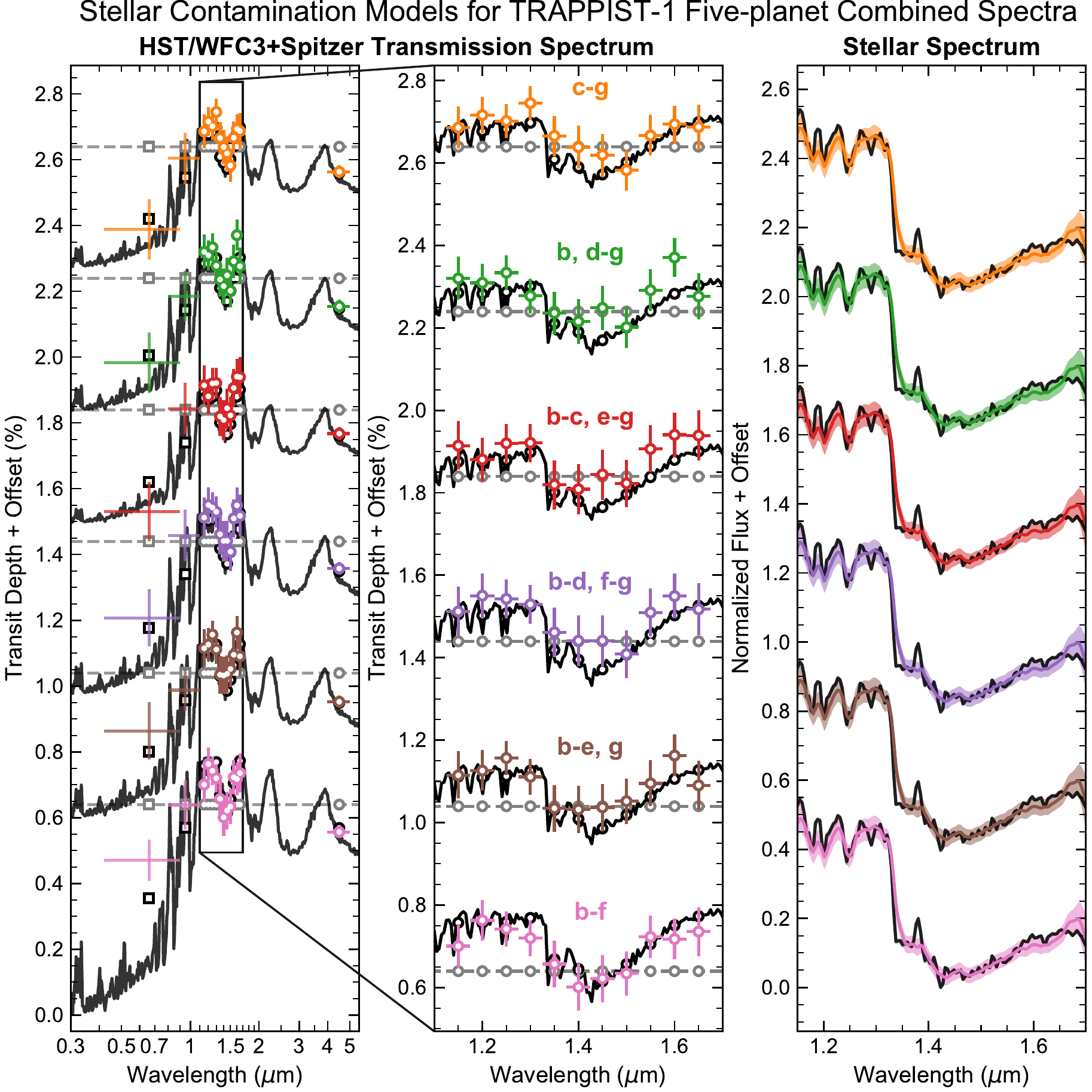}
\caption{
Stellar contamination models jointly fit to HST+Spitzer \trone{} five-planet combined transmission spectra and observed \textit{HST} stellar spectrum. 
The unfitted \textit{K2} (0.42--0.9~$\micron$) and \textit{I+z} (0.8--1.1~$\micron$) combined transit depths \citep{Ducrot2018} are overplotted as transparent colored crosses, along with the predictions for the \texttt{flat} and \texttt{contamination} models (gray and black squares, respectively) for comparison.
The remaining figure elements are the same as those in Figure~\ref{fig:CPAT_fits_five_planets_all_data}. 
}
\end{figure*}

\begin{figure*}[htbp]
\label{fig:CPAT_corner}
\plottwo{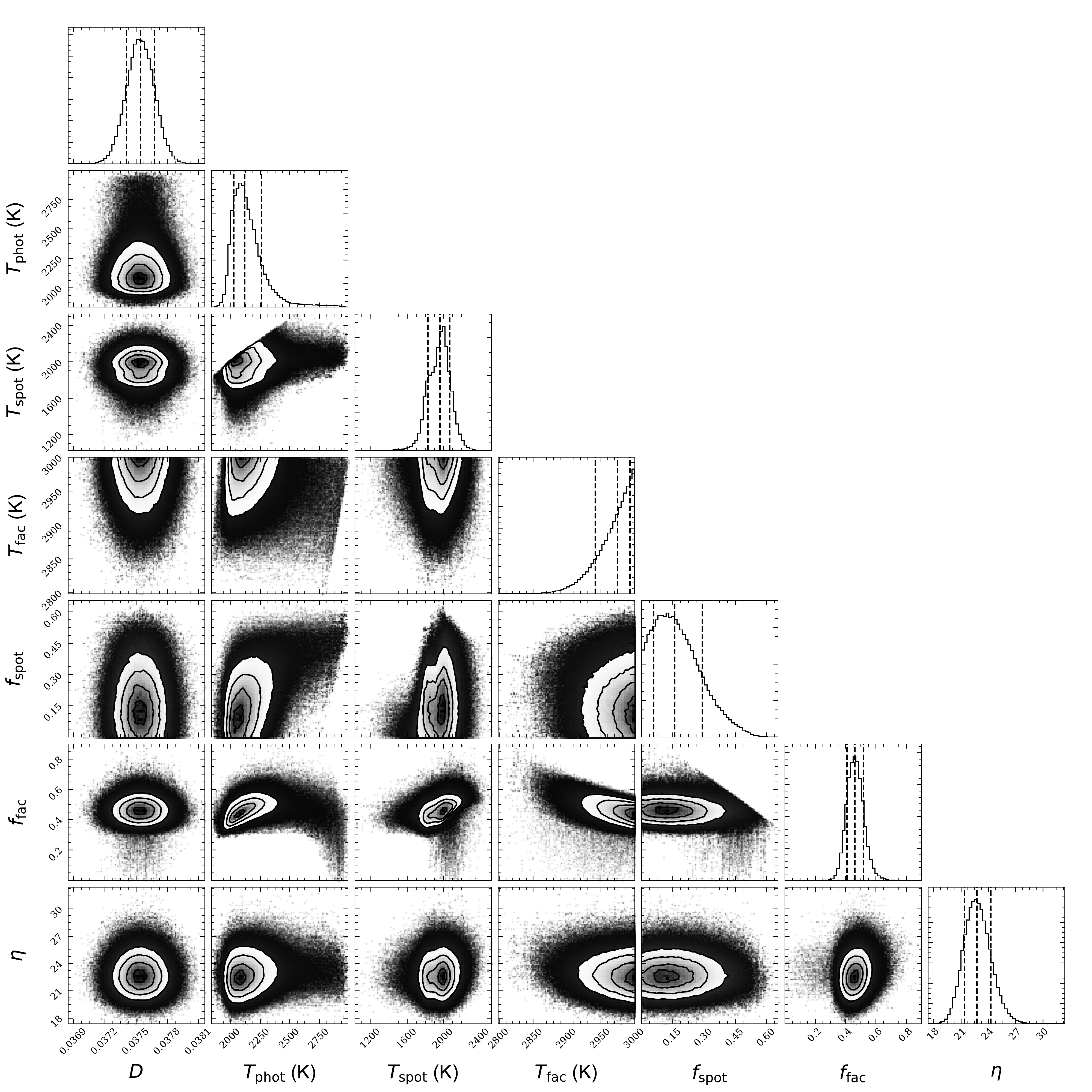}{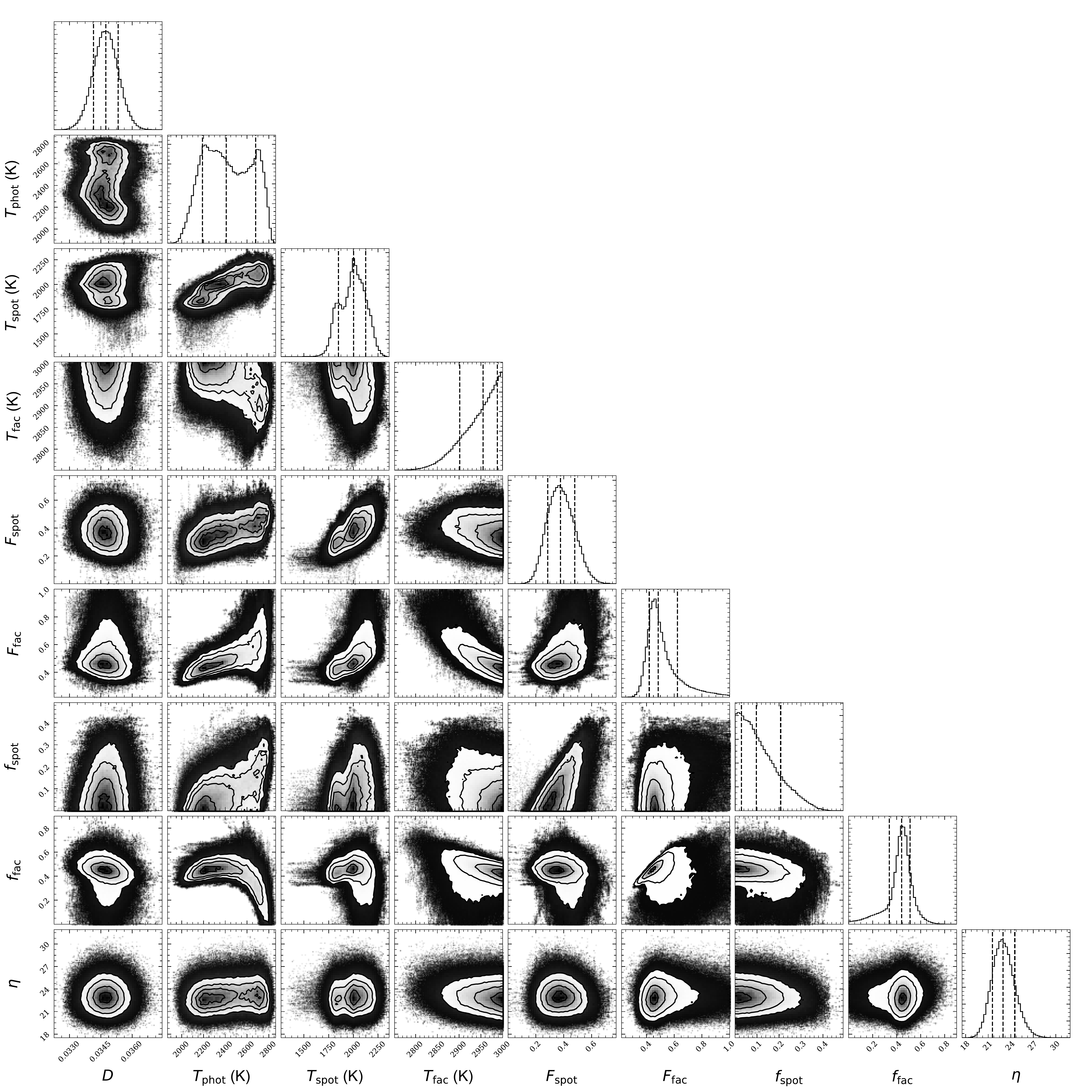}
\caption{
Posterior distributions of free parameters in CPAT model fits to HST+Spitzer combined transmission spectrum of \trone{} b--g and observed \textit{HST} stellar spectrum. 
Results are shown for the \texttt{flat} and \texttt{contamination} models in the left and right panels, respectively.
None of the parameter distributions differ significantly from those in Figure~\ref{fig:CPAT_corner_all_data}, which were fitted using the complete K2+SSO+HST+Spitzer dataset.
}
\end{figure*}

Here we summarize the results of the same stellar contamination analysis described in Section~\ref{Section:StellarContamination} as originally performed using transmission spectra comprised of only the HST transit depths from this work and the Spitzer 4.5~$\micron$ transit depths from \citet{Delrez2018}.
Tables~\ref{tab:results_CPAT_single} and \ref{tab:results_CPAT_combined} summarize the results for the single-planet and combined transmission spectra, respectively.
As with the full analysis in Section~\ref{Section:StellarContamination}, the \texttt{contamination} models are preferred for the \trone{}~d single-planet spectrum and all variations of the combined spectra.
The predictions for \textit{K2} and \textit{I+z} transit depths from the best-fit \texttt{flat} and \texttt{contamination} to the combined transmission spectra are provided in Table~\ref{tab:predictions}.
These are shown in Figures~\ref{fig:CPAT_oot_spectra} and \ref{fig:CPAT_fits_five_planets}, along with the HST+Spitzer data, best-fit models, and \textit{K2} and \textit{I+z} depths from \citet{Ducrot2018}.
Examples of the posterior distributions for the HST+Spitzer fits are provided in Figure~\ref{fig:CPAT_corner}.

\begin{deluxetable*}{lcccccccccccccccc}[bp]
\tablenum{15}
\tabletypesize{\scriptsize}
\tablecaption{
This table has been moved from the main text. Columns for $\Delta$\,AIC and $\Delta$\,BIC were added.
Results of Stellar Contamination Model Fits to Single-Planet Spectra Using Only HST and Spitzer Data \label{tab:results_CPAT_single}}
\tablehead{\colhead{Dataset}                    & 
           \colhead{Model}                      & 
           \multicolumn{9}{c}{Fitted Parameter} & 
           \colhead{AIC$_\textrm{c}$}           &
           \colhead{$\Delta$\,AIC$_\textrm{c}$} &
           \colhead{BIC}                        &
           \colhead{$\Delta$ BIC}                        &
           \colhead{$\chi^{2}$}                 &
           \colhead{$p$}                        \\
           \cline{3-11} 
           \colhead{}                           & 
           \colhead{}                           & 
           \colhead{$D$ (\%)}                   & 
           \colhead{$T_\mathrm{phot}$ (K)}      & 
           \colhead{$T_\mathrm{spot}$ (K)}      & 
           \colhead{$T_\mathrm{fac}$ (K)}       & 
           \colhead{$F_\mathrm{spot}$}          & 
           \colhead{$F_\mathrm{fac}$}           & 
           \colhead{$f_\mathrm{spot}$}          & 
           \colhead{$f_\mathrm{fac}$}           & 
           \colhead{$\eta$}                     & 
           \colhead{}                           & 
           \colhead{}                           &
           \colhead{}                           & 
           \colhead{}                           & 
           \colhead{}                           & 
           \colhead{}}
         
\startdata
b                       & \texttt{flat}  & ${0.745}^{+0.006}_{-0.006}$ & ${2117}^{+85}_{-128}$  & ${1961}^{+108}_{-134}$ & ${2974}^{+26}_{-14}$ & ${16}^{+7}_{-15}$  & ${46}^{+5}_{-6}$   & -                 & -                  & ${23}^{+1}_{-2}$ & -707.8 & -8.6 & -687.9 & -14.0 & 148.3 & 0.21 \\
b                       & \texttt{cont.} & ${0.675}^{+0.021}_{-0.023}$ & ${2210}^{+190}_{-241}$ & ${1980}^{+119}_{-136}$ & ${2964}^{+36}_{-20}$ & ${29}^{+16}_{-16}$ & ${47}^{+5}_{-8}$   & ${10}^{+4}_{-10}$ & ${52}^{+11}_{-9}$  & ${23}^{+1}_{-2}$ & -699.2 & -    & -673.9 & -     & 151.5 & 0.13 \\
c                       & \texttt{flat}  & ${0.705}^{+0.005}_{-0.005}$ & ${2118}^{+87}_{-126}$  & ${1961}^{+107}_{-134}$ & ${2974}^{+26}_{-14}$ & ${16}^{+7}_{-15}$  & ${46}^{+5}_{-6}$   & -                 & -                  & ${23}^{+1}_{-2}$ & -718.6 & -6.8 & -698.8 & -12.2 & 139.5 & 0.38 \\
c                       & \texttt{cont.} & ${0.663}^{+0.020}_{-0.020}$ & ${2227}^{+154}_{-236}$ & ${1973}^{+98}_{-160}$  & ${2965}^{+35}_{-19}$ & ${29}^{+15}_{-13}$ & ${47}^{+5}_{-7}$   & ${9}^{+4}_{-9}$   & ${48}^{+9}_{-9}$   & ${23}^{+1}_{-2}$ & -711.8 & -    & -686.6 & -     & 140.4 & 0.31 \\
d                       & \texttt{flat}  & ${0.388}^{+0.003}_{-0.003}$ & ${2117}^{+86}_{-127}$  & ${1961}^{+109}_{-132}$ & ${2974}^{+26}_{-14}$ & ${16}^{+7}_{-15}$  & ${46}^{+5}_{-6}$   & -                 & -                  & ${23}^{+1}_{-2}$ & -696.4 & 25.7 & -676.5 & 20.4  & 178.5 & 0.01 \\
d                       & \texttt{cont.} & ${0.309}^{+0.015}_{-0.016}$ & ${2653}^{+151}_{-98}$  & ${2030}^{+91}_{-62}$   & ${2931}^{+62}_{-31}$ & ${53}^{+11}_{-8}$  & ${54}^{+12}_{-16}$ & ${9}^{+4}_{-9}$   & ${42}^{+19}_{-18}$ & ${23}^{+1}_{-2}$ & -722.1 & -    & -696.9 & -     & 149.5 & 0.16 \\
e (T5)\tablenotemark{a} & \texttt{flat}  & ${0.478}^{+0.004}_{-0.004}$ & ${2117}^{+87}_{-125}$  & ${1961}^{+109}_{-133}$ & ${2974}^{+26}_{-14}$ & ${16}^{+7}_{-15}$  & ${46}^{+5}_{-6}$   & -                 & -                  & ${23}^{+1}_{-2}$ & -732.6 & -4.2 & -712.8 & -9.6  & 139.6 & 0.38 \\
e (T5)\tablenotemark{a} & \texttt{cont.} & ${0.493}^{+0.021}_{-0.021}$ & ${2123}^{+82}_{-120}$  & ${1982}^{+119}_{-120}$ & ${2973}^{+27}_{-15}$ & ${15}^{+8}_{-12}$  & ${47}^{+5}_{-6}$   & ${12}^{+5}_{-12}$ & ${44}^{+6}_{-7}$   & ${23}^{+1}_{-2}$ & -728.4 & -    & -703.2 & -     & 139.0 & 0.34 \\
e (T7)\tablenotemark{b} & \texttt{flat}  & ${0.506}^{+0.004}_{-0.004}$ & ${2117}^{+86}_{-125}$  & ${1960}^{+105}_{-136}$ & ${2974}^{+26}_{-14}$ & ${16}^{+7}_{-15}$  & ${46}^{+5}_{-5}$   & -                 & -                  & ${23}^{+1}_{-2}$ & -721.4 & 6.6  & -701.5 & 1.3   & 152.0 & 0.15 \\
e (T7)\tablenotemark{b} & \texttt{cont.} & ${0.447}^{+0.021}_{-0.022}$ & ${2559}^{+187}_{-135}$ & ${2030}^{+113}_{-81}$  & ${2941}^{+59}_{-27}$ & ${46}^{+11}_{-10}$ & ${51}^{+9}_{-12}$  & ${10}^{+4}_{-10}$ & ${39}^{+14}_{-12}$ & ${23}^{+1}_{-2}$ & -728.0 & -    & -702.8 & -     & 139.2 & 0.34 \\
e\tablenotemark{c}      & \texttt{flat}  & ${0.494}^{+0.003}_{-0.003}$ & ${2116}^{+88}_{-124}$  & ${1960}^{+108}_{-133}$ & ${2974}^{+26}_{-14}$ & ${16}^{+7}_{-15}$  & ${46}^{+5}_{-6}$   & -                 & -                  & ${23}^{+1}_{-2}$ & -742.8 & -7.5 & -723.0 & 13.0  & 137.7 & 0.42 \\
e\tablenotemark{c}      & \texttt{cont.} & ${0.479}^{+0.021}_{-0.023}$ & ${2214}^{+127}_{-178}$ & ${1970}^{+98}_{-161}$  & ${2968}^{+32}_{-18}$ & ${27}^{+11}_{-11}$ & ${47}^{+5}_{-7}$   & ${9}^{+4}_{-9}$   & ${45}^{+6}_{-7}$   & ${23}^{+1}_{-2}$ & -735.3 & -    & -710.0 & -     & 139.1 & 0.34 \\
f                       & \texttt{flat}  & ${0.643}^{+0.006}_{-0.005}$ & ${2116}^{+85}_{-126}$  & ${1961}^{+109}_{-132}$ & ${2974}^{+26}_{-14}$ & ${16}^{+7}_{-15}$  & ${46}^{+5}_{-6}$   & -                 & -                  & ${23}^{+1}_{-2}$ & -728.5 & -8.3 & -708.6 & -13.7 & 134.8 & 0.49 \\
f                       & \texttt{cont.} & ${0.625}^{+0.025}_{-0.025}$ & ${2196}^{+116}_{-193}$ & ${1984}^{+112}_{-141}$ & ${2969}^{+31}_{-17}$ & ${24}^{+12}_{-13}$ & ${48}^{+5}_{-6}$   & ${10}^{+4}_{-10}$ & ${46}^{+7}_{-7}$   & ${23}^{+1}_{-2}$ & -720.2 & -    & -694.9 & -     & 137.1 & 0.39 \\
g                       & \texttt{flat}  & ${0.776}^{+0.006}_{-0.006}$ & ${2117}^{+86}_{-128}$  & ${1961}^{+111}_{-131}$ & ${2974}^{+26}_{-14}$ & ${16}^{+7}_{-15}$  & ${46}^{+5}_{-6}$   & -                 & -                  & ${23}^{+1}_{-2}$ & -726.3 & -4.7 & -706.4 & -10.3 & 134.7 & 0.49 \\
g                       & \texttt{cont.} & ${0.747}^{+0.027}_{-0.028}$ & ${2138}^{+90}_{-138}$  & ${1981}^{+118}_{-124}$ & ${2971}^{+29}_{-15}$ & ${18}^{+9}_{-13}$  & ${47}^{+5}_{-6}$   & ${11}^{+5}_{-11}$ & ${49}^{+7}_{-7}$   & ${23}^{+1}_{-2}$ & -721.6 & -    & -696.3 & -     & 134.2 & 0.45 \\
\enddata

\tablecomments{
The table elements are the same as those in Table~\ref{tab:results_CPAT_single_all_data}.
Each model has 142 data points (12 for the transmission spectrum and 130 for the stellar spectrum). 
\texttt{Flat} models have 135 degrees of freedom and \texttt{contamination} models have 133.}
\tablenotetext{a}{Transit 5}
\tablenotetext{b}{Transit 7}
\tablenotetext{c}{Weighted mean of Transits 5 and 7}
\end{deluxetable*}


\begin{deluxetable*}{lcccccccccccccccc}
\tablenum{16}
\tabletypesize{\scriptsize}
\tablecaption{
Results of Stellar Contamination Model Fits to Combined Spectra Using HST+Spitzer Data Only 
\label{tab:results_CPAT_combined}}
\tablehead{\colhead{Combination}                & 
           \colhead{Model}                      & 
           \multicolumn{9}{c}{Fitted Parameter} & 
           \colhead{AIC$_\textrm{c}$}           &
           \colhead{$\Delta$\,AIC$_\textrm{c}$} &
           \colhead{BIC}                        &
           \colhead{$\Delta$\,BIC}               & 
           \colhead{$\chi^{2}$}                 & 
           \colhead{$p$} \\
           \cline{3-11} 
           \colhead{}                           & 
           \colhead{}                           & 
           \colhead{$D$ (\%)}                   & 
           \colhead{$T_\mathrm{phot}$ (K)}      & 
           \colhead{$T_\mathrm{spot}$ (K)}      & 
           \colhead{$T_\mathrm{fac}$ (K)}       & 
           \colhead{$F_\mathrm{spot}$}          & 
           \colhead{$F_\mathrm{fac}$}           & 
           \colhead{$f_\mathrm{spot}$}          & 
           \colhead{$f_\mathrm{fac}$}           & 
           \colhead{$\eta$}                     & 
           \colhead{}                           & 
           \colhead{}                           &
           \colhead{}                           & 
           \colhead{}                           & 
           \colhead{}                           & 
           \colhead{}}
         
\startdata
b--g       & \texttt{flat}  & ${3.754}^{+0.013}_{-0.013}$ & ${2119}^{+88}_{-125}$  & ${1961}^{+110}_{-133}$ & ${2974}^{+26}_{-14}$ & ${16}^{+7}_{-14}$  & ${46}^{+5}_{-6}$  & -                 & -                 & ${23}^{+1}_{-2}$ & -671.9 & 30.5 & -652.1 & 25.0 & 169.0 & 0.03 \\
b--g       & \texttt{cont.} & ${3.474}^{+0.059}_{-0.059}$ & ${2408}^{+235}_{-252}$ & ${2002}^{+142}_{-134}$ & ${2955}^{+45}_{-24}$ & ${38}^{+9}_{-10}$  & ${48}^{+7}_{-10}$ & ${10}^{+5}_{-10}$ & ${44}^{+9}_{-7}$  & ${23}^{+1}_{-2}$ & -702.4 & -    & -677.1 & -    & 131.3 & 0.52 \\
c--g       & \texttt{flat}  & ${3.005}^{+0.012}_{-0.011}$ & ${2116}^{+86}_{-126}$  & ${1960}^{+106}_{-135}$ & ${2974}^{+26}_{-14}$ & ${16}^{+7}_{-15}$  & ${46}^{+5}_{-6}$  & -                 & -                 & ${23}^{+1}_{-2}$ & -687.9 & 17.2 & -668.0 & 11.9 & 157.2 & 0.09 \\
c--g       & \texttt{cont.} & ${2.796}^{+0.054}_{-0.054}$ & ${2351}^{+175}_{-250}$ & ${1987}^{+120}_{-157}$ & ${2958}^{+42}_{-22}$ & ${36}^{+9}_{-10}$  & ${48}^{+6}_{-9}$  & ${9}^{+4}_{-9}$   & ${45}^{+8}_{-6}$  & ${23}^{+1}_{-2}$ & -705.1 & -    & -679.9 & -    & 132.8 & 0.49 \\
b, d--g    & \texttt{flat}  & ${3.048}^{+0.012}_{-0.012}$ & ${2118}^{+88}_{-126}$  & ${1962}^{+112}_{-130}$ & ${2974}^{+26}_{-14}$ & ${16}^{+7}_{-15}$  & ${46}^{+5}_{-6}$  & -                 & -                 & ${23}^{+1}_{-2}$ & -680.4 & 20.2 & -660.6 & 14.7 & 163.9 & 0.05 \\
b, d--g    & \texttt{cont.} & ${2.810}^{+0.054}_{-0.055}$ & ${2377}^{+198}_{-275}$ & ${1997}^{+117}_{-163}$ & ${2957}^{+43}_{-23}$ & ${37}^{+10}_{-10}$ & ${48}^{+7}_{-9}$  & ${9}^{+4}_{-9}$   & ${45}^{+9}_{-7}$  & ${23}^{+1}_{-2}$ & -700.6 & -    & -675.3 & -    & 136.7 & 0.40 \\
b--c, e--g & \texttt{flat}  & ${3.372}^{+0.013}_{-0.013}$ & ${2116}^{+88}_{-126}$  & ${1959}^{+108}_{-132}$ & ${2974}^{+26}_{-14}$ & ${16}^{+7}_{-15}$  & ${46}^{+5}_{-6}$  & -                 & -                 & ${23}^{+1}_{-2}$ & -687.1 & 14.4 & -667.2 & 9.0  & 154.7 & 0.12 \\
b--c, e--g & \texttt{cont.} & ${3.168}^{+0.057}_{-0.054}$ & ${2294}^{+152}_{-223}$ & ${1956}^{+97}_{-165}$  & ${2962}^{+38}_{-21}$ & ${33}^{+8}_{-10}$  & ${47}^{+5}_{-8}$  & ${8}^{+4}_{-8}$   & ${45}^{+7}_{-6}$  & ${23}^{+1}_{-2}$ & -701.5 & -    & -676.2 & -    & 133.5 & 0.47 \\
b--d, f--g & \texttt{flat}  & ${3.259}^{+0.013}_{-0.013}$ & ${2116}^{+85}_{-126}$  & ${1961}^{+109}_{-131}$ & ${2974}^{+26}_{-14}$ & ${16}^{+7}_{-15}$  & ${46}^{+5}_{-6}$  & -                 & -                 & ${23}^{+1}_{-2}$ & -675.9 & 25.8 & -656.0 & 20.4 & 165.6 & 0.04 \\
b--d, f--g & \texttt{cont.} & ${3.007}^{+0.054}_{-0.054}$ & ${2392}^{+223}_{-269}$ & ${1997}^{+128}_{-151}$ & ${2956}^{+44}_{-24}$ & ${37}^{+10}_{-10}$ & ${48}^{+7}_{-10}$ & ${9}^{+5}_{-9}$   & ${45}^{+9}_{-7}$  & ${23}^{+1}_{-2}$ & -701.7 & -    & -676.4 & -    & 133.0 & 0.48 \\
b--e, g    & \texttt{flat}  & ${3.110}^{+0.012}_{-0.012}$ & ${2115}^{+88}_{-122}$  & ${1959}^{+108}_{-132}$ & ${2974}^{+26}_{-14}$ & ${16}^{+7}_{-15}$  & ${46}^{+5}_{-6}$  & -                 & -                 & ${23}^{+1}_{-2}$ & -673.1 & 30.9 & -653.2 & 25.5 & 170.1 & 0.02 \\
b--e, g    & \texttt{cont.} & ${2.850}^{+0.051}_{-0.053}$ & ${2406}^{+230}_{-238}$ & ${2001}^{+140}_{-127}$ & ${2954}^{+46}_{-25}$ & ${38}^{+9}_{-10}$  & ${48}^{+7}_{-9}$  & ${9}^{+4}_{-9}$   & ${45}^{+9}_{-7}$  & ${23}^{+1}_{-2}$ & -704.0 & -    & -678.7 & -    & 131.9 & 0.51 \\
b--f       & \texttt{flat}  & ${2.976}^{+0.012}_{-0.012}$ & ${2118}^{+87}_{-126}$  & ${1961}^{+112}_{-128}$ & ${2974}^{+26}_{-14}$ & ${16}^{+7}_{-15}$  & ${46}^{+5}_{-6}$  & -                 & -                 & ${23}^{+1}_{-2}$ & -672.2 & 31.3 & -652.3 & 26.0 & 171.4 & 0.02 \\
b--f       & \texttt{cont.} & ${2.733}^{+0.050}_{-0.052}$ & ${2502}^{+259}_{-153}$ & ${2028}^{+151}_{-91}$  & ${2950}^{+50}_{-25}$ & ${41}^{+10}_{-10}$ & ${50}^{+8}_{-10}$ & ${10}^{+5}_{-10}$ & ${43}^{+11}_{-8}$ & ${23}^{+1}_{-2}$ & -703.5 & -    & -678.3 & -    & 132.7 & 0.49 \\
\enddata

\tablecomments{
The table elements are the same as those in Table~\ref{tab:results_CPAT_combined_all_data}.
Each model has 142 data points (12 for the transmission spectrum and 130 for the stellar spectrum). 
\texttt{Flat} models have 135 degrees of freedom and \texttt{contamination} models have 133.}
\end{deluxetable*}

\begin{deluxetable*}{lccc}
\tablenum{17}
\tabletypesize{\small}
\tablecaption{Predictions for \textit{K2} (0.42--0.9~$\micron$) and \textit{I+z} (0.8--1.1~$\micron$) Transit Depths from Fits to Combined HST+Spitzer Transmission Spectra 
\label{tab:predictions}}
\tablehead{\colhead{Combination}                    & 
           \colhead{Model}                      & 
           \colhead{\textit{K2} Transit Depth (\%)} &
           \colhead{\textit{I+z} Transit Depth (\%)} \\
           \colhead{} &
           \colhead{} &
           \colhead{(0.42--0.9~$\micron$)} &
           \colhead{(0.8--1.1~$\micron$)}
           }
\startdata
b--g       & \texttt{flat}  & 3.75 & 3.75 \\
b--g       & \texttt{cont.} & 3.43 & 3.63 \\
c--g       & \texttt{flat}  & 3.00 & 3.00 \\
c--g       & \texttt{cont.} & 2.79 & 2.91 \\
b, d--g    & \texttt{flat}  & 3.05 & 3.05 \\
b, d--g    & \texttt{cont.} & 2.81 & 2.95 \\
b--c, e--g & \texttt{flat}  & 3.37 & 3.37 \\
b--c, e--g & \texttt{cont.} & 3.15 & 3.27 \\
b--d, f--g & \texttt{flat}  & 3.26 & 3.26 \\
b--d, f--g & \texttt{cont.} & 3.00 & 3.16 \\
b--e, g    & \texttt{flat}  & 3.11 & 3.11 \\
b--e, g    & \texttt{cont.} & 2.87 & 3.03 \\
b--f       & \texttt{flat}  & 2.98 & 2.98 \\
b--f       & \texttt{cont.} & 2.69 & 2.91 \\
\enddata
\end{deluxetable*}


\end{document}